\documentclass[12pt]{article}
\pdfoutput=1

\usepackage{epsfig}
\usepackage{psfrag}
\usepackage{latexsym}
\usepackage{indentfirst}
\usepackage{fancyhdr}
\usepackage{dsfont}
\usepackage{amsmath}
\usepackage{amssymb}
\usepackage{amsfonts}
\usepackage{mathrsfs}
\usepackage{amsthm}
\usepackage{pifont}
\usepackage{dsfont}
\usepackage{multirow}
\usepackage{array}
\usepackage{chngpage}
\usepackage{longtable}
\usepackage{diagbox}
\usepackage{cite}
\usepackage{bbold}
\usepackage{color}
\usepackage{colordvi}
\usepackage{fancybox}
\usepackage[footnotesize]{caption2}
\usepackage{graphicx}
\usepackage[center,footnotesize,hang]{subfigure}
\usepackage{bbm}
\usepackage[colorlinks, linkcolor=red, anchorcolor=black, citecolor=green]{hyperref}
\usepackage{multirow}
\usepackage{array}
\usepackage{textcomp}  
\usepackage{ulem}
\usepackage{enumitem}
\usepackage{mathrsfs}  
\newcommand{\PreserveBackslash}[1]{\let\temp=\\#1\let\\=\temp}
\newcolumntype{C}[1]{>{\PreserveBackslash\centering}p{#1}}
\newcolumntype{R}[1]{>{\PreserveBackslash\raggedleft}p{#1}}
\newcolumntype{L}[1]{>{\PreserveBackslash\raggedright}p{#1}}
\allowdisplaybreaks  

\newcommand{\bq}{\begin{eqnarray}}
\newcommand{\nq}{\end{eqnarray}}

\makeatletter
\@addtoreset{equation}{section}
\makeatother

\begin{document}

\title{\hfill ~\\[0mm]
        \textbf{Lepton and Quark Mixing Patterns from Finite Flavor Symmetries}}
\date{}

\author{\\[1mm]Chang-Yuan Yao\footnote{E-mail: {\tt phyman@mail.ustc.edu.cn}}~,~Gui-Jun Ding\footnote{E-mail: {\tt dinggj@ustc.edu.cn}}\\ \\
\it{\small Department of Modern Physics, University of Science and
    Technology of China,}\\
  \it{\small Hefei, Anhui 230026, China}\\[4mm] }
\maketitle

\begin{abstract}

We perform a systematical and analytical study of lepton mixing which can be derived from the subgroups of $SU(3)$ under the assumption that neutrinos are Dirac particles. We find that type D groups can predict lepton mixing patterns compatible with the experimental data at $3\sigma$ level. The lepton mixing matrix turns out to be of the trimaximal form, and the Dirac $CP$ violating phase is trivial. Moreover, we extend the flavor symmetry to the quark sector. The Cabibbo mixing between the first two generations of quarks can be generated by type D groups. Since all the finite subgroups of $U(3)$ that are not the subgroups of $SU(3)$ have not been classified, an exhaustive scan over all finite discrete groups up to order 2000 is performed with the help of the computer algebra system \texttt{GAP}. We find that only 90 (10) groups for Dirac (Majorana) neutrinos can generate the lepton mixing angles in the experimentally preferred ranges. The lepton mixing matrix is still the trimaximal pattern and the Dirac $CP$ phase remains trivial. The smallest groups that lead to viable mixing angles are $[162, 10]$, $[162, 12]$ and $[162, 14]$. For quark flavor mixing, the correct order of magnitude of the Cabibbo-Kobayashi-Maskawa matrix elements can not be generated. Only the Cabibbo mixing is allowed even if we impose very loose constraints $0.1\leq |V_{us}|\leq0.3$ and $|V_{ub}|\leq |V_{cb}|<|V_{us}|$. The group $\Delta(6\times7^2)$ can predict a Cabibbo angle $\theta_q=\pi/14$ in good agreement with the best fit value. The observed Cabibbo mixing angle can easily be accommodated if the first two left-handed quark fields are assigned to a doublet. The groups that can give rise to both phenomenologically viable lepton mixing angles and acceptable Cabibbo angles are discussed, and the groups $\Delta(6\times9^2)$, $[648, 259]$, $[648, 260]$, $[648, 266]$, and $\Delta(6\times14^2)$ are especially promising in the case of the triplet assignment for both quark and lepton sectors. The three groups $\left[496, 19\right]$, $\left[496, 21\right]$, and $\left[496, 23\right]$ are interesting candidates for flavor symmetry if the left-handed lepton and quark fields are assigned to triplet and doublet plus singlet, respectively.

\end{abstract}
\thispagestyle{empty}
\vfill

\newpage
\setcounter{page}{1}

\section{Introduction}

It has been firmly established that flavor mixing exists in both quark and lepton sectors, and the quark and lepton mixing structures are described by the Cabibbo-Kobayashi-Maskawa (CKM) matrix and the Pontecorvo-Maki-Nakagawa-Sakata (PMNS) matrix, respectively~\cite{pdg}. It is well known that the CKM matrix is nearly diagonal and its off-diagonal elements are rather small. On the other hand, each entry of the PMNS matrix is of order one except the (13) element, which is related with the reactor mixing angle $\theta_{13}$~\cite{Capozzi:2013csa,Forero:2014bxa,Gonzalez-Garcia:2014bfa}. Furthermore, the $CP$ violating phase in the quark sector has been precisely measured, while the value of the leptonic $CP$ phase is not known so far. Obviously the mixing patterns in quark and lepton sectors are quite different. This is the so-called flavor puzzle of the standard model, yet the underlying origin is not revealed.

There have been many attempts in the literature to explain the observed flavor mixing structures. The most popular approach is to invoke some flavor symmetry acting on different families to describe the observed patterns. In particular, the finite discrete symmetry group was extensively explored to understand the leptonic mixing angles (please see Refs.~\cite{Altarelli:2010gt,Ishimori:2010zr,Grimus:2011fk,King:2013eh,King:2014nza}  for some recent reviews). In this setup, a discrete flavor symmetry group is broken down to different subgroups in the neutrino and charged lepton sectors and the mismatch between the two subgroups determines the PMNS matrix. Conversely, the flavor group can be reconstructed from the residual symmetries of the mass matrices. This approach was first motivated by the distinctive tri-bimaximal mixing pattern ~\cite{Lam:2007qc}. After the reactor angle $\theta_{13}$ is precisely measured to be nonvanishing by T2K~\cite{Abe:2011sj}, MINOS~\cite{Adamson:2011qu}, Double-Chooz~\cite{Abe:2011fz}, RENO~\cite{Ahn:2012nd}, and DAYA-BAY~\cite{An:2012eh}, much effort has been devoted to searching for finite flavor symmetry groups that can give rise to neutrino mixing angles in agreement with experimental data. Under the assumption of Majorana neutrinos, a comprehensive scan of $SU(3)$ subgroups with order less than 512~\cite{Lam:2012ga} as a flavor symmetry has been performed. A more complete search for $U(3)$ subgroups of order smaller than 1536 has been undertaken~\cite{Holthausen:2012wt}. It is amazing that only three groups $\Delta(6\times 10^2)$, $(Z_{18}\times Z_6)\rtimes S_3$, and $\Delta(6\times 16^2)$ can lead to acceptable mixing patterns~\cite{Holthausen:2012wt}. Discrete groups of order up to 200~\cite{Holthausen:2013vba} and the ``exceptional'' finite groups~\cite{Hagedorn:2013nra} have also been investigated with the assumption of Dirac neutrinos. The discrete flavor symmetry paradigm is further extended to the quark sector to predict quark mixing angles~\cite{Holthausen:2013vba,Araki:2013rkf}. Moreover, the variant scenario of only one row or one column of the PMNS matrix constrained by flavor symmetry is discussed~\cite{Lavoura:2014kwa}. Now there appears to be some consensus that at least the $\Delta(6n^{2})$ groups are viable candidates for flavor symmetry~\cite{King:2013vna}. Instead of exploiting finite non-Abelian flavor symmetries, a systematic group space scan of discrete Abelian flavor symmetries has been performed in Ref.~\cite{Plentinger:2008up}.

Recent notable progress is that the complete lists of lepton mixing matrices that can be obtained from finite flavor symmetries have been derived for Majorana neutrinos~\cite{Fonseca:2014koa}. It is found that the mixing patterns are strongly restricted, the second column (in absolute values) of the PMNS matrix has to be $(1, 1, 1)^{T}/\sqrt{3}$ in order to accommodate the experimental data, and the Dirac $CP$ phase is trivial. In this work, we want to explore whether new mixing patterns can be achieved if neutrinos are Dirac particles instead of Majorana particles. We analytically compute all the possible mixing patterns in the case that the flavor symmetry group is a finite subgroup of $SU(3)$ or it belongs to $\Sigma(3N^3)$, which is a known subgroup series of $U(3)$. In particular, the mixing patterns that can be derived from the discrete groups $\left(Z_{3n}\times Z_{n}\right)\rtimes S_3$ will be systematically studied for the first time. As the vast majority of finite subgroups of $U(3)$ are not subgroups of $SU(3)$, and a complete classification of all discrete subgroups of $U(3)$ is not available, we perform a numerical scan of mixing patterns derived from finite discrete groups up to order 2000 with the help of powerful computer algebra system \texttt{GAP}~\cite{GAP,SmallGroups,repsn}. Another motivation of this work is to scrutinize whether viable quark mixing angles together with leptonic mixing angles can be reproduced from the same discrete flavor symmetry group, and which group is the smallest one capable of achieving this.

The remaining parts of this paper are organized as follows. In Sec.~\ref{sec:framework} we review the idea of discrete flavor symmetry from both bottom-up and top-down approaches, and the method of extracting the flavor mixing matrix from residual flavor symmetry is presented. In Sec.~\ref{sec:mixing_from_SU(3)}, we study the mixing patterns that can be derived from the subgroups of $SU(3)$ under the assumption that neutrinos are Dirac particles. We find that the subgroups of type D can give rise to lepton mixing patterns that are compatible with experimental data, the PMNS matrix is of the trimaximal form, and the Dirac $CP$ violating phase is trivial. Moreover, the flavor symmetry is extended to the quark sector, and the Cabibbo mixing can be generated by the type D groups. In the same fashion, the mixing patterns arising from the $U(3)$ subgroup series $\Sigma(3N^3)$ are discussed in Sec~\ref{sec:mixing_from_Sigma}. However, no mixing pattern in the experimentally preferred $3\sigma$ range is found unless the remnant symmetries are partially or completely accidental. Since there are a lot of finite groups that are subgroups of $U(3)$ but not subgroups of $SU(3)$, in Sec.~\ref{sec:scan_groups} we perform an exhaustive scan of finite discrete groups with order less than 2000 with the help of the powerful group software \texttt{GAP}. The possible residual symmetries and the corresponding mixing matrices as well as mixing parameters for each group are available at the Website~\cite{webdata}. We find that the lepton flavor mixing in accordance with experimental data is still of the trimaximal pattern, and only the Cabibbo mixing is obtained even if we impose very loose constraints on the CKM matrix elements
$0.1\leq |V_{us}|\leq0.3$ and $|V_{ub}|\leq |V_{cb}|<|V_{us}|$. We summarize our results and conclude this paper in Sec.~\ref{sec:conclusion}. Finally the procedure of how to convert a nonunitary representation into a unitary representation is presented in Appendix~\ref{sec:Appd_unitarize}.

\section{\label{sec:framework}Basic framework}

In this section, we briefly review the basic idea of flavor symmetry and how the mixing matrices can be predicted from the structure of flavor symmetry group $\mathcal{G_F}$ and the assumed symmetry breaking pattern. In the bottom-up approach, the mass terms of the Lagrangian for quarks and charged leptons is of the form
\begin{equation}
\mathcal{L}^{ql}_{mass}=-\overline{U}_{R}m_{U}U_{L}-\overline{D}_{R}m_{D}D_{L}-\overline{l}_{R}m_{l}l_{L}+H.c.\,,
\end{equation}
where $U_L \equiv (u_L, c_L, t_L)^T$, $D_L \equiv (d_L, s_L, b_L)^T$, $l_L \equiv (e_L,\mu_L,\tau_L)^T$, and $U_R \equiv (u_R, c_R, t_R)^T$, $D_R \equiv (d_R, s_R, b_R)^T$, $l_R \equiv (e_R,\mu_R,\tau_R)^T$ represent the three generations of left-handed and right-handed up type quark, down type quark, and charged lepton fields. If neutrinos are Dirac particles, the mass terms are
\begin{equation}
\mathcal{L}^{\nu}_{mass}=-\overline{\nu}_{R}m_{\nu}\nu_{L}+\text{H.c.}\,,
\end{equation}
where $\nu_L \equiv (\nu_{eL}, \nu_{\mu L}, \nu_{\tau L})^T$ and $\nu_R \equiv (\nu_{eR}, \nu_{\mu R}, \nu_{\tau R})^T$ are the left- and right-handed neutrino fields. For Majorana neutrinos, it is
\begin{equation}
\label{eq:nu_mass_term}\mathcal{L}^{\nu}_{mass}=\frac{1}{2}\nu^{T}_{L}C^{-1}m_{\nu}\nu_{L}+\text{H.c.}\,,
\end{equation}
where $C$ is the charge-conjugation matrix. Both quark and lepton mass matrices are diagonalized by
\begin{eqnarray}
\nonumber&&V^{\dagger}_{U}m^{\dagger}_{U}m_{U}V_{U}=\textrm{diag}\left(m^2_{u}, m^2_{c}, m^2_{t}\right),\\ \nonumber&&V^{\dagger}_{D}m^{\dagger}_{D}m_{D}V_{D}=\textrm{diag}\left(m^2_{d}, m^2_{s}, m^2_{b}\right),\\ \label{eq:mcdmc_diag}&&V^{\dagger}_{l}m^{\dagger}_{l}m_{l}V_{l}=\textrm{diag}\left(m^2_{e}, m^2_{\mu}, m^2_{\tau}\right)\,,
\end{eqnarray}
and
\begin{eqnarray}
\nonumber&&V^{\dagger}_{\nu}m^{\dagger}_{\nu}m_{\nu}V_{\nu}=\textrm{diag}\left(m^2_{1}, m^2_{2}, m^2_{3}\right),\quad \text{for Dirac neutrinos}\,,\\
\label{eq:mnu_diag}&&V^{T}_{\nu}m_{\nu}V_{\nu}=\textrm{diag}\left(m_{1}, m_{2}, m_{3}\right),\quad \text{for Majorana neutrinos}\,.
\end{eqnarray}
The CKM and PMNS matrices are defined as
\begin{equation}
V_{\text{CKM}}=V^{\dagger}_{U}V_{D},\qquad U_{\text{PMNS}}=V^{\dagger}_{l}V_{\nu}\,.
\end{equation}
One can straightforwardly check that the Hermitian combinations $m^{\dagger}_{U}m_{U}$, $m^{\dagger}_{D}m_{D}$, and $m^{\dagger}_{l}m_{l}$ are invariant under the unitary transformations,
\begin{equation}
U_{L}\rightarrow T_{U}U_{L},\qquad D_{L}\rightarrow T_{D}D_{L},\qquad l_{L}\rightarrow T_{l}l_{L}\,,
\end{equation}
with
\begin{equation}
\label{eq:Tc}T_{c}=V_{c}\,\textrm{diag}\left(e^{i\alpha_{c}}, e^{i\beta_{c}}, e^{i\gamma_{c}}\right)V^{\dagger}_{c}\,,
\end{equation}
where the subscript $c\in\left\{U, D, l\right\}$, $\alpha_{c}$, $\beta_{c}$, and $\gamma_{c}$ are arbitrary real parameters, i.e.,
\begin{equation}
T^{\dagger}_{c}m^{\dagger}_{c}m_{c}T_{c}=m^{\dagger}_{c}m_{c}\,.
\end{equation}
In the language of group theory, each of the quark and charged lepton mass terms has a full $U(1)\times U(1)\times U(1)$ symmetry. In the same fashion, for the Dirac neutrino, we have
\begin{equation}
\nu_{L}\rightarrow T_{\nu}\nu_{L},\qquad m^{\dagger}_{\nu}m_{\nu}\rightarrow T^{\dagger}_{\nu}m^{\dagger}_{\nu}m_{\nu}T_{\nu}=m^{\dagger}_{\nu}m_{\nu}\,,
\end{equation}
with
\begin{equation}
\label{Eq:Tnu_Dirac}T_{\nu}=V_{\nu}\textrm{diag}\left(e^{i\alpha_{\nu}}, e^{i\beta_{\nu}}, e^{i\gamma_{\nu}}\right)V^{\dagger}_{\nu}\,,
\end{equation}
where $\alpha_{\nu}$, $\beta_{\nu}$, and $\gamma_{\nu}$ are real. In the case of the Majorana neutrino, the symmetries of the neutrino mass term in Eq.~\eqref{eq:nu_mass_term} are strongly constrained. It has been established that the symmetry transformation of $m_{\nu}$ is~\cite{Lam:2007qc}
\begin{equation}
\nu_{L}\rightarrow T_{\nu}\nu_{L},\qquad m_{\nu}\rightarrow T^{T}_{\nu}m_{\nu}T_{\nu}=m_{\nu}\,,
\end{equation}
with
\begin{equation}
\label{Eq:Tnu_Majorana}T_{\nu}=V_{\nu}\textrm{diag}\left(\pm1, \pm1, \pm1\right)V^{\dagger}_{\nu}\,.
\end{equation}
Note that if the lightest neutrino is massless, the corresponding diagonal entry ``$\pm1$'' could be an arbitrary phase factor. The eight possible values of $T_{\nu}$ in Eq.~\eqref{Eq:Tnu_Majorana} constitute the symmetry group $Z_2\times Z_2\times Z_2$. However, only $Z_2\times Z_2$ is effective since the third $Z_2$ just contributes an overall ``$-1$'' factor that can be absorbed by redefining the neutrino fields. In other words, the effective symmetry of the neutrino mass term is the $Z_2\times Z_2$ Klein four group for the Majorana neutrino. Inversely, given the symmetry transformations that leave the quark and lepton mass terms invariant, one is able to reconstruct the explicit form of their mass matrices.

In the top-down approach, the above mentioned symmetries of the quark and lepton mass terms originate from some flavor symmetry group $\mathcal{G_F}$ at high energy scale. The full Lagrangian is invariant under $\mathcal{G_F}$, and then subsequently $\mathcal{G_F}$ is broken down to different subgroups $\mathcal{G_{\nu}}$ and $\mathcal{G}_{l}$ in the neutrino and charged lepton sectors, and to $\mathcal{G}_{U}$ and $\mathcal{G}_{D}$ in the up quark and down quark sectors, respectively. From Eqs.~(\ref{eq:Tc}, \ref{Eq:Tnu_Dirac}, \ref{Eq:Tnu_Majorana}), we know
\begin{eqnarray}
\nonumber&&\mathcal{G}_{c}\subseteq U(1)\times U(1) \times U(1),\\
\nonumber&&\mathcal{G_{\nu}}\subseteq U(1)\times U(1) \times U(1),\quad \text{for Dirac neutrinos}\,,\\
&&\mathcal{G_{\nu}}\subseteq Z_2\times Z_2 \times Z_2,\quad \text{for Majorana neutrinos}\,.
\end{eqnarray}
That is to say, the residual symmetries $\mathcal{G}_{l}$, $\mathcal{G}_{U}$, and $\mathcal{G}_{D}$ can be any Abelian subgroups of $\mathcal{G_{F}}$, and similarly $\mathcal{G_{\nu}}$ can be any Abelian subgroup of $\mathcal{G_{F}}$ if neutrinos are Dirac particles while $\mathcal{G_{\nu}}$ can only be effectively (or a subgroup of) a Klein four subgroup for Majorana neutrinos. In this work, we consider the scenario that $\mathcal{G_F}$ is a non-Abelian finite discrete group. As usual, the three generations of left-handed lepton doublets are assigned to be an irreducible faithful three-dimensional representation $\rho_{\mathbf{3}}$ of $\mathcal{G_F}$~\cite{Lam:2012ga,Holthausen:2012wt,Holthausen:2013vba,Hagedorn:2013nra,Lavoura:2014kwa,King:2013vna}. Analogously the three left-handed quark doublets are also embedded into the same three-dimensional representations~\cite{Holthausen:2013vba,Araki:2013rkf}. For the residual symmetries $\mathcal{G}_{\nu, l, U, D}$ to hold, the quark and lepton mass matrices have to satisfy the following constraints:
\begin{equation}
\label{eq:inv_mc}\rho^{\dagger}_{\mathbf{3}}(g_c)m^{\dagger}_{c}m_{c}\rho_{\mathbf{3}}(g_c)=m^{\dagger}_{c}m_{c},\qquad g_{c}\in\mathcal{G}_{c}\,,
\end{equation}
and
\begin{eqnarray}
\nonumber&&\rho^{\dagger}_{\mathbf{3}}(g_{\nu})m^{\dagger}_{\nu}m_{\nu}\rho_{\mathbf{3}}(g_{\nu})=m^{\dagger}_{\nu}m_{\nu},\quad \text{for Dirac neutrinos}\,, \qquad g_{\nu}\in \mathcal{G}_{\nu},\\
\label{eq:inv_mnu}&&\rho^{T}_{\mathbf{3}}(g_{\nu})m_{\nu}\rho_{\mathbf{3}}(g_{\nu})=m_{\nu},\quad \text{for Majorana neutrinos}\,, \qquad g_{\nu}\in \mathcal{G}_{\nu}\,.
\end{eqnarray}
Notice that it is sufficient to impose the invariant conditions of Eqs.~(\ref{eq:inv_mc}, \ref{eq:inv_mnu}) on the generators of $\mathcal{G}_c$ and $\mathcal{G_{\nu}}$. Furthermore, we see that Eqs.~(\ref{eq:inv_mc}, \ref{eq:inv_mnu}) imply
\begin{equation}
\left[\rho_{\mathbf{3}}(g_c), m^{\dagger}_{c}m_{c}\right]=0,\qquad \left[\rho_{\mathbf{3}}(g_{\nu}), m^{\dagger}_{\nu}m_{\nu}\right]=0\,.
\end{equation}
As $\rho_{\mathbf{3}}(g_c)$ and $m^{\dagger}_{c}m_{c}$ commute with each other, the unitary transformation $V_{c}$ that diagonalizes the mass matrix $m^{\dagger}_{c}m_{c}$ also diagonalizes $\rho_{\mathbf{3}}(g_c)$ up to permutations and phases of columns, as shown in Eqs.~\eqref{eq:mcdmc_diag} and \eqref{eq:Tc}. Similarly $\rho_{\mathbf{3}}(g_{\nu})$ and $m^{\dagger}_{\nu}m_{\nu}$ (or $m_{\nu}$) are diagonalized by the same matrix $V_{\nu}$, as displayed in Eqs.~(\ref{eq:mnu_diag}, \ref{Eq:Tnu_Dirac}, \ref{Eq:Tnu_Majorana}). In other words, given a group representation, the unitary transformations $V_{c}$ and $V_{\nu}$ (and eventually the CKM matrix $V_{\text{CKM}}$ and the PMNS matrix $U_{\text{PMNS}}$) can be obtained by diagonalizing the representation matrices of the generators of $\mathcal{G}_{c}$ and $\mathcal{G_{\nu}}$, respectively. In practice, we only need to find the eigenvectors of $\rho_{\mathbf{3}}(g_c)$ and $\rho_{\mathbf{3}}(g_{\nu})$ that form the column vectors of $V_{c}$ and $V_{\nu}$. In this approach, it is not necessary to construct the explicit form of the mass matrices $m^{\dagger}_{c}m_{c}$ and $m^{\dagger}_{\nu}m_{\nu}$ (or $m_{\nu}$) although this can be accomplished straightforwardly. Since the order of both quark and lepton masses cannot be pinned down in this framework, particularly the neutrino mass spectrum can be either normal ordering (NO) or inverted ordering (IO), the unitary matrices $V_{c}$ and $V_{\nu}$ are uniquely fixed up to permutations and phases of their column vectors. As a result, both $U_{\text{PMNS}}$ and $V_{\text{CKM}}$ are determined up to independent row and column permutations and arbitrary phase matrices multiplied from the left-handed and right-handed sides. Notice that if we switch the roles of the subgroups $\mathcal{G}_{\nu}$ and $\mathcal{G}_{l}$, the PMNS matrix $U_{\text{PMNS}}$ would become its Hermitian conjugate. Analogously, the exchange of $\mathcal{G}_{U}$ and $\mathcal{G}_{D}$ would lead to $V_{\text{CKM}}$ being Hermitian conjugated.

It is remarkable that the PMNS matrix $U_{\text{PMNS}}$ stems from the mismatch between $\mathcal{G}_{\nu}$ and $\mathcal{G}_{l}$ and their relative embedding into $\mathcal{G_F}$, and $V_{\text{CKM}}$ from the misalignment of $\mathcal{G}_{U}$ and $\mathcal{G}_{D}$ in this framework. Physical results only depend on the structure of flavor symmetry group $\mathcal{G_F}$ and the assumed symmetry breaking patterns, and they are independent of how the required residual symmetries are dynamically realized.

\section{\label{sec:mixing_from_SU(3)}Flavor mixing achievable from subgroups of $SU(3)$ }

The finite subgroups of $SU(3)$ have been systematically classified nearly 100 years ago~\cite{Miller_book}. They are further studied in Refs.~\cite{Grimus:2010ak,Ludl:2010bj,Merle:2011vy,Grimus:2011fk,Grimus:2013apa} in recent years, and the complete list can be found, for example, in \cite{Grimus:2011fk,Grimus:2013apa}. Now it has been firmly established that all discrete subgroups of $SU(3)$ can be divided into five categories: type A, type B, type C, type D, and type E. In the following, we shall investigate the viability of each type of these groups as a flavor symmetry for Dirac neutrinos and further study the predictions for the mixing matrix. The groups of type A are Abelian groups, and they are isomorphic to $Z_m\times Z_n$, where $n$ is a divisor of $m$. Obviously type A subgroups are not suitable as flavor symmetry groups due to the Abelian characterization. Groups of type B are finite subgroups of $U(2)$, and up to basis transformations, such $SU(3)$ matrices have the form
\begin{equation}
\left(\begin{array}{cc}
\det M^{\ast} ~&~ 0_{1\times2} \\
0_{2\times1} ~&~ M
\end{array}\right)\quad
\text{with} \quad M \in U(2)\,.
\end{equation}
This means that the three-dimensional representations are reducible. As a consequence, at least four entries of the mixing matrix would be vanishing such that the reactor mixing angle $\theta_{13}$ would be zero. Therefore type B subgroups can-not give phenomenologically viable lepton mixing patterns. Type E subgroups are the so-called ``exceptional'' finite groups of $SU(3)$, and they include $\Sigma(60)$, $\Sigma(60)\times{Z}_3$, $\Sigma(168)$, $\Sigma(168)\times{Z}_3$, $\Sigma(36\times3)$, $\Sigma(72\times3)$, $\Sigma(216\times3)$, and $\Sigma(360\times3)$~\cite{Grimus:2011fk,Grimus:2010ak}. The mixing patterns that can be derived from these exceptional groups
have been comprehensively studied in Ref.~\cite{Hagedorn:2013nra}. The results of~\cite{Hagedorn:2013nra} are confirmed by our systematical numerical analysis for the groups up to order 2000 in the next section. Interestingly, the groups of type~C comprise infinitely many series, two of which are the well-known $\Delta(3n^2)$ and $T_m$ series. Nevertheless, all groups of type D are described by just two infinite series where one is the well-known series $\Delta(6n^2)$. Notice that $\Delta(6n^2)$ as flavor symmetry groups have been widely discussed~\cite{
Delta_96,Lam:2013ng,King:2013vna,Ding:2014ora,Hagedorn:2014wha}, and they remain viable candidates for flavor symmetry~\cite{Holthausen:2012wt,Holthausen:2013vba,Talbert:2014bda}. In this section, we shall present a complete analytical study of mixing patterns obtained from type C and type D groups.

As shown in Refs.~\cite{Ludl:2011gn,Grimus:2013apa}, the structures of groups of type C and D are given by
\begin{equation}
\label{eq:typeCD}
\begin{split}
\text{type C}:&\quad \mathcal{G}\cong(Z_m\times Z_n)\rtimes Z_3\,,\\
\text{type D}:&\quad \mathcal{G}\cong(Z_m\times Z_n)\rtimes S_3\,,\\
\end{split}
\end{equation}
where $S_3$ is isomorphic to $Z_3\rtimes Z_2$, and $n$ is a divisor of $m$. It is worth mentioning that the above group structure crucially depends on the definition of the semidirect product. A new group could be obtained if the definition of the semidirect product is changed. We really find that some $U(3)$ subgroups apparently have the same structure as the type D groups while they are, in fact, different because of different semidirect structures. The groups of type C can be conveniently expressed in terms of three generators $a$, $c$, and $d$, while type D groups involve one more generator $b$, and the explicit forms of these generators are
\begin{equation}
\label{eq:3dimrep}
a=\left(
\begin{array}{ccc}
 0 & 1 & 0 \\
 0 & 0 & 1 \\
 1 & 0 & 0 \\
\end{array}
\right),~~
b=\left(
\begin{array}{ccc}
 -1 & 0 & 0 \\
 0 & 0 & -1 \\
 0 & -1 & 0 \\
\end{array}
\right),~~
c=\left(
\begin{array}{ccc}
 \eta  & 0 & 0 \\
 0 & \eta^k  & 0 \\
 0 & 0 & \eta^{-k-1} \\
\end{array}
\right),~~
d=\left(
\begin{array}{ccc}
 1 & 0 & 0 \\
 0 & \eta^{-r} & 0 \\
 0 & 0 & \eta^r \\
\end{array}
\right)\,,
\end{equation}
where $\eta=e^{2\pi i/m}$. In fact, the matrices in Eq.~\eqref{eq:3dimrep} form a faithful irreducible representation of the type C and type D groups. One can check that the following multiplication rules are satisfied:
\begin{equation}
\label{eq:relator}
\begin{split}
a^3=&b^2=(ab)^2=c^m=d^n=1,\quad cd=dc,\\
aca&^{-1}=c^kd^l,\quad ada^{-1}=c^{-r}d^{-(k+1)},\\ \quad &bcb^{-1}=cd^{l'},\quad bdb^{-1}=d^{-1}\,,\\
\end{split}
\end{equation}
where $r=m/n$, and the last two lines define the semidirect product. For groups of both type C and type D, the integers $k$, $l$, and $r$ fulfill the relation:
\begin{equation}
\label{eq:rela1}1+k+k^2=lr\,.
\end{equation}
For groups of type D, the parameters $k$, $l'$, and $r$ satisfy
\begin{equation}
\label{eq:rela2}1+2k=l'r\,.
\end{equation}
As we can see, groups of type C and type D are completely fixed by the parameters $m$, $n$ and $k$ according to the multiplication rules in Eq.~\eqref{eq:relator}. Therefore, we can denote these groups by the symbols $C_{m,n}^{(k)}$ and $D_{m,n}^{(k)}$, respectively.

\begin{table}[t!]
\begin{center}
\begin{tabular}{|c|c|}
\hline\hline
Group series & Constraints on parameters  \\
\hline
$C_{rn,n}^{(k)}$ & $m=rn$, $3 \nmid r$
\rule{0mm}{5.5mm}\\
$C_{3r^{\prime}n,n}^{(k)}\cong{Z}_3\times C_{r^{\prime}n,n}^{(k)}$ & $m=3r^{\prime}n$, $r=3r^{\prime}$, $3\nmid r^{\prime}$, $3\nmid n$
\\
$C_{9r^{\prime}n^{\prime},3n^{\prime}}^{(k)}$ & $m=9r^{\prime}n^{\prime}$, $n=3n^{\prime}$, $r=3r^{\prime}$, $3 \nmid r^{\prime}$
\\[1mm]
\hline
$D_{n,n}^{(0)} \equiv \Delta(6n^2)$ & $m=n$, $r=1$
\rule{0mm}{5.5mm}\\
$D_{3n,n}^{(1)}\cong{Z}_3\times\Delta(6n^2)$ & $m=3n$, $r=3$, $3 \nmid n$
\\
$D_{9n^{\prime},3n^{\prime}}^{(1)}$ & $m=9n^{\prime}$, $n=3n^{\prime}$, $r=3$
\\[1mm]
\hline\hline
\end{tabular}
\end{center}
\caption{\label{tab:CDsummary}Summary of the groups of type C and type D. The notation $\nmid$ denotes ``does not divide.'' For the type C groups, the parameters $r$ and $r^{\prime}$ can be any product of primes of the form $6z+1$ with non-negative integer $z$. The value of $k$ is fixed by Eq.~\eqref{eq:rela1}.}
\end{table}

The group structures of type C and type D groups have been studied in Ref.~\cite{Grimus:2013apa}. The complete classification of these groups is summarized in Table~\ref{tab:CDsummary}. Notice that there is an infinite series $C_{rn,n}^{(k)}$ for every pair $(r,k)$. Therefore infinitely many type C series are obtained since there are infinitely many admissible $r$. In Table~\ref{tab:r-table} we have listed all allowed values of $r$ with $r<100$ and the corresponding solution(s) of $k$. Some well-known type C groups are included in table~\ref{tab:r-table}. For example, if we fix $r=1$ and $k=0$, we shall get the dihedral-like groups
\begin{equation}
\label{eq:delta3n2}C_{n,n}^{(0)}\equiv \Delta(3n^2)\cong (Z_n\times Z_n)\rtimes Z_3\,.
\end{equation}
If we set $n=1$, $r=m$ is the product of primes of the form $6z+1$ with $z$ being a positive integer, and then the so-called $T_m$ group is obtained,
\begin{equation}
C_{m,1}^{(k)}\equiv T_m\cong Z_m\rtimes Z_3\,,
\end{equation}
where $m$ and $k$ are related by
\begin{equation}
1+k+k^2=0\hskip-0.08in\pmod m\,.
\end{equation}
From Table~\ref{tab:r-table}, we see that these groups exist only for specific integers $m=7$, 13, 19, 31, 37, 43, 49, 61, 67, 73, 79, 91, 97, $\ldots$, and there are two solutions of $k=9$ and 16 for $m=91$. Consequently there are two nonisomorphic groups $T_{91}$ and $T^{\prime}_{91}$.

For type D groups, the values of the parameters $r$ and $k$
are strongly constrained. It is found that only two cases are admissible: $r=1$, $k=0$, $m=n$ or $r=3$, $k=1$, $m=3n$. Therefore all type D groups can be classified into two series $D^{(0)}_{n,n}$ and $D^{(1)}_{3n,n}$. Notice that $D^{(0)}_{n,n}$ is exactly the well-known $\Delta(6n^2)$ group~\cite{Escobar:2008vc}, and the chosen generators $a^{\prime}$, $b^{\prime}$, $c^{\prime}$, and $d^{\prime}$ of Ref.~\cite{Escobar:2008vc} are related to the present ones by
\begin{equation}
\label{eq:relation}  a^{\prime}=a,\quad b^{\prime}=ab,\quad c^{\prime}=cd,\quad d^{\prime}=d^{-1}\,.
\end{equation}
Furthermore, $D^{(1)}_{3n,n}$ is isomorphic to $Z_3\times\Delta(6n^2)$ if $n$ is not divisible by 3~\cite{Grimus:2013apa}. As a result, all type D groups can be genuinely described by $\Delta(6n^2)$ and $D^{(1)}_{9n^{\prime},3n^{\prime}}\cong(Z_{9n'}\times Z_{3n'})\rtimes S_3$.

\begin{table}[t!]
\begin{center}
  \begin{tabular}{|r|r||r|r|}
    \hline\hline
$r$ & $k$ &  $r$ & $k$ \\\cline{1-2} \cline{3-4}
\rule{0mm}{5.5mm}
1 & 0                   & $7 \times 7 = 49$ & 18 \\\hline
3 & 1                  & $3 \times 19 = 57$ & 7 \\\hline
7 & 2                  & 61 & 13 \\\hline
13 & 3               & 67 & 29 \\\hline
19 & 7                 & 73 & 8 \\\hline
$3\times 7 = 21$ & 4    & 79 & 23 \\\hline
31 & 5                  & $7 \times 13 = 91$ & 9,\,16 \\\hline
37 & 10                 & $3 \times 31 = 93$ & 25 \\\hline
$3 \times 13 = 39$ & 16 & 97 & 35 \\\hline
43 & 6                  &&\\\hline
\hline
\end{tabular}
\end{center}
\caption{\label{tab:r-table}List of all allowed $r< 100$ and the corresponding solution(s) $k$ of Eq.~\eqref{eq:rela1}. }
\end{table}

\subsection{\label{subsec:residual_sym_diag}Remnant symmetries and their diagonalization}

In this section, we shall discuss the possible remnant symmetry groups and the corresponding unitary diagonalization matrices, if the flavor symmetry group $\mathcal{G_F}$ is a group of type C or type D. As shown in Sec.~\ref{sec:framework}, both residual symmetries $\mathcal{G}_{\nu}$ and $\mathcal{G}_{l}$ are Abelian subgroups of $\mathcal{G_F}$ if neutrinos are Dirac particles. The same is true for $\mathcal{G}_{U,D}$ in the quark sector. First we consider the situation that the remnant symmetry group $\mathcal{G}_{r}$ with $r=\nu, l, U, D$ is generated by a single element.
If the generator of $\mathcal{G}_{r}$ is represented by a matrix with nondegenerate eigenvalues, the diagonalization matrix $V_{r}$ of $\mathcal{G}_{r}$ is uniquely determined by its eigenvectors up to permutations and phases. On the other hand, if two of the three eigenvalues are degenerate, only one generation can be distinguished from the other two
generations and consequently only one column of $V_{r}$ can be fixed by the postulated residual symmetry. Such a column will be called ``mixing vector'' hereafter. The generator of $\mathcal{G}_{r}$ can be written as $c^{s}d^{t}$, $bc^{s}d^{t}$, $ac^{s}d^{t}$, $a^{2}c^{s}d^{t}$, $abc^{s}d^{t}$, or $a^{2}bc^{s}d^{t}$ with $s=0, 1, \ldots, m-1$, $t=0, 1, \ldots, n-1$. The corresponding unitary transformation $V_{r}$ can be straightforwardly calculated as follows:

\begin{itemize}[labelindent=-0.7em, leftmargin=1.6em]
\item $\mathcal{G}_r=\langle c^sd^t\rangle$

The representation matrix of the generator $c^{s}d^{t}$ is given by
\begin{equation}
\label{eq:rho1}
\rho_{\mathbf{3}}(c^sd^t)=\left(
\begin{array}{ccc}
 \eta^s ~&~ 0 ~&~ 0 \\
 0 ~&~ \eta^{ks-rt} ~&~ 0 \\
 0 ~&~ 0 ~&~ \eta^{rt-(k+1)s}
\end{array}
\right)\,.
\end{equation}
Its eigenvalues are
\begin{equation}
\label{eq:eig1}
\eta^s,\quad \eta^{ks-rt},\quad \eta^{rt-(k+1)s}\,.
\end{equation}
The unitary matrix $V_{r}$, which diagonalizes $\rho_{\mathbf{3}}(c^sd^t)$, is of the form
\begin{equation}
\label{eq:N1}\mathcal{N}_1:~ V_{r}=\left(\begin{array}{ccc} 1&0&0\\ 0&1&0\\ 0&0&1\\ \end{array}\right)\,.
\end{equation}
Nondegeneracy of the eigenvalues imposes the following constraints on the parameters $s$ and $t$:
\begin{eqnarray}
\nonumber&&(2k+1)s-2rt\neq0\hskip-0.08in\pmod m,\\
\nonumber&&(k+2)s-rt\neq0\hskip-0.08in\pmod m,\\
\label{eq:constraint1}&&(k-1)s-rt\neq0\hskip-0.08in\pmod m\,.
\end{eqnarray}
In the case of two degenerate eigenvalues, only one column of $V_{r}$ can be fixed. This column vector would be denoted by $v_{r}$ henceforth. To be more precise, we have
{\small
\begin{eqnarray}
\nonumber\mathcal{D}_1:&&v_{r}=\left(1,0,0\right)^{T}~\text{for}~(2k+1)s-2rt=0 \hskip-0.08in\pmod{m},~~(k+2)s-rt\neq0\hskip-0.08in\pmod{m},\\
\nonumber\mathcal{D}_2:&&v_{r}=\left(0,1,0\right)^{T}~\text{for}~ (k+2)s-rt=0\hskip-0.08in\pmod{m},~~(k-1)s-rt\neq0\hskip-0.08in\pmod{m},\\
\label{eq:D1}\mathcal{D}_3:&&v_{r}=\left(0,0,1\right)^{T}~\text{for}~ (k-1)s-rt=0\hskip-0.08in\pmod{m},~~(k+2)s-rt\neq0\hskip-0.08in\pmod{m}\,.
\end{eqnarray}}

\item $\mathcal{G}_r=\langle bc^s d^t\rangle$

\begin{equation}
\label{eq:rho2}
\rho_{\mathbf{3}}(bc^s d^t)=\left(
\begin{array}{ccc}
 -\eta^s & 0 & 0 \\
 0 & 0 & -\eta^{rt-(k+1)s} \\
 0 & -\eta^{ks-rt} & 0
\end{array}
\right)\,,
\end{equation}
with the eigenvalues
\begin{equation}
\label{eq:e2}-\eta^{-s/2},\quad \eta^{-s/2},\quad -\eta^s\,.
\end{equation}
The diagonalization matrix $V_{r}$ is given by
\begin{equation}
\label{eq:N2}\mathcal{N}_2:~ V_{r}=\frac{1}{\sqrt{2}}\left(
\begin{array}{ccc}
 0 & 0 & \sqrt{2} \\
 \eta ^{rt-(k+\frac{1}{2})s} & -\eta ^{rt-(k+\frac{1}{2})s} & 0 \\
 1 & 1 & 0
\end{array}
\right)\,.
\end{equation}
To avoid degeneracy among the eigenvalues, we obtain the constraint
\begin{equation}
\label{eq:constraint2}3s\neq0\hskip-0.08in\pmod{m}\,.
\end{equation}
If two of the eigenvalues are degenerate, the fixed column vector is
\begin{equation}
\label{eq:D2}
\begin{split}
\mathcal{D}_4:&~ \frac{1}{\sqrt{2}}\left(0,\eta ^{rt-(k+\frac{1}{2}) s},1\right)^{T}~~\text{for}~~ 3s=m\hskip-0.08in\pmod{2m},\\
\mathcal{D}_5:&~ \frac{1}{\sqrt{2}}\left(0,-\eta ^{rt-(k+\frac{1}{2}) s},1\right)^{T}~~\text{for}~~ 3s=0\hskip-0.08in\pmod{2m}\,.
\end{split}
 \end{equation}
\item $\mathcal{G}_r=\langle ac^sd^t\rangle$

The order of the generator $ac^sd^t$ is always three; therefore residual symmetry $\mathcal{G}_r=\langle ac^sd^t\rangle$ is a $Z_3$ subgroup with
\begin{equation}
\label{eq:rho3}\rho_{\mathbf{3}}(ac^sd^t)=
\left(
\begin{array}{ccc}
 0 & \eta ^{k s-r t} & 0 \\
 0 & 0 & \eta ^{r t-(k+1) s} \\
 \eta ^s & 0 & 0
\end{array}
\right)\,,
\end{equation}
which is diagonalized by
\begin{equation}
\label{eq:N3}\mathcal{N}_3:~ V_{r}=\frac{1}{\sqrt{3}}
\left(\begin{array}{ccc}
 \eta ^{-s} & \omega \eta ^{-s} &  \omega ^2 \eta ^{-s} \\
 \eta ^{rt-(k+1)s} &  \omega ^2 \eta ^{rt-(k+1)s} & \omega  \eta ^{rt-(k+1)s} \\
 1 & 1 & 1
\end{array}
\right)\,,
\end{equation}
where $\omega=e^{2i\pi/3}$. The eigenvalues are 1, $\omega$, and $\omega^2$ for any values of $s$ and $t$.

\item $\mathcal{G}_r=\langle a^2c^sd^t\rangle$

In this case, $\mathcal{G}_r=\langle a^2c^sd^t\rangle$ is also a $Z_3$ subgroup for any values of $s$ and $t$,
\begin{equation}
\label{eq:rho3tilde}\rho_{\mathbf{3}}(a^2c^sd^t)=
\left(
\begin{array}{ccc}
 0 & 0 & \eta ^{r t-(k+1) s} \\
 \eta ^s & 0 & 0 \\
 0 & \eta ^{k s-r t} & 0
\end{array}
\right)\,,
\end{equation}
whose eigenvalues are 1, $\omega$, and $\omega^2$. Since the eigenvalues are not degenerate, the unitary matrix $V_{r}$ is completely fixed
\begin{equation}
\label{eq:N3tilde} \mathcal{N}^{\prime}_3:~ V_{r}=\frac{1}{\sqrt{3}}
\left(\begin{array}{ccc}
 \eta ^{rt-(k+1)s} & \omega \eta ^{rt-(k+1)s}  & \omega ^2 \eta ^{rt-(k+1)s} \\
 \eta ^{r t-k s} &  \omega ^2 \eta ^{r t-k s} & \omega \eta ^{r t-k s}  \\
 1 & 1 & 1
\end{array}
\right)\,.
\end{equation}
Note that this $V_{r}$ can be obtained from the one in Eq.~\eqref{eq:N3} via the replacement $s\to (k+1)s-rt$, $t\to s(1+k+k^2)/r-kt$. This fact is because of the relation $(a^2c^sd^t)^2=ac^{(k+1)s-rt}d^{s(1+k+k^2)/r-kt}$.

\item $\mathcal{G}_r=\langle abc^sd^t\rangle$

\begin{equation}
\label{eq:rho4}\rho_{\mathbf{3}}(abc^sd^t)=
\left(\begin{array}{ccc}
 0 & 0 & -\eta^{rt-(k+1)s} \\
 0 & -\eta^{ks-rt} & 0 \\
 -\eta^s & 0 & 0
\end{array}
\right)\,,
\end{equation}
which is diagonalized by the unitary transformation
\begin{equation}
\label{eq:N4}\mathcal{N}_4:~  V_{r}=\frac{1}{\sqrt{2}}\left(
\begin{array}{ccc}
 0 ~&~ \eta^{\frac{1}{2}(rt-(k+2)s)} ~&~ -\eta^{\frac{1}{2}(rt-(k+2)s)} \\
 \sqrt{2} ~&~ 0 ~&~ 0 \\
 0 ~&~ 1 ~&~ 1
\end{array}
\right)\,,
\end{equation}
with the eigenvalues
\begin{equation}
\label{eq:e4}-\eta^{ks-rt},\quad -\eta^{\frac{1}{2}(rt-ks)},\quad \eta^{\frac{1}{2}(rt-ks)}\,.
 \end{equation}
The nondegeneracy of the solutions requires
\begin{equation}
\label{eq:constraint4}3(rt-ks)\neq0\hskip-0.08in\pmod{m}\,.
\end{equation}
The degenerate solutions are given by
\begin{equation}
\label{eq:D4}\begin{split}
\mathcal{D}_6:&~ v_r=\frac{1}{\sqrt{2}}\left(\eta^{\frac{1}{2} (rt-(k+2)s)},0,1\right)^{T}~~\text{for}~~ 3(rt-ks)=m\hskip-0.08in\pmod{2m},\\
\mathcal{D}_7:&~ v_r=\frac{1}{\sqrt{2}}\left(-\eta^{\frac{1}{2} (rt-(k+2)s)},0,1\right)^{T}~~\text{for}~~ 3(rt-ks)=0\hskip-0.08in\pmod {2m}\,.
\end{split}
\end{equation}

\item $\mathcal{G}_r=\langle a^2bc^sd^t\rangle$

\begin{equation}
\label{eq:rho5}\rho_{\mathbf{3}}(a^2bc^sd^t)=
\left(\begin{array}{ccc}
 0 & -\eta^{ks-rt} & 0 \\
 -\eta^s & 0 & 0 \\
 0 & 0 & -\eta^{rt-(k+1)s}
\end{array}
\right)\,.
\end{equation}
The unitary diagonalization matrix $V_{r}$ is determined to be
\begin{equation}
\label{eq:N5}\mathcal{N}_5:~ V_{r}=\frac{1}{\sqrt{2}}\left(
\begin{array}{ccc}
 0 ~&~ \eta^{\frac{1}{2}((k-1)s-rt)} ~&~ -\eta^{\frac{1}{2}((k-1)s-rt)} \\
 0 ~&~ 1 ~&~ 1 \\
 \sqrt{2} ~&~ 0 ~&~ 0
\end{array}
\right)\,.
\end{equation}
The corresponding eigenvalues are
\begin{equation}
\label{eq:e5}-\eta^{rt-(k+1)s},\quad -\eta^{\frac{1}{2}((k+1)s-rt)},\quad \eta^{\frac{1}{2}((k+1)s-rt)}\,,
\end{equation}
with the nondegenerate condition
\begin{equation}
\label{eq:constraint5}3((k+1)s-rt)\neq0\hskip-0.08in\pmod{m}\,.
\end{equation}
The partial degeneracy of the eigenvalues leads to
\begin{equation}
\label{eq:D5}\begin{split}
\mathcal{D}_{8}:&~ v_r=\frac{1}{\sqrt{2}}\left(\eta^{\frac{1}{2}((k-1)s-rt)},1,0\right)^{T}~~\text{for}~~ 3((k+1)s-rt)=m\hskip-0.08in\pmod{2m},\\
\mathcal{D}_{9}:&~ v_r=\frac{1}{\sqrt{2}}\left(-\eta^{\frac{1}{2}((k-1)s-rt)},1,0\right)^{T}~~\text{for}~~ 3((k+1)s-rt)=0\hskip-0.08in\pmod{2m}\,.\\
\end{split}
\end{equation}
\end{itemize}

As shown above, only one row of $V_{r}$ can be fixed for the cases with degenerate eigenvalues such as $\mathcal{D}_1$, $\ldots$, $\mathcal{D}_{9}$. To resolve this degeneracy and fully determine the unitary matrix $V_{r}$, we extend the remnant symmetry group to be the product of cyclic groups, e.g., $\mathcal{G}_{r}=\mathcal{G}_1\times\mathcal{G}_2$. Here we follow the minimality principle in which either $\mathcal{G}_1$ or $\mathcal{G}_2$ alone is insufficient to distinguish the three generations of fermions.\footnote{If $\mathcal{G}_1$ (or $\mathcal{G}_2$) is sufficient to distinguish among the generations, the presence of the group $\mathcal{G}_2$ (or $\mathcal{G}_1$) would not add any information as far as lepton mixing is concerned, and the unitary matrix $V_{r}$ would be of the form of Eqs.~\eqref{eq:N1}, \eqref{eq:N2}, \eqref{eq:N3}, \eqref{eq:N4}, or \eqref{eq:N5}.} That is to say, the generators of both $\mathcal{G}_1$ and $\mathcal{G}_2$ are represented by matrices that have two degenerate eigenvalues; in such instances, one of $\mathcal{G}_1$ and $\mathcal{G}_2$ alone determines only a column in $V_{r}$ and the third column can be fixed by the unitary condition. As a consequence, neither $\mathcal{G}_1$ nor $\mathcal{G}_2$ can be the group generated by $ac^sd^t$ or $a^2c^sd^t$. Moreover, by definition, the generators of $\mathcal{G}_1$ and $\mathcal{G}_2$ should be commutable with each other. All admissible combinations of $\mathcal{G}_1$ and $\mathcal{G}_2$ satisfying the above requirements are summarized in Table~\ref{tab:extension}.

For the cases of $\mathcal{D}_{1,2,3}$, the degeneracy can be eliminated by taking $\mathcal{G}_{1}=\langle c^{s}d^{t}\rangle$ and $\mathcal{G}_{2}=\langle c^{s^{\prime}}d^{t^{\prime}}\rangle$, where $s$, $t$ and $s^{\prime}$, $t^{\prime}$ should fulfill any two different conditions shown in Eq.~\eqref{eq:D1}. Obviously the unitary transformation $V_{r}$, which diagonalizes both $\mathcal{G}_{1}$ and $\mathcal{G}_{2}$, would be a trivial unit matrix up to permutations and phases of rows and columns. Then we turn to the scenario of $\mathcal{D}_{4,5}$. For $\mathcal{G}_{1}=\langle bc^{s}d^{t}\rangle$ with $3s=l_2m\pmod{2m}$ and $l_2=0,1$, only one column of $V_{r}$ would be determined to be $\frac{1}{\sqrt{2}}\left(0, -(-1)^{l_2}\eta^{rt-(k+\frac{1}{2}) s},1\right)^{T}$. We can enlarge the residual symmetry group to $\mathcal{G}_r=\mathcal{G}_1\times\mathcal{G}_2$ with $\mathcal{G}_{2}=\langle c^{s^{\prime}}d^{t^{\prime}}\rangle$ or $\mathcal{G}_{2}=\langle bc^{s^{\prime}}d^{t^{\prime}}\rangle$. In case of $\mathcal{G}_{2}=\langle c^{s^{\prime}}d^{t^{\prime}}\rangle$, the commutability between $\mathcal{G}_{1}$ and $\mathcal{G}_{2}$ leads to the condition $(2k+1)s^{\prime}-2rt^{\prime}=0\pmod{m}$. Consequently another column fixed by $\mathcal{G}_2$ is $\left(1, 0, 0\right)^{T}$, which is exactly $\mathcal{D}_1$. Therefore the unitary transformation $V_r$ is of the form of $\mathcal{N}_2$ while the constraint $3s\neq0\pmod{m}$ is removed. For the second possibility $\mathcal{G}_{2}=\langle bc^{s^{\prime}}d^{t^{\prime}}\rangle$ with $3s^{\prime}=l_3m\pmod{2m}$ and $l_3=0,1$, the product $\mathcal{G}_1\times\mathcal{G}_2$ is well defined if and only if we have $(2k+1)(s-s^{\prime})-2r(t-t^{\prime})=l_1m\pmod{2m}$ with $l_1=0, 1$. Thus the mixing vector dictated by this $\mathcal{G}_2$ becomes $\frac{1}{\sqrt{2}}\left(0,-(-1)^{l_3}\eta^{rt^{\prime}-(k+\frac{1}{2}) s^{\prime}},1\right)^{T}=\frac{1}{\sqrt{2}}\left(0, -(-1)^{l_1+l_3}\eta^{rt-(k+\frac{1}{2}) s},1\right)^{T}$. Both mixing vectors $\mathcal{D}_4$ and $\mathcal{D}_5$ would be obtained for $l_1+l_2+l_3=1, 3$ such that $V_{r}$ is still of the form of $\mathcal{N}_2$. In the same fashion, the degeneracy in $\mathcal{D}_{6,7}$ and $\mathcal{D}_{8,9}$ can also be resolved by extending $\mathcal{G}_{r}$ from $\langle a^{\alpha}bc^{s}d^{t}\rangle$ to its product with $\langle c^{s^{\prime}}d^{t^{\prime}}\rangle$ or $\langle a^{\alpha}bc^{s^{\prime}}d^{t^{\prime}}\rangle$ with $\alpha=0, 1$, and the unitary matrix $V_{r}$ is still given by $\mathcal{N}_4$ and $\mathcal{N}_5$, respectively, as shown in Table~\ref{tab:extension}. In short, for type D groups, the diagonalization matrix $V_{r}$ enforced by residual symmetry can take only five distinct forms $\mathcal{N}_{1,2,3,4,5}$ in Eqs.~(\ref{eq:N1}, \ref{eq:N2}, \ref{eq:N3}, \ref{eq:N4}, \ref{eq:N5}) where both $s$ and $t$ are free of any constraints. For type C groups, the generator $b$ is absent; therefore $V_{r}$ can only be of the form of $\mathcal{N}_{1,3}$.

\begin{table}[t!]
\begin{center}
\renewcommand\arraystretch{1.3}
\begin{tabular}{|c|c|c|c|}
\hline\hline
$\mathcal{G}_1$ & $\mathcal{G}_2$ & Constraints on group parameters  & Form of $V_{r}$\\\hline
$\langle c^sd^t\rangle$   & $\langle c^{s^{\prime}}d^{t^{\prime}}\rangle$                   & $\begin{array}{l}
\left\{\begin{array}{c}
(2k+1)s-2rt=0\pmod{m}\\
(k+2)s^{\prime}-rt^{\prime}=0\pmod{m}
\end{array}\right.\\[0.15in]
\hskip-0.16in\text{or} \left\{\begin{array}{c}
(2k+1)s-2rt=0\pmod{m}\\
(k-1)s^{\prime}-rt^{\prime}=0\pmod{m}
\end{array}\right.\\[0.15in]
\hskip-0.16in\text{or}
\left\{\begin{array}{c}
(k+2)s-rt=0\pmod{m}\\
(k-1)s^{\prime}-rt^{\prime}=0\pmod{m}
\end{array}\right.\\[0.1in]
\qquad\quad(s\leftrightarrow s^{\prime},~~ t\leftrightarrow t^{\prime})
\end{array}$  &   $\mathcal{N}_1$\\\hline

\multirow{2}{*}[-12pt]{$\langle bc^sd^t\rangle$}  &   $\langle c^{s^{\prime}}d^{t^{\prime}}\rangle$                  & $\begin{array}{c}
(2k+1)s^{\prime}-2rt^{\prime}=0\pmod{m}\\
3s=0\pmod{m}
\end{array}$  &  \multirow{2}{*}[-12pt]{$\mathcal{N}_2$} \\\cline{2-3}

   &   $\langle bc^{s^{\prime}}d^{t^{\prime}}\rangle$                  &  $\begin{array}{c}
(2k+1)(s-s^{\prime})-2r(t-t^{\prime})=l_1m\pmod{2m}\\
3s=l_2m\pmod{2m},\quad 3s^{\prime}=l_3m\pmod{2m}
\end{array}$ &  \\\hline

\multirow{2}{*}[-22pt]{$\langle abc^sd^t\rangle$}   & $\langle c^{s^{\prime}}d^{t^{\prime}}\rangle$                & $\begin{array}{c}
(k+2)s^{\prime}-rt^{\prime}=0\pmod{m}\\
3(rt-ks)=0\pmod{m}
\end{array}$ & \multirow{2}{*}[-22pt]{$\mathcal{N}_4$} \\\cline{2-3}

  &   $\langle abc^{s^{\prime}}d^{t^{\prime}}\rangle$  & $\begin{array}{c}
(k+2)(s-s^{\prime})-r(t-t^{\prime})=l_1m\pmod{2m}\\
3(rt-ks)=l_2m\pmod{2m}\\
3(rt^{\prime}-ks^{\prime})=l_3m\pmod{2m}
\end{array}$  &  \\\hline

\multirow{2}{*}[-22pt]{$\langle a^2bc^sd^t\rangle$}   & $\langle c^{s^{\prime}}d^{t^{\prime}}\rangle$     & $\begin{array}{c}
(k-1)s^{\prime}-rt^{\prime}=0\pmod{m}\\
3((k+1)s-rt)=0\pmod{m}
\end{array}$ &  \multirow{2}{*}[-22pt]{$\mathcal{N}_5$}\\\cline{2-3}

   &   $\langle a^2bc^{s^{\prime}}d^{t^{\prime}}\rangle$               & $\begin{array}{c}
(k-1)(s-s^{\prime})-r(t-t^{\prime})=l_1m\pmod{2m}\\
3((k+1)s-rt)=l_2m\pmod{2m}\\
3((k+1)s^{\prime}-rt^{\prime})=l_3m\pmod{2m}
\end{array}$ &   \\\hline\hline
\end{tabular}
\end{center}
\caption{\label{tab:extension}The possible extension of the remnant flavor symmetry $\mathcal{G}_r=\mathcal{G}_1\times\mathcal{G}_2$ and the corresponding unitary transformation $V_{r}$. Here we confine
ourselves to the minimal residual symmetry $\mathcal{G}_{r}$ that allows the distinction of three generations. In other words, $\mathcal{G}_1$ or $\mathcal{G}_2$ alone fixes only a column in $V_r$, and the column vectors associated with $\mathcal{G}_1$ or $\mathcal{G}_2$ should be different. As a result, we have the parameters $l_{1,2,3}=0,1$ and $l_1+l_2+l_3=1,3$.}
\end{table}

\subsection{\label{subsec:PMNS_analytical_Dirac}Predictions of lepton mixing matrix for Dirac neutrino}

If neutrinos are Dirac particles, both remnant symmetries $\mathcal{G}_{l}$ and $\mathcal{G}_{\nu}$ can be any Abelian subgroups that are sufficient to distinguish the three generations of leptons. As a result, the charged lepton and neutrino diagonalization matrices $V_{l}$ and $V_{\nu}$ can be any ones of $\mathcal{N}_{1,2,3,4,5}$ when the flavor symmetry is a group of type D. The PMNS matrix is defined as
\begin{equation}
\label{eq:Upmns} U_{\text{PMNS}}=V_l^{\dagger}V_{\nu}\,.
\end{equation}
There are altogether $5\times5=25$ possible combinations of $V_{l}$ and $V_{\nu}$, and they can be further divided into seven classes as follows. Here we would like to emphasize again that the PMNS matrix is determined up to independent row and column permutations and the multiplication of arbitrary phase matrices from the left and the right sides. In the following, $s$ and $t$ denote the exponents of the generators $c$ and $d$, respectively in the residual symmetry $\mathcal{G}_l$, and $p$ and $q$ are the exponents of $c$ and $d$ in $\mathcal{G}_{\nu}$.
\begin{itemize}[labelindent=-0.7em, leftmargin=1.6em]

\item{$(V_{l}, V_{\nu})=(\mathcal{N}_1,\mathcal{N}_1)$}
\begin{equation}
\label{eq:couple1_gen}
U_{\text{PMNS}}=\left(
\begin{array}{ccc}
 1 & 0 & 0 \\
 0 & 1 & 0 \\
 0 & 0 & 1
\end{array}
\right)\,.
\end{equation}

\item{$(V_{l},V_{\nu})=(\mathcal{N}_1,\mathcal{N}_3)$, $(\mathcal{N}_3,\mathcal{N}_1)$}
\begin{equation}
\label{eq:couple2_gen}
U_{\text{PMNS}}=\frac{1}{\sqrt{3}}
\left(
\begin{array}{ccc}
 1 & 1 & 1 \\
 1 & \omega  & \omega ^2 \\
 1 & \omega ^2 & \omega
\end{array}
\right)\,,
\end{equation}
which is the so-called democratic mixing matrix in which all elements have the same absolute value~\cite{Cabibbo:1977nk}.

\item{$(V_{l}, V_{\nu})=(\mathcal{N}_1,\mathcal{N}_2),(\mathcal{N}_2,\mathcal{N}_1),(\mathcal{N}_1,\mathcal{N}_4),(\mathcal{N}_4,\mathcal{N}_1),(\mathcal{N}_1,\mathcal{N}_5),(\mathcal{N}_5,\mathcal{N}_1),(\mathcal{N}_2,\mathcal{N}_2),(\mathcal{N}_4,\mathcal{N}_4),(\mathcal{N}_5,\mathcal{N}_5)$}
\begin{equation}
\label{eq:couple3_gen}
U_{\text{PMNS}}=\left(
\begin{array}{ccc}
 1 & 0 & 0 \\
 0 & ~\cos\theta &~ -\sin\theta \\
 0 & ~\sin\theta &~ \cos\theta
\end{array}
\right)\,,
\end{equation}
where
\begin{equation}
\label{eq:theta3}
\theta=\left\{
\begin{array}{cl}
\frac{\pi}{4}, &~~ (\mathcal{N}_1,\mathcal{N}_2),(\mathcal{N}_1,\mathcal{N}_4),(\mathcal{N}_1,\mathcal{N}_5),\\
 &~~(\mathcal{N}_2,\mathcal{N}_1),(\mathcal{N}_4,\mathcal{N}_1),(\mathcal{N}_5,\mathcal{N}_1)\\[0.04in]
\frac{(2k+1)(s-p)-2r(t-q)}{2m}\pi, &~~ (\mathcal{N}_2,\mathcal{N}_2)\\[0.04in]
\frac{(k+2)(s-p)-r(t-q)}{2m}\pi, & ~~ (\mathcal{N}_4,\mathcal{N}_4)\\[0.04in]
\frac{(1-k)(s-p)+r(t-q)}{2m}\pi, & ~~(\mathcal{N}_5,\mathcal{N}_5)
\end{array}
\right.
\end{equation}

\item {$(V_{l}, V_{\nu})=(\mathcal{N}_2,\mathcal{N}_4)$, $(\mathcal{N}_4,\mathcal{N}_2)$, $(\mathcal{N}_2,\mathcal{N}_5)$, $(\mathcal{N}_5,\mathcal{N}_2)$, $(\mathcal{N}_4,\mathcal{N}_5)$, $(\mathcal{N}_5,\mathcal{N}_4)$}
\begin{equation}
\label{eq:couple4_gen}
U_{\text{PMNS}}=\frac{1}{2}\left(
\begin{array}{ccc}
\sqrt{2} & \sqrt{2} & 0 \\
-1 & 1 &  \sqrt{2} \\
1& -1 & \sqrt{2}
\end{array}
\right)\,.
\end{equation}
Obviously all the above four possible forms of $U_{\text{PMNS}}$ in Eqs.~(\ref{eq:couple1_gen}, \ref{eq:couple2_gen}, \ref{eq:couple3_gen}, \ref{eq:couple4_gen}) are not compatible with the present experimental data~\cite{Gonzalez-Garcia:2014bfa}.

\item{$(V_{l}, V_{\nu})=(\mathcal{N}_2,\mathcal{N}_3)$, $(\mathcal{N}_4,\mathcal{N}_3)$, $(\mathcal{N}_5,\mathcal{N}_3)$}
\begin{equation}
\label{eq:couple5_gen}
U_{\text{PMNS}}=
\frac{1}{\sqrt{3}}\left(
\begin{array}{ccc}
\sqrt{2} \cos\theta &~ -\sqrt{2}\cos(\theta-\frac{\pi}{3}) &~ -\sqrt{2}\cos(\theta+\frac{\pi}{3}) \\
 1 &~ 1 &~ 1 \\
-\sqrt{2}\sin\theta &~ \sqrt{2}\sin(\theta-\frac{\pi}{3}) &~\sqrt{2}\sin(\theta+\frac{\pi}{3})
\end{array}
\right)\,,
\end{equation}
where
\begin{equation}
\label{eq:theta5}
\theta=\left\{
\begin{array}{cl}
\frac{-2(k+1)p+(2k+1)s-2r(t-q)}{2m}\pi, &~~ (\mathcal{N}_2,\mathcal{N}_3)\\[0.04in]
\frac{(k+2)s-2p-rt}{2m}\pi, &~~ (\mathcal{N}_4,\mathcal{N}_3)\\[0.04in]
\frac{-2kp-(1-k)s-r(t-2q)}{2m}\pi, &~~ (\mathcal{N}_5,\mathcal{N}_3)
\end{array}
\right.
\end{equation}
We see that one row of the PMNS matrix is $(1, 1, 1)/\sqrt{3}$ in this case. The lepton mixing angles can easily be read out as
{\small
\begin{equation}
\label{eq:mixing_angles_gen}\sin^2\theta_{13}=\frac{2}{3}\cos^2(\theta+\frac{\pi}{3}),~\sin^2\theta_{12}=\frac{2\cos^2(\theta-\frac{\pi}{3})}{1+2\sin^2(\theta+\frac{\pi}{3})},~\sin^2\theta_{23}=\frac{1}{1+2\sin^2(\theta+\frac{\pi}{3})}\,.
\end{equation}}
Eliminating the parameter $\theta$, correlations among the three mixing angles follow immediately:
\begin{equation}
\sin^2\theta_{12}=\frac{1}{2}\pm\frac{1}{2}\tan\theta_{13}\sqrt{2-\tan^2\theta_{13}},\qquad 3\sin^2\theta_{23}\cos^2\theta_{13}=1\,.
\end{equation}
Inputting the $3\sigma$ range $0.0188\leq\sin^2\theta_{13}\leq0.0251$~\cite{Gonzalez-Garcia:2014bfa}, we obtain $0.387\leq\sin^2\theta_{12}\leq0.403$ (or $0.597\leq\sin^2\theta_{12}\leq0.613$ for the plus sign) and $0.340\leq\sin^3\theta_{23}\leq0.342$, which are not favored by the present data~\cite{Gonzalez-Garcia:2014bfa}. Alternatively, we can exchange the second and third rows of the PMNS matrix in Eq.~\eqref{eq:couple5_gen}, the atmospheric mixing angle changes from $\theta_{23}$ to $\pi/2-\theta_{23}$, and consequently we have $0.658\leq\sin^3\theta_{23}\leq0.660$. Although the mixing pattern in Eq.~\eqref{eq:couple5_gen} fails to describe the data well, small corrections to the leading order results are generally expected to exist in concrete models~\cite{Altarelli:2010gt,King:2013eh,King:2014nza}. As a consequence, this pattern could possibly be brought into agreement with the experimental data after higher order corrections are included. This mixing pattern should be interesting from the model building perspective.

\item{$(V_{l}, V_{\nu})=(\mathcal{N}_3,\mathcal{N}_3)$}
\begin{equation}
\label{eq:couple6_gen}
U_{\text{PMNS}}=\frac{1}{3}\left(
\begin{array}{ccc}
1+2e^{i\delta}\cos\theta &~ 1-2e^{i\delta}\cos\left(\theta-\frac{\pi}{3}\right) &~ 1-2e^{i\delta}\cos\left(\theta+\frac{\pi}{3}\right) \\
1-2e^{i\delta}\cos\left(\theta+\frac{\pi}{3}\right) &~ 1+2e^{i\delta}\cos\theta  &~  1-2e^{i\delta}\cos\left(\theta-\frac{\pi}{3}\right) \\
1-2e^{i\delta}\cos\left(\theta-\frac{\pi}{3}\right) & ~1-2e^{i\delta}\cos\left(\theta+\frac{\pi}{3}\right) &
~1+2e^{i\delta}\cos\theta
\end{array}
\right)\,,
\end{equation}
where
\begin{equation}
\label{eq:theta6}
\theta=\frac{k(s-p)-r(t-q)}{m}\pi,\qquad\delta=\frac{(k+2)(s-p)-r(t-q)}{m}\pi\,.
\end{equation}
The possible exchange of the second and third rows of the PMNS matrix leads to another different mixing pattern. The lepton mixing matrices arising from other permutations of rows and columns can be obtained from these two patterns by redefining the parameter $\theta$. We notice that the solar and the atmospheric mixing angles are closely related as $\sin^2\theta_{12}=\sin^2\theta_{23}$ or $\sin^2\theta_{12}=1-\sin^2\theta_{23}$, which are not compatible with the preferred values from global fits at the $3\sigma$ level~\cite{Gonzalez-Garcia:2014bfa}. Furthermore, we perform a numerical analysis where both $\theta$ and $\delta$ are treated as random numbers in the interval of $[-\pi, \pi]$. The correlation and the allowed region of $\sin\theta_{13}$ and $\sin^2\theta_{12}$ are plotted in Fig.~\ref{fig:theta13_23_N3N3}. Obviously the observed values of $\sin\theta_{13}$ and $\sin^2\theta_{12}$ cannot be accommodated simultaneously. Therefore this mixing pattern is not viable.

\begin{figure}[t!]
\begin{center}
\includegraphics[width=0.6\textwidth]{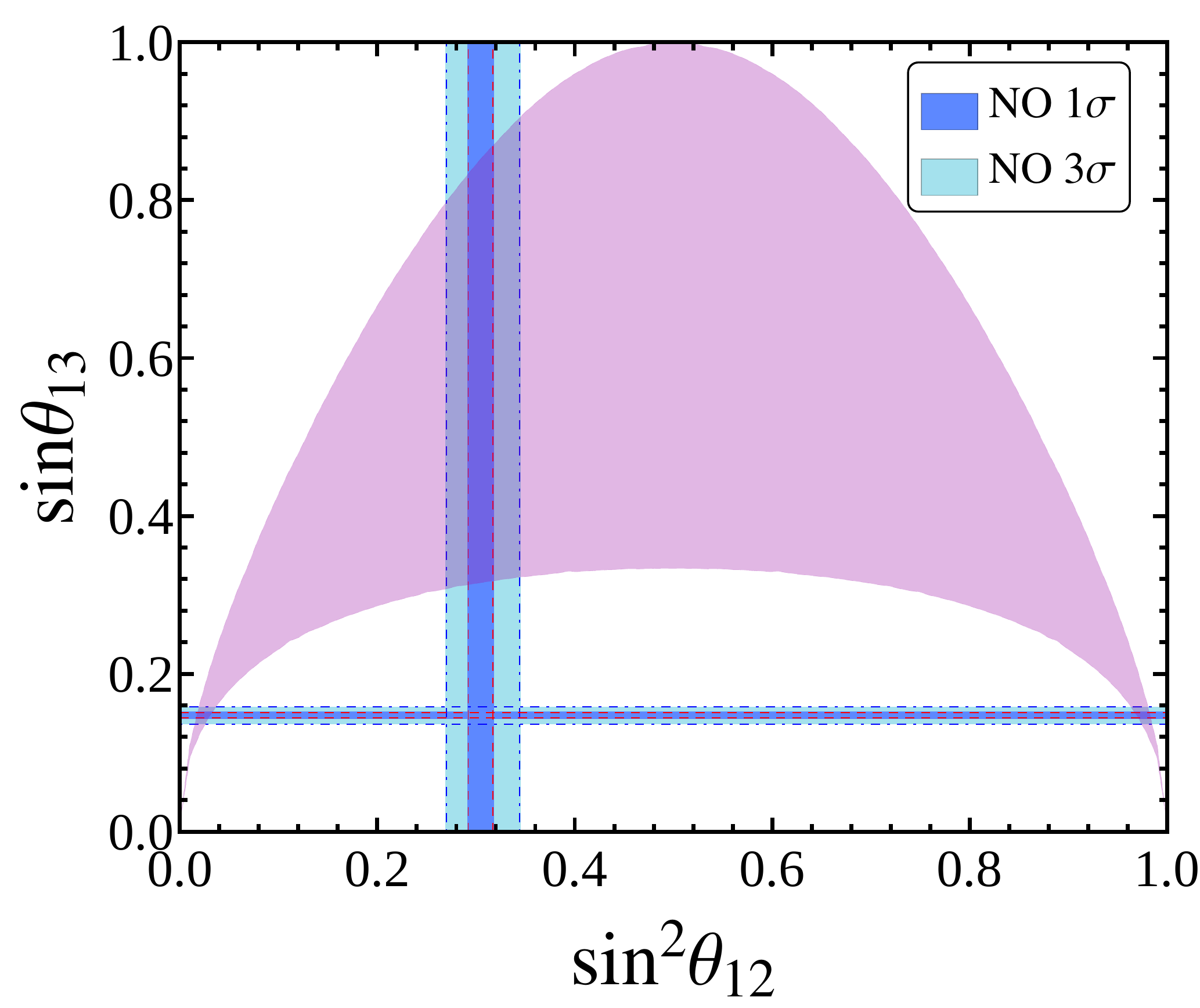}
\caption{\label{fig:theta13_23_N3N3}The correlation between $\sin\theta_{13}$ and $\sin^2\theta_{12}$ for the mixing pattern in Eq.~\eqref{eq:couple6_gen}. The pink region represents the
possible values of the mixing angles when $\theta$ and $\delta$ freely vary in the region of $[-\pi, \pi]$. The $1\sigma$ and $3\sigma$ bounds  are adopted from Ref.~\cite{Gonzalez-Garcia:2014bfa}.}
\end{center}
\end{figure}

\item{$(V_{l}, V_{\nu})=(\mathcal{N}_3,\mathcal{N}_2)$, $(\mathcal{N}_3,\mathcal{N}_4)$, $(\mathcal{N}_3,\mathcal{N}_5)$}
\begin{equation}
\label{eq:couple7_gen}
U_{\text{PMNS}}=\frac{1}{\sqrt{3}}\left(
\begin{array}{ccc}
\sqrt{2}\cos\theta &~ 1 &~ -\sqrt{2}\sin\theta \\
-\sqrt{2}\cos\left(\theta-\frac{\pi}{3}\right) &~ 1 &~ \sqrt{2}\sin\left(\theta-\frac{\pi}{3}\right) \\
-\sqrt{2}\cos\left(\theta+\frac{\pi}{3}\right) &~ 1 &~ \sqrt{2}\sin\left(\theta+\frac{\pi}{3}\right)
\end{array}
\right)\,,
\end{equation}
where
\begin{equation}
\label{eq:theta7}
\theta=\left\{
\begin{array}{cl}
\frac{-(2k+1)p+2(k+1)s-2r(t-q)}{2m}\pi, &~~ (\mathcal{N}_3,\mathcal{N}_2)\\[0.04in]
\frac{-(k+2)p+qr+2s}{2m}\pi, &~~ (\mathcal{N}_3,\mathcal{N}_4)\\[0.04in]
\frac{(1-k)p+2ks+r(q-2t)}{2m}\pi, &~~ (\mathcal{N}_3,\mathcal{N}_5)
\end{array}
\right.
\end{equation}
We see that one column of the PMNS matrix in Eq.~\eqref{eq:couple7_gen} is $(1, 1, 1)^{T}/\sqrt{3}$. This mixing pattern is known as the trimaximal mixing. To investigate the permutations of rows and columns of $U_{\text{PMNS}}$, it is appropriate to use the representation of the permutations $\mathcal{P}\in S_3$ as permutation matrices,
\begin{equation}
\label{eq:permrep}
\hskip-0.04in \mathcal{P}\to M(\mathcal{P})=(e_{\mathcal{P}(1)},e_{\mathcal{P}(2)},e_{\mathcal{P}(3)})~~\text{with}~~ e_1=(1,0,0)^T, e_2=(0,1,0)^T, e_3=(0,0,1)^T\,.
\end{equation}
After performing row permutation $\mathcal{P}$ and column permutation $\mathcal{P}^{\prime}$, the PMNS matrix becomes
\begin{equation}
\label{eq:perm}U_{\text{PMNS}}\rightarrow M(\mathcal{P})U_{\text{PMNS}}M(\mathcal{P}^{\prime})\,.
\end{equation}
The constant vector $(1, 1, 1)^{T}/\sqrt{3}$  should be placed in the second column to accommodate the experimentally measured mixing angles, and the first and the third columns can be interchanged. On the other hand, the three rows can be permutated in any way. For all the admissible permutations, we find the resulting PMNS matrix can be obtained from the original one in Eq.~\eqref{eq:couple7_gen} by redefinition of the parameter $\theta$ up to an irrelevant diagonal matrix multiplied from the right-hand side with entries $\pm1$. The replacement rules are listed in Table~\ref{tab:mapping_viable}. Without loss of generality, we can extract the predictions for the mixing parameters from the PMNS matrix in Eq.~\eqref{eq:couple7_gen}. Since each matrix element is real, the Dirac $CP$ phase is trivial, i.e., $\sin\delta_{CP}=0$. The lepton mixing angles are given by
\begin{equation}
\label{eq:mixing_angles_gen}\sin^2\theta_{13}=\frac{2}{3}\sin^2\theta,\quad\sin^2\theta_{12}=\frac{1}{1+2\cos^2\theta},\quad\sin^2\theta_{23}=\frac{2\sin^2(\theta-\frac{\pi}{3})}{1+2\cos^2\theta}\,.
\end{equation}
\begin{table}[t!]
\renewcommand{\arraystretch}{0.9}
\centering
\begin{tabular}{|c|c|c|c|c|c|c|}\hline\hline
\diagbox{$\mathcal{P}^{\prime}$}{$\mathcal{P}$} & $(123)$ & $(213)$ & $(321)$ & $(132)$ & $(312)$ & $(231)$\\\hline
 & & & & & & \\[-0.15in]
$(123)$ & $\displaystyle{\theta\rightarrow\theta}$ & $\displaystyle{\theta\rightarrow-\theta+\frac{\pi}{3}}$ & $\displaystyle{\theta\rightarrow-\theta-\frac{\pi}{3}}$ & $\displaystyle{\theta\rightarrow-\theta}$ &  $\displaystyle{\theta\rightarrow\theta-\frac{\pi}{3}}$ & $\displaystyle{\theta\rightarrow\theta+\frac{\pi}{3}}$\\[0.10in]\hline
 & & & & & & \\[-0.15in]
$(321)$ & $\displaystyle{\theta\rightarrow\theta+\frac{\pi}{2}}$ & $\displaystyle{\theta\rightarrow-\theta-\frac{\pi}{6}}$ & $\displaystyle{\theta\rightarrow-\theta+\frac{\pi}{6}}$ & $\displaystyle{\theta\rightarrow-\theta-\frac{\pi}{2}}$ & $\displaystyle{\theta\rightarrow\theta+\frac{\pi}{6}}$& $\displaystyle{\theta\rightarrow\theta-\frac{\pi}{6}}$ \\[0.10in]\hline\hline
\end{tabular}
\caption{\label{tab:mapping_viable}The replacement rules by which the permutated PMNS matrix can be obtained from the initial one in Eq.~\eqref{eq:couple7_gen}, where $\mathcal{P}$ and $\mathcal{P}^{\prime}$ refer to permutation of rows and columns respectively. Here a generic permutation $(1, 2, 3)\rightarrow (n_1, n_2, n_3)$ is simply denoted by $(n_1 n_2 n_3)$.}
\end{table}
As the three mixing angles depend on a single parameter $\theta$, they are strongly correlated,
\begin{equation}
\label{eq:correlation_trimaximal}3\sin^2\theta_{12}\cos^2\theta_{13}=1,\qquad \sin^2\theta_{23}=\frac{1}{2}\pm\frac{1}{2}\tan\theta_{13}\sqrt{2-\tan^2\theta_{13}}\,,
\end{equation}
which are displayed in Fig.~\ref{fig:corr_gen}. The solar mixing angle has a lower bound given by $\sin^2\theta_{12}\geq1/3$ due to the first relation in Eq.~\eqref{eq:correlation_trimaximal}. The measurement of the reactor mixing angle $\sin^2\theta_{13}\simeq0.0218$ yields $\sin^2\theta_{12}\simeq0.341$, $\sin^2\theta_{23}\simeq0.395$ or 0.605 which are within the experimentally preferred $3\sigma$ ranges although the solar mixing angle is somewhat close to the present $3\sigma$ upper limit~\cite{Gonzalez-Garcia:2014bfa}.

\end{itemize}

The groups of type C can be generated by three generators $a$, $c$, and $d$, and the generator $b$ is dropped. Consequently the diagonalization matrices $V_{l}$ and $V_{\nu}$ can be $\mathcal{N}_1$ and $\mathcal{N}_3$ shown in Eqs.~(\ref{eq:N1}, \ref{eq:N3}). As a result, the lepton flavor mixing matrix would be a unit matrix for $(V_{l}, V_{\nu})=(\mathcal{N}_1,\mathcal{N}_1)$, democratic mixing for $(V_{l}, V_{\nu})=(\mathcal{N}_1,\mathcal{N}_3)$, $(\mathcal{N}_3,\mathcal{N}_1)$ or a form of Eq.~\eqref{eq:couple6_gen} for $(V_{l}, V_{\nu})=(\mathcal{N}_3,\mathcal{N}_3)$. As discussed above, all three mixing patterns are excluded by neutrino oscillation data. Therefore if lepton flavor mixing is completely determined by residual symmetries that are subgroups of flavor symmetry, type C groups such as the typical $T_n$ and $\Delta(3n^2)$ series are not suitable as a flavor symmetry. However, in the indirect approach~\cite{King:2013eh} where flavor symmetry is
completely broken, viable lepton mixing may be obtained from a flavor symmetry group of type C for some special vacuum alignments.

\begin{figure}[t!]
\begin{center}
\includegraphics[width=0.98\textwidth]{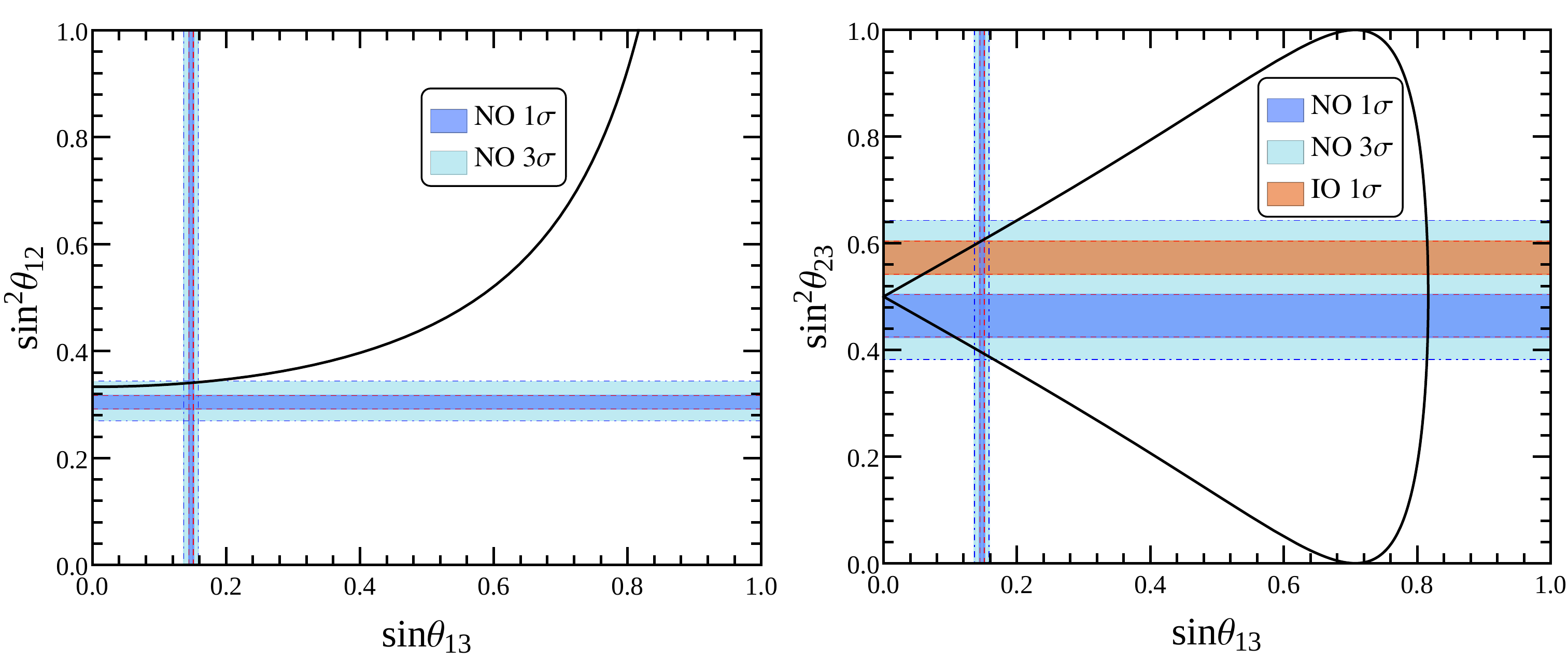}
\caption{\label{fig:corr_gen}Correlations between the mixing angles for the viable mixing pattern in Eq.~\eqref{eq:couple7_gen}, where the flavor symmetry is a group of type D. }
\end{center}
\end{figure}

\subsection{\label{subsec:PMNS_analytical_Maj}Predictions of lepton mixing matrix for Majorana neutrino}

If neutrinos are assumed to be Majorana particles, the residual subgroup $\mathcal{G}_{\nu}$ should be a Klein group $Z_2\times Z_2$, while $\mathcal{G}_{l}$ can still be any Abelian subgroups of $\mathcal{G_F}$ with order not less than 3. For a type D flavor symmetry group, its $Z_2$ elements are given by
\begin{equation}
\label{eq:Z2_1} abc^{x_1}d^{y_1},\quad a^2bc^{x_2}d^{y_2},\quad bd^{y_3}\,,
\end{equation}
where the integers $x_i$, $y_i$ ($i=1,2$) satisfy
\begin{equation}
\label{eq:xyrelation}kx_1-ry_1=0\hskip-0.08in\pmod{m},\quad (k+1)x_2-ry_2=0\hskip-0.08in\pmod{m}\,.
  \end{equation}
There are another three order 2 elements
\begin{equation}
\label{eq:Z2_2} d^{n/2},\quad c^{m/2}d^{kn/2},\quad c^{m/2}d^{(k+1)n/2}
\end{equation}
for even $n$. Then $m$ should be even as well since $r=m/n=1$ or 3. It is wellknown that a Klein group is generated by two order 2 elements. We find that groups of type D have the following four Klein subgroups if $n$ is divisible by 2.
\begin{eqnarray}
\nonumber&&K_4^{(1)}=\{1,d^{n/2},c^{m/2}d^{kn/2},c^{m/2}d^{(k+1)n/2}\},\\
\nonumber&&K_4^{(2)}=\{1,d^{n/2},bd^{y_3},bd^{y_3+n/2}\},\\
\nonumber&&K_4^{(3)}=\{1,c^{m/2}d^{kn/2},abc^{x_1}d^{y_1},abc^{x_1+m/2}d^{y_1+kn/2}\},\\
&&K_4^{(4)}=\{1,c^{m/2}d^{(k+1)n/2},a^2bc^{x_2}d^{y_2},a^2bc^{x_2+m/2}d^{y_2+(k+1)n/2}\}\,.
\end{eqnarray}
Notice that $K^{(1)}_4$ is a normal subgroup while the latter three Klein subgroups are conjugate to each other as
\begin{equation}
K^{(3)}_4=a^2K^{(2)}_4a,\quad K^{(4)}_4=aK^{(2)}_4a^2\,.
\end{equation}
If the full Klein symmetry  is preserved by the neutrino mass matrix, the unitary transformation $V_{\nu}$ that diagonalizes the neutrino mass matrix can easily be obtained by evaluating the eigenvalues and eigenvectors of the representation matrices. The explicit form of $V_{\nu}$ for each residual Klein symmetry is given by
\begin{eqnarray}
\nonumber&&K_4^{(1)}:  V_{\nu}=\left(
\begin{array}{ccc}
 1 & 0 & 0 \\
 0 & 1 & 0 \\
 0 & 0 & 1
\end{array}
\right)\mapsto\mathcal{N}_1,\\
\nonumber&&K_4^{(2)}:  V_{\nu}=\frac{1}{\sqrt{2}}
\left(\begin{array}{ccc}
 0 & 0 & \sqrt{2} \\
 \eta^{ry_3} &~ -\eta^{ry_3} & 0 \\
 1 & 1 & 0
\end{array}
\right)
\mapsto\mathcal{N}_2(s=0,t=y_3),\\
\nonumber&&K_4^{(3)}:  V_{\nu}=\frac{1}{\sqrt{2}}
\left(\begin{array}{ccc}
 0 &~ \eta^{-x_1} &~ -\eta^{-x_1} \\
 \sqrt{2} & 0 & 0 \\
 0 & 1 & 1
\end{array}
\right)\mapsto\mathcal{N}_4(s=x_1,t=y_1),\\
&&K_4^{(4)}:  V_{\nu}=\frac{1}{\sqrt{2}}
\left(\begin{array}{ccc}
0 &~ \eta^{-x_2} &~ -\eta^{-x_2} \\
0 & 1 & 1 \\
\sqrt{2} & 0 & 0
\end{array}
\right)\mapsto\mathcal{N}_5(s=x_2, t=y_2)\,,
\end{eqnarray}
where the constraints in Eq.~\eqref{eq:xyrelation} are taken into account. From the general analysis of Sec~\ref{subsec:PMNS_analytical_Dirac}, we see that the PMNS would be a unit matrix, a democratic mixing matrix, or a rotation matrix in the (23)-plane if the neutrino residual symmetry $\mathcal{G}_{\nu}=K^{(1)}_4$. The predictions in agreement with the experimental data can be achieved for $\mathcal{G}_{l}=\langle ac^{s}d^{t}\rangle$ and $\mathcal{G}_{\nu}=K^{(2)}_4$, $K^{(3)}_4$, or $K^{(4)}_4$, and the PMNS matrix is of the trimaximal form
\begin{equation}
\label{eq:majoranaUpmns}U_{\text{PMNS}}=\frac{1}{\sqrt{3}}\left(
\begin{array}{ccc}
\sqrt{2}\cos\theta &~ 1 &~ -\sqrt{2}\sin\theta \\
-\sqrt{2}\cos\left(\theta-\frac{\pi}{3}\right) &~  1 & ~\sqrt{2}\sin\left(\theta-\frac{\pi}{3}\right) \\
-\sqrt{2}\cos\left(\theta+\frac{\pi}{3}\right) &~ 1 & ~\sqrt{2}\sin\left(\theta+\frac{\pi}{3}\right)
\end{array}
\right)\,,
\end{equation}
with the angle
\begin{equation}
\label{eq:theta_Maj}
\theta=\left\{
\begin{array}{cl}
\frac{(k+1)s-r(t-y_3)}{m}\pi, &~~\mathcal{G}_{\nu}=K^{(2)}_4 \\[0.04in]
\frac{s-x_1}{m}\pi, &~~\mathcal{G}_{\nu}=K^{(3)}_4 \\[0.04in]
\frac{ks-rt+x_2}{m}\pi, &~~\mathcal{G}_{\nu}=K^{(4)}_4
\end{array}
\right.
\end{equation}
where $m=rn$ and $n$ is even. The corresponding predictions for the mixing angles are given in Eq.~\eqref{eq:mixing_angles_gen}. Then we proceed to discuss the case that the flavor symmetry is a finite $SU(3)$ subgroup of type C. The Klein subgroup would be uniquely $K^{(1)}_4$, the charged lepton diagonalization matrix $V_{l}$ can be $\mathcal{N}_1$ or $\mathcal{N}_3$, and therefore the lepton flavor mixing is a trivially identity matrix or the democratic mixing pattern, both of which are not phenomenologically viable.

\subsection{\label{subsec:numerical_results}Numerical results}

As we have shown in Secs.~\ref{subsec:PMNS_analytical_Dirac} and \ref{subsec:PMNS_analytical_Maj}, among all the possible mixing patterns derived from the type D groups as a flavor symmetry, only the trimaximal mixing can be in agreement with experimental data. However, no realistic mixing patterns can be obtained from the type C groups unless the remnant symmetries $\mathcal{G}_{l}$ and $\mathcal{G}_{\nu}$ of the lepton mass matrices are partially contained in the flavor symmetry group or are purely accidental.
The explicit form of the PMNS matrix $U_{\text{PMNS}}$ is presented in Eq.~\eqref{eq:couple7_gen}, and the predictions for the mixing angles are shown in Eq.~\eqref{eq:mixing_angles_gen}. Obviously both $U_{\text{PMNS}}$ and mixing angles periodically depend on a single real parameter $\theta$ with period $2\pi$. Without loss of generality, the fundamental interval of the parameter $\theta$ is chosen to be $(-\pi,\pi~]$ in the following. $\theta$ can take a series of discrete values, and the expressions of $\theta$ values are given in Eqs.~\eqref{eq:theta7} and \eqref{eq:theta_Maj} for Dirac and Majorana neutrinos, respectively. Notice that permutations of rows and columns of the PMNS matrix are equivalent to a redefinition of $\theta$ listed in Table~\ref{tab:mapping_viable}.

Depending on the values of the parameters $k$ and $r$, the groups of type D  can be classified into two categories: $D^{(0)}_{n,n}$ for $k=0$, $r=1$ and $D^{(1)}_{3n,n}$ for $k=1$, $r=3$, as shown in Table~\ref{tab:CDsummary}. Since $D^{(1)}_{3n,n}$ is isomorphic to $Z_3\times D^{(0)}_{n,n}$ if $n$ cannot be divisible by 3, the representation matrices of $D^{(1)}_{3n,n}$ can be obtained from those of $D^{(0)}_{n,n}$ by multiplying $1$, $e^{2\pi{i}/3}$, or $e^{-2\pi{i}/3}$. Therefore $D^{(1)}_{3n,n}$ and $D^{(0)}_{n,n}$ give rise to the same sets of PMNS matrices for $3\nmid{n}$. It is sufficient to focus only on the $D^{(0)}_{n,n}$ and $D^{(1)}_{9n^{\prime},3n^{\prime}}$ series.

\begin{description}
\item[a)] \textbf{For groups of} $D_{n,n}^{(0)}\cong \Delta(6n^2)$
\end{description}

\begin{figure}[htb!]
\begin{center}
\includegraphics[width=0.98\textwidth]{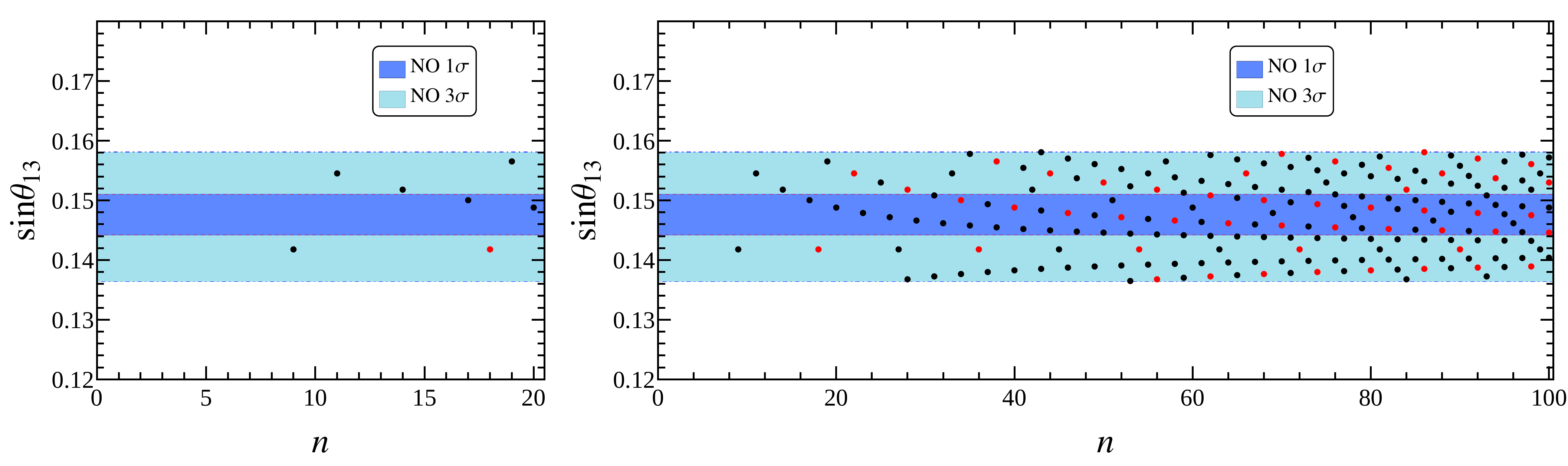}
\includegraphics[width=0.98\textwidth]{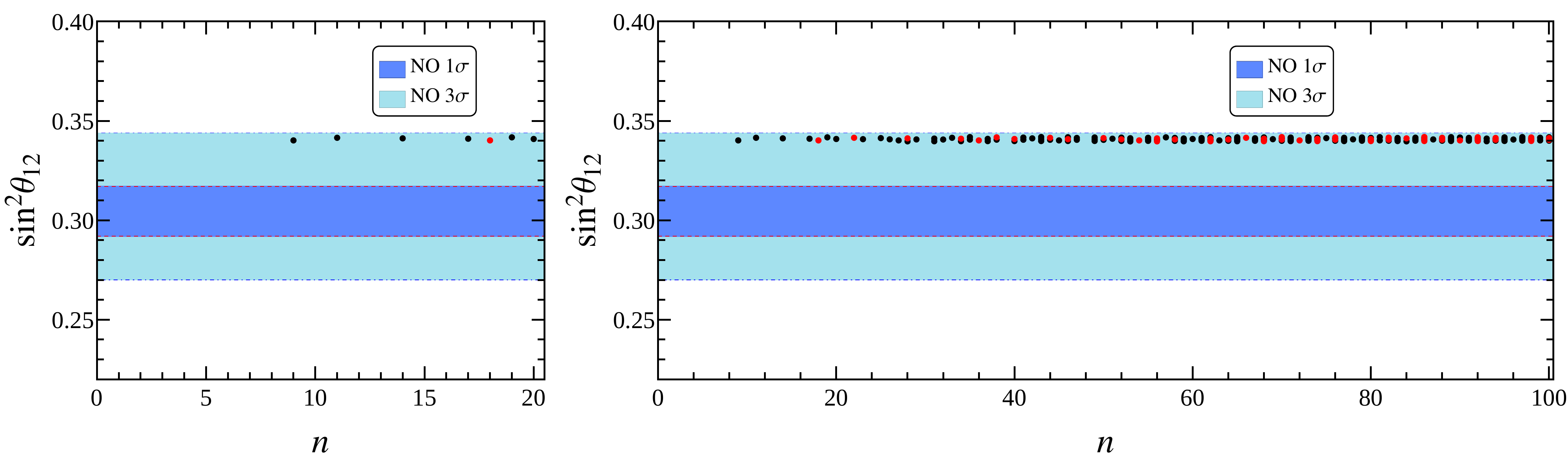}
\includegraphics[width=0.98\textwidth]{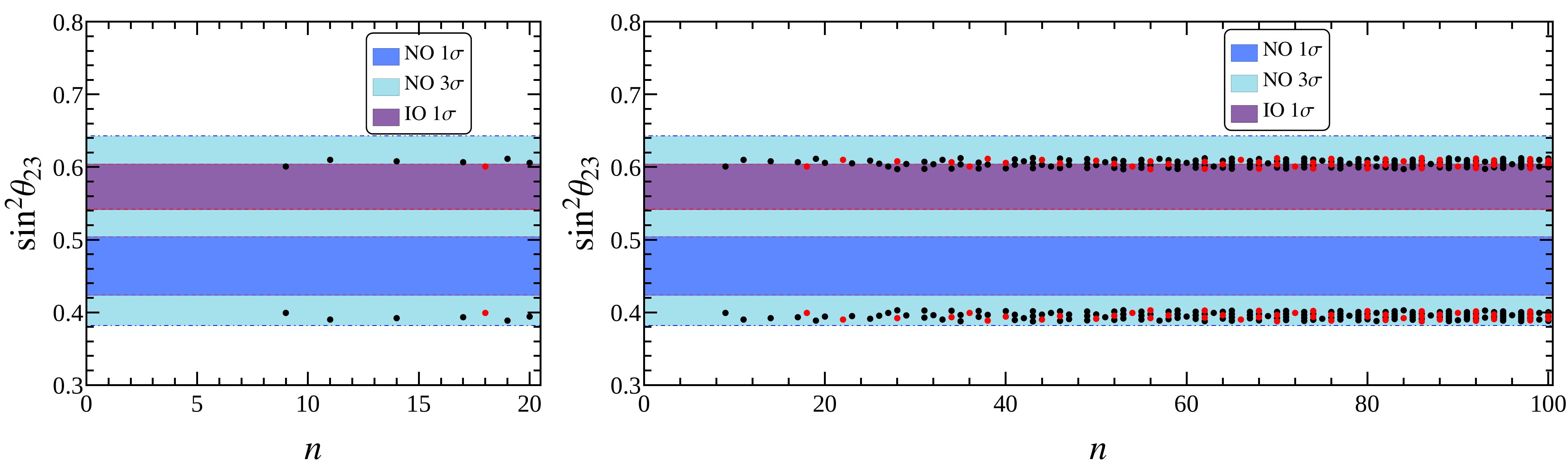}
\caption{\label{fig:Dtype1}The viable values of $\sin\theta_{13}$, $\sin^2\theta_{12}$, and $\sin^2\theta_{23}$ for $D_{n,n}^{(0)}$ flavor symmetry with respect to the group index $n$, where the predicted mixing angles that fall outside of the $3\sigma$ limits are not shown. To distinguish the predictions for the Dirac and Majorana neutrinos, we present the data in different colors, all the red and black points refer to the predictions for Dirac neutrinos while the red points are for Majorana neutrinos. The experimentally preferred $1\sigma$ and $3\sigma$ regions are adapted from Ref.~\cite{Gonzalez-Garcia:2014bfa}.}
\end{center}
\end{figure}

We have $k=0$ and $r=1$ in this case. The parameter $\theta$ is determined by the residual symmetries as follows:
\begin{eqnarray}
\nonumber&&\theta=\frac{2s-2t-p+2q}{2n}\pi,~\frac{2s-2p+q}{2n}\pi,~\mathrm{or}~\frac{-2t+p+q}{2n}\pi,\quad \text{for Dirac neutrinos},\\
&&\theta=\frac{s-t+y_3}{n}\pi,~\frac{s-x_1}{n}\pi,~\mathrm{or}~\frac{-t+x_2}{n}\pi,\quad \text{for Majorana neutrinos}\,,
\end{eqnarray}
where $s, t, p, q, x_1, x_2, y_3=0, 1, \ldots, n-1$. Obviously $\theta$ can take the values
\begin{eqnarray}
\nonumber&&\theta\hskip-0.08in\pmod{2\pi}=-\pi+\frac{1}{2n}\pi, -\pi+\frac{2}{2n}\pi,\ldots,\pi-\frac{1}{2n}\pi, \pi,\quad \text{for Dirac neutrinos},\\
&&\theta\hskip-0.08in\pmod{2\pi}=-\pi+\frac{1}{n}\pi, -\pi+\frac{2}{n}\pi,\ldots,\pi-\frac{1}{n}\pi, \pi,\quad \text{for Majorana neutrinos}\,,
\end{eqnarray}
where $n$ has to be an even number in the case of Majorana neutrinos. We see that $\theta$ can take $4n$ and $2n$ discrete values for Dirac and Majorana neutrinos, respectively. In other words, only half of the $\theta$ values allowed for Dirac neutrinos can be achievable for Majorana neutrinos. We perform a scan from $n=2$ to $n=100$, and the possible predictions for the mixing angles $\sin\theta_{13}$, $\sin^2\theta_{12}$, and $\sin^2\theta_{23}$ for each $D_{n,n}^{(0)}$ group are plotted in Fig.~\ref{fig:Dtype1}, where we require all three mixing angles to be within the $3\sigma$ intervals~\cite{Gonzalez-Garcia:2014bfa}. To clearly see what we can obtain from flavor symmetry groups of relative small order, the subfigures with $n$ ranging from 2 to 20 are picked out. Here $n=1$ is excluded because $D_{1,1}^{(0)}\cong{S}_3$ does not have three dimensional irreducible representations. We indicate the results for Majorana neutrinos with red points. Note that the same results can also be achieved for Dirac neutrinos; i.e., there are always black points that overlap with red points. From Fig.~\ref{fig:Dtype1}, we see that only five groups of order less than 2000 can give rise to mixing angles that are in agreement with
experimental data at the $3\sigma$ level. Since the five points are located at $n=9, 11, 14, 17, 18$, the corresponding flavor symmetry groups are $\Delta(486)$, $\Delta(726)$, $\Delta(1176)$, $\Delta(1734)$, and $\Delta(1944)$, respectively. The value of $\theta$ should be fixed at $\pm\frac{\pi}{18}$, $\pm\frac{2\pi}{33}$, $\pm\frac{5\pi}{84}$, $\pm\frac{\pi}{17}$, and $\pm\frac{\pi}{18}$, respectively to obtain viable mixing angles. Furthermore, we notice that only the last group $\Delta(1944)$ allows neutrinos to be Majorana particles. Although $\Delta(1176)$ has Klein subgroups, it gives rise to the lepton mixing angles $\sin^2\theta_{13}\simeq0.0148$, $\sin^2\theta_{12}\simeq0.338$, and $\sin^2\theta_{23}\simeq0.414$ or 0.586 for Majorana neutrinos, and the reactor mixing angle $\theta_{13}$ is outside the experimentally preferred $3\sigma$ range~\cite{Gonzalez-Garcia:2014bfa}. Therefore, neutrinos should be Dirac particles in order to obtain a viable lepton mixing pattern from $\Delta(1176)$ flavor symmetry.

\begin{description}
\item[b)] \textbf{For groups of} $D_{9n^{\prime}, 3n^{\prime}}^{(1)}\cong(Z_{9n^{\prime}}\times Z_{3n^{\prime}})\rtimes S_3$
\end{description}

\begin{figure}[htb!]
\begin{center}
\includegraphics[width=0.98\textwidth]{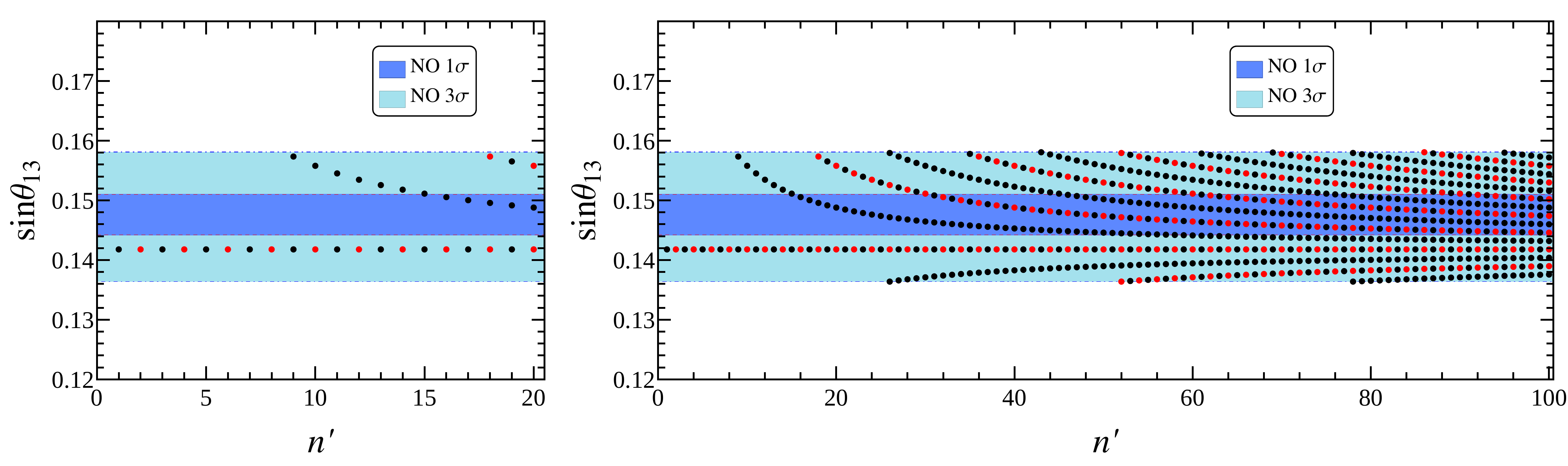}
\includegraphics[width=0.98\textwidth]{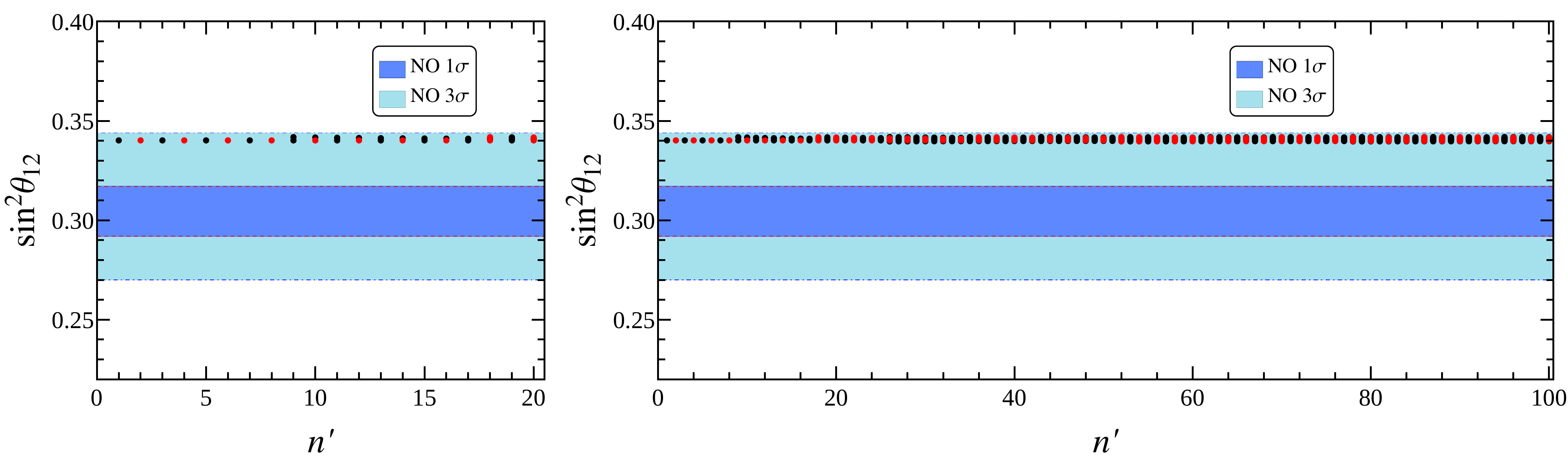}
\includegraphics[width=0.98\textwidth]{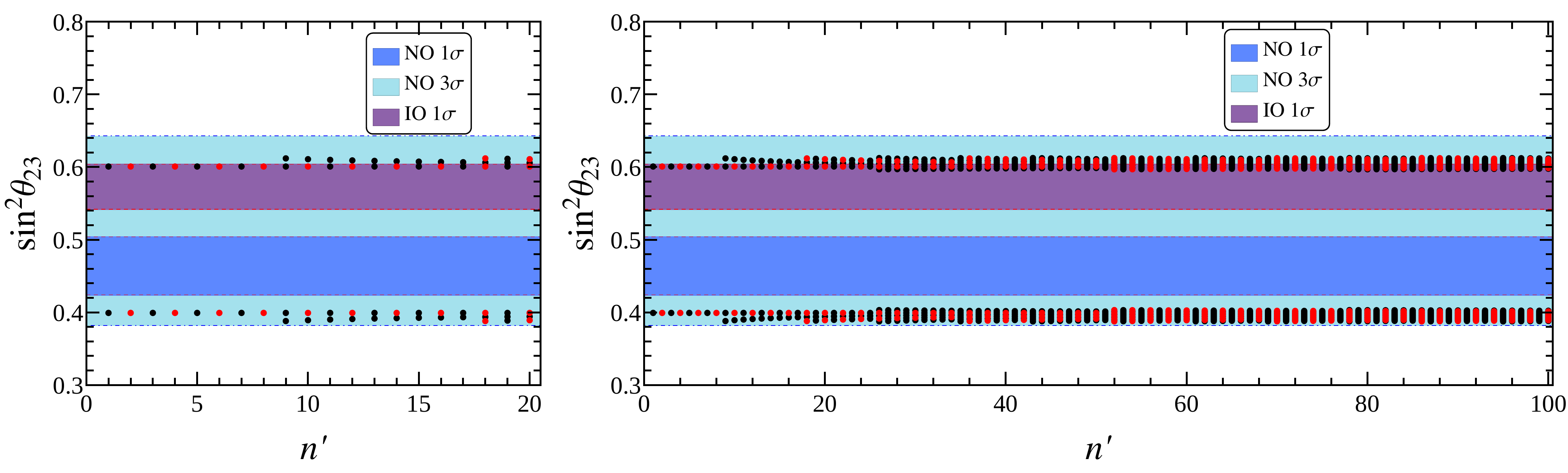}
\caption{\label{fig:Dtype2}The viable values of $\sin\theta_{13}$, $\sin^2\theta_{12}$, and $\sin^2\theta_{23}$ for $D_{9n^{\prime}, 3n^{\prime}}^{(1)}\cong(Z_{9n^{\prime}}\times Z_{3n^{\prime}})\rtimes S_3$ flavor symmetry with respect to the group index $n$, where the predicted mixing angles that fall outside of the $3\sigma$ limits are not shown. To distinguish the predictions for the Dirac and Majorana neutrinos, we present the data in different colors; all the red and black points refer to the predictions for Dirac neutrinos while the red points are for Majorana neutrinos. The experimentally preferred $1\sigma$ and $3\sigma$ regions are adapted from Ref.~\cite{Gonzalez-Garcia:2014bfa}. }
\end{center}
\end{figure}

The parameters are fixed to be $k=1$ and $r=3$. The analytical expressions for $\theta$ are given by
\begin{eqnarray}
\nonumber&&\theta=\frac{4s-6t-3p+6q}{18n^{\prime}}\pi,~\frac{2s-3p+3q}{18n^{\prime}}\pi,~\mathrm{or}~\frac{2s-6t+3q}{18n^{\prime}}\pi,\quad \text{for Dirac neutrinos},\\
&&\theta=\frac{2s-3t+3y_3}{9n'}\pi,~\frac{s-x_1}{9n'}\pi,~\mathrm{or}~\frac{s-3t+x_2}{9n'}\pi,\quad \text{for Majorana neutrinos},
\end{eqnarray}
where $s, p, x_1, x_2=0, 1,\ldots, 9n^{\prime}-1$ and $t, q, y_3=0, 1,\ldots 3n^{\prime}-1$. Hence $\theta$ can take the following discrete values:
\begin{eqnarray}
\nonumber&&\hskip-0.4in\theta\hskip-0.08in\pmod{2\pi}=-\pi+\frac{1}{18n^{\prime}}\pi, -\pi+\frac{2}{18n^{\prime}}\pi,\ldots, \pi-\frac{1}{18n^{\prime}}\pi, \pi,\quad \text{for Dirac neutrinos},\\
&&\hskip-0.4in\theta\hskip-0.08in\pmod{2\pi}=-\pi+\frac{1}{9n^{\prime}}\pi, -\pi+\frac{2}{9n^{\prime}}\pi,\ldots, \pi-\frac{1}{9n^{\prime}}\pi, \pi,\quad \text{for Majorana neutrinos}\,,
\end{eqnarray}
where $n^{\prime}$ should be divisible by two as well in the case of Majorana neutrinos. Again only half of the $\theta$ values for Dirac neutrinos can survive in the Majorana case. The numerical results are plotted in Fig.~\ref{fig:Dtype2}. We see that the mixing angles in accordance with experimental data can always be achieved for any value of $n^{\prime}$. Moreover, the first eight $D_{9n^{\prime}, 3n^{\prime}}^{(1)}$ groups with $n^{\prime}=1, 2, \ldots, 8$ lead to the same physical predictions, and accordingly the parameter $\theta$ is fixed at $\pm\frac{\pi}{18}$.
For the first three groups $D_{9,3}^{(1)}\cong(Z_{9}\times Z_{3})\rtimes S_3$, $D_{18,6}^{(1)}\cong(Z_{18}\times Z_{6})\rtimes S_3$ and $D_{27,9}^{(1)}\cong(Z_{27}\times Z_{9})\rtimes S_3$ whose orders are less than 2000, only $D_{18,6}^{(1)}\cong(Z_{18}\times Z_{6})\rtimes S_3$ comprises Klein subgroups, and neutrinos could be Majorana particles. The same values of the mixing angles as those of Refs.~\cite{Holthausen:2012wt,Talbert:2014bda} are obtained. In contrast with $\Delta(486)$ for $\mathcal{G_{F}}=\Delta(6n^2)$, $D_{9,3}^{(1)}\cong(Z_{9}\times Z_{3})\rtimes S_3$ with order 162 is the smallest group that can accommodate the experimental data.

\subsection{\label{subsec:quark_mixing_analytical}Predictions of quark mixing}

Analogous to what we have done for the lepton sector, the three generations of left-handed quarks are assigned to a faithful triplet of the flavor symmetry group $\mathcal{G_F}$, which is broken down to remnant subgroups $\mathcal{G}_{U}$ and $\mathcal{G}_{D}$ in the up quark and down quark sectors, respectively. Both $\mathcal{G}_{U}$ and $\mathcal{G}_{D}$ can be any Abelian subgroups that can distinguish among the three generations. Therefore the possible forms of the lepton mixing matrix worked out in Sec.~\ref{subsec:PMNS_analytical_Dirac} for Dirac neutrinos are also applicable to the quark flavor mixing. It is well known that the CKM matrix is predominantly a unit matrix, and the largest mixing is the Cabibbo mixing between the first and the second generations~\cite{pdg}. As a result, a trivial identity CKM matrix is a good first order approximation, and it can always be obtained from any flavor symmetry group once the residual symmetries $\mathcal{G}_{U}$ and $\mathcal{G}_{D}$ coincide. It is remarkable that a better approximation can be achieved if we choose $(V_{U}, V_{D})=(\mathcal{N}_2,\mathcal{N}_2)$, $(\mathcal{N}_4,\mathcal{N}_4)$, or $(\mathcal{N}_5,\mathcal{N}_5)$, and the corresponding CKM matrix is determined (up to permutations of rows and columns) to be of the form
\begin{equation}
\label{eq:couplequark}
V_{\text{CKM}}=\left(
\begin{array}{ccc}
 \cos\theta & ~-\sin\theta &~ 0 \\
 \sin\theta &~ \cos\theta &~ 0 \\
 0 &~ 0 &~ 1
\end{array}
\right)\,,
\end{equation}
where
\begin{equation}
\label{eq:theta_quark}
\theta=\left\{
\begin{array}{cl}
\frac{(2k+1)(s-p)-2r(t-q)}{2m}\pi, &~~ (\mathcal{N}_2,\mathcal{N}_2)\\[0.04in]
\frac{(k+2)(s-p)-r(t-q)}{2m}\pi, & ~~ (\mathcal{N}_4,\mathcal{N}_4)\\[0.04in]
\frac{(1-k)(s-p)+r(t-q)}{2m}\pi, & ~~ (\mathcal{N}_5,\mathcal{N}_5)
\end{array}
\right.
\end{equation}
Here we have assumed that the residual subgroups $\mathcal{G}_{U}$ and $\mathcal{G}_{D}$ are generated by $a^{\alpha}bc^{s}d^{t}$ and $a^{\alpha}bc^{p}d^{q}$, respectively, or their direct product extensions. Apparently the Cabibbo mixing angle can be correctly produced for an appropriate value of $\theta$. The first two rows and columns of Eq.~\eqref{eq:couplequark} can be permutated in this framework, and the resulting CKM matrix can be obtained by redefining the parameter $\theta$ in Eq.~\eqref{eq:couplequark} as $\theta\rightarrow-\theta$, $\theta\rightarrow\pi/2-\theta$ or $\theta\rightarrow\theta-\pi/2$. If the flavor symmetry group is of type C, $V_{U}$ and $V_{D}$ can be only $\mathcal{N}_1$ or $\mathcal{N}_3$ given in Eqs.~(\ref{eq:N1},\ref{eq:N3}), the CKM matrix would be a unit matrix, democratic one, or one of the form of Eq.~\eqref{eq:couple6_gen}. Only the unit case is viable as a leading order approximation. Notice that the Cabibbo mixing of Eq.~\eqref{eq:couplequark} can also be generated from type B flavor symmetry groups~\cite{Lam:2007qc_Cabibbo,Blum:2007jz}, if the first and the second generations of left-handed quark fields are embedded into a doublet. Under this assignment, we have considered all the possible finite groups with faithful two-dimensional irreducible representation up to order 1000 in Sec.~\ref{sec:scan_groups}, and the phenomenologically viable Cabibbo angles that can be achieved are listed in Table~\ref{tab:quark_mixing_2plus1} below. Notice that the type B groups do not have faithful three-dimensional irreducible representations. In the following, we shall focus on the type D groups, and the series of $\theta$ values can easily be determined for fixed values of $r$ and $k$.

\begin{figure}[htb!]
\centering
\includegraphics[width=0.95\textwidth]{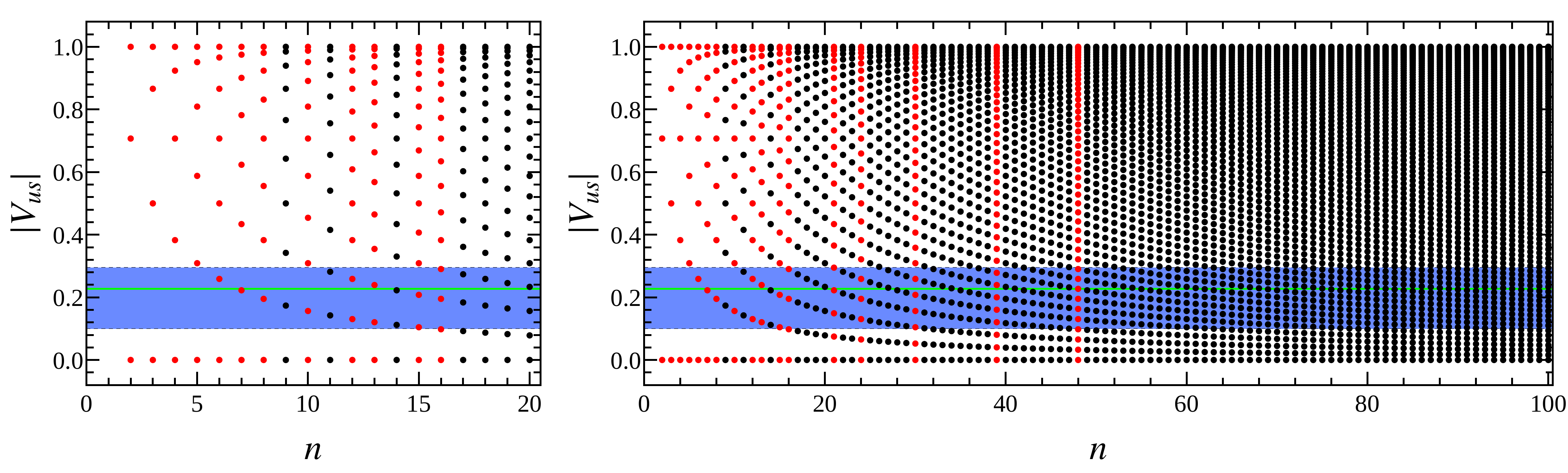}
\caption{\label{fig:D1quark}The possible values of $|V_{us}|=|\sin\theta|$
for $D_{n,n}^{(0)}\cong\Delta(6n^2)$ flavor symmetry group. The green band represents the regions allowed by the experimental data~\cite{pdg}, and the blue band denotes the region $0.1\leq|V_{us}|\leq0.3$. If the present experimental data of lepton mixing angles cannot be accommodated by a $\Delta(6n^2)$ group, the corresponding predictions for $|V_{us}|$ are plotted with red points. }
\end{figure}

\begin{description}
\item[a)] \textbf{For groups of} $D_{n,n}^{(0)}\cong \Delta(6n^2)$
\end{description}
The parameters are fixed to be $k=0$ and $r=1$ in this case, and then $\theta$ can take the forms
\begin{equation}
\theta=\frac{s-2t-p+2q}{2n}\pi,\quad\frac{2s-t-2p+q}{2n}\pi,~~\mathrm{or}~~\frac{s+t-p-q}{2n}\pi\,,
\end{equation}
where $s, t, p, q=0,1, \ldots, n-1$. Therefore, the possible values of $\theta$ are
\begin{equation}
\theta\hskip-0.08in\pmod{2\pi}=-\pi+\frac{1}{2n}\pi, -\pi+\frac{2}{2n}\pi,\ldots, \pi-\frac{1}{2n}\pi, \pi\,.
\end{equation}
We show all possible predictions for the CKM matrix element $|V_{us}|=|\sin\theta|$ for each $\Delta(6n^2)$ group in Fig.~\ref{fig:D1quark}. The matrix element $V_{us}$ has been quite precisely measured with $|V_{us}|=0.2253\pm0.0008$~\cite{pdg}, such that the currently allowed region appears to be a narrow green line. If a $\Delta(6n^2)$ flavor symmetry group cannot give rise to lepton mixing angles compatible with experimental data, the corresponding predictions for $|V_{us}|$ are indicated with red points. For $n=7$ or $n=14$, we can predict the Cabibbo mixing angle to fulfill $|V_{us}|=\sin\frac{\pi}{14}\simeq0.2225$, which is close to the experimental result. Furthermore, the $\Delta(6n^2)$ group with $n=14$ can generate lepton mixing angles within the experimentally preferred $3\sigma$ regions while it cannot for $n=7$.

\begin{figure}[htb!]
\centering
\includegraphics[width=0.95\textwidth]{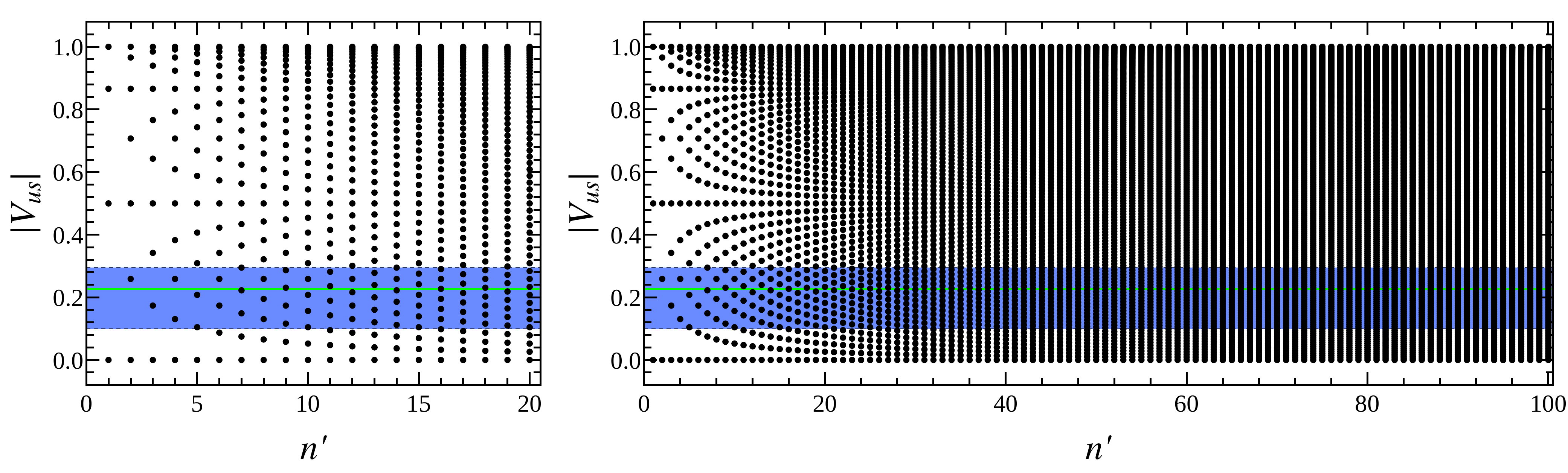}
\caption{\label{fig:D2quark}
The possible values of $|V_{us}|=|\sin\theta|$
for $D_{9n^{\prime},3n^{\prime}}^{(1)}$ flavor symmetry group. The green band represents the regions allowed by the experimental data~\cite{pdg}, and the blue band denotes the region $0.1\leq|V_{us}|\leq0.3$.
}
\end{figure}

\begin{description}
\item[b)] \textbf{For groups of} $D_{9n^{\prime}, 3n^{\prime}}^{(1)}\cong(Z_{9n^{\prime}}\times Z_{3n^{\prime}})\rtimes S_3$
\end{description}

We have $k=1$ and $r=3$ in this case, and the expressions for $\theta$ are
\begin{equation}
\theta=\frac{s-2t-p+2q}{6n^{\prime}}\pi,\quad \frac{s-t-p+q}{6n^{\prime}}\pi,~~\mathrm{or}~~\frac{t-q}{6n^{\prime}}\pi\,,
\end{equation}
where $s, p=0, 1, \ldots, 6n^{\prime}-1$ and $t, q=0, 1,\ldots, 3n^{\prime}-1$. The series of the $\theta$ values are given by
\begin{equation}
\theta\hskip-0.08in\pmod{2\pi}=-\pi+\frac{1}{6n^{\prime}}\pi, -\pi+\frac{2}{6n^{\prime}}\pi,\ldots, \pi-\frac{1}{6n^{\prime}}\pi, \pi\,.
\end{equation}
The results for $|V_{us}|=|\sin\theta|$ are displayed in Fig.~\ref{fig:D2quark}. We see that the $D_{9n^{\prime}, 3n^{\prime}}^{(1)}$ group with $n^{\prime}=7$ allows us to predict $|V_{us}|=\sin\frac{\pi}{14}\simeq0.2225$, and meanwhile the observed leptonic mixing angles can be accommodated.

\section{\label{sec:mixing_from_Sigma}$U(3)$ subgroup series $\Sigma(3N^3)$ and mixing patterns}

The present knowledge on finite subgroups of $U(3)$ that are not the subgroups of $SU(3)$ is partial. Therefore, it is not possible to perform a systematical analysis of mixing patterns achievable from all $U(3)$ discrete subgroups as we have done in Sec.~\ref{sec:mixing_from_SU(3)}. In this section, we shall focus on
the $\Sigma(3N^3)$ group that is a known series of $U(3)$ finite subgroups~\cite{Ishimori:2010zr}. The discrete group $\Sigma(3N^3)$ is defined by three $Z_N$ symmetries $Z_N, Z^{\prime}_N$, and $Z^{\prime\prime}_N$, which commute with each other, and their $Z_3$ permutations. $\Sigma(3N^3)$ is isomorphic to $(Z_{N}\times Z^{\prime}_{N}\times Z^{\prime\prime}_{N})\rtimes Z_3$. If we denote the generators of $Z_N, Z^{\prime}_N$, and $Z^{\prime\prime}_N$ by $a$, $a^{\prime}$, and $a^{\prime\prime}$, respectively, and the $Z_3$ generator by $b$, then the multiplication rules of the $\Sigma(3N^3)$ group are given by
\begin{equation}
\left.\begin{array}{c}
a^N=a'^N=a''^N=1,\quad b^3=1,\\
aa'=a'a,\quad aa''=a''a,\quad a'a''=a''a',\\
b^2ab=a'', \quad b^2a'b=a,\quad b^2a''b=a'\,.
\end{array}
\right.
\end{equation}
Consequently all the elements of $\Sigma(3N^3)$ can be written as
\begin{equation}
g=b^ka^ma'^na''^l\,,
\end{equation}
with $k=0,1,2$ and $m, n, l=0,\ldots,N-1$. In a specific irreducible faithful three-dimensional representation, the generators are presented by the following matrices:
\begin{equation}
a=\left(
\begin{array}{ccc}
 1 & 0 & 0 \\
 0 & 1 & 0 \\
 0 & 0 & \varrho
\end{array}
\right),~~
a^{\prime}=\left(
\begin{array}{ccc}
 1 & 0 & 0 \\
 0 & \varrho  & 0 \\
 0 & 0 & 1
\end{array}
\right),~~
a^{\prime\prime}=\left(
\begin{array}{ccc}
 \varrho  & 0 & 0 \\
 0 & 1 & 0 \\
 0 & 0 & 1
\end{array}
\right), ~~
b=\left(
\begin{array}{ccc}
 0 & 1 & 0 \\
 0 & 0 & 1 \\
 1 & 0 & 0
\end{array}
\right)\,,
\end{equation}
where $\varrho=e^{2\pi i/N}$. Next, we proceed to discuss the Abelian subgroups of $\Sigma(3N^3)$ as residual flavor symmetries and the corresponding diagonalization matrices.
\begin{itemize}[labelindent=-0.7em, leftmargin=1.6em]
\item{$\mathcal{G}_r=\langle ba^ma'^na''^l\rangle$}

The representation matrix of the generator is
\begin{equation}
\rho_{\mathbf{3}}(ba^ma'^na''^l)=\left(
\begin{array}{ccc}
 0 & \varrho^n & 0 \\
 0 & 0 & \varrho^m \\
 \varrho^l & 0 & 0
\end{array}
\right)\,,
\end{equation}
with eigenvalues
\begin{equation}
\varrho^{\frac{1}{3}(l+m+n)},\quad \omega^2\varrho^{\frac{1}{3}(l+m+n)},\quad
\omega\rho^{\frac{1}{3}(l+m+n)}\,.
\end{equation}
Since the eigenvalues are not degenerate, the corresponding diagonalization matrix is uniquely fixed and is given by
\begin{equation}
\widetilde{\mathcal{N}}_1:~ V_{r}=\frac{1}{\sqrt{3}}\left(
\begin{array}{ccc}
\varrho^{\frac{1}{3}(-2l+m+n)} &  \omega^2\varrho^{\frac{1}{3}(-2l+m+n)}  &  \omega\varrho^{\frac{1}{3}(-2l+m+n)}  \\
\varrho^{\frac{1}{3}(-l+2m-n)} &  \omega\varrho^{\frac{1}{3}(-l+2m-n)}  &
\omega^2\varrho^{\frac{1}{3}(-l+2m-n)}  \\
 1 & 1 & 1
\end{array}
\right)\,,
\end{equation}
up to permutations and phases of the column vectors.

\item{$\mathcal{G}_r=\langle b^2a^ma'^na''^l\rangle$}

\begin{equation}
\rho_{\mathbf{3}}(b^2a^ma'^na''^l)=\left(
\begin{array}{ccc}
 0 & 0 & \varrho^m \\
 \varrho^l & 0 & 0 \\
 0 & \varrho^n & 0
\end{array}
\right)\,.
\end{equation}
The eigenvalues are given by
\begin{equation}
\varrho^{\frac{1}{3}(l+m+n)},\quad \omega^2\varrho^{\frac{1}{3}(l+m+n)},\quad \omega\varrho^{\frac{1}{3}(l+m+n)}\,,
\end{equation}
which are also not degenerate, and the unitary transformation $V_{r}$ is determined to be
\begin{equation}
\widetilde{\mathcal{N}}_2:~ V_{r}=\frac{1}{\sqrt{3}}\left(
\begin{array}{ccc}
\varrho^{\frac{1}{3}(-l+2m-n)} &  \omega\varrho^{\frac{1}{3}(-l+2m-n)}  & \omega^2\varrho^{\frac{1}{3}(-l+2m-n)}  \\
\varrho^{\frac{1}{3}(l+m-2n)} & \omega^2\varrho^{\frac{1}{3}(l+m-2n)}  & \omega\varrho^{\frac{1}{3}(l+m-2n)}  \\
 1 & 1 & 1
\end{array}
\right)\,.
\end{equation}

\item{$\mathcal{G}_r=\langle a^ma'^na''^l\rangle$}

The element $a^ma'^na''^l$ is represented by
\begin{equation}
\rho_{\mathbf{3}}(a^ma'^na''^l)=\left(
\begin{array}{ccc}
 \varrho^l & 0 & 0 \\
 0 & \varrho^n & 0 \\
 0 & 0 & \varrho^m
\end{array}
\right)\,.
\end{equation}
It has the eigenvalues
\begin{equation}
\label{eq:eigenvalues_Sigma3N3}\varrho^l,\quad \varrho^n,\quad \varrho^m\,.
\end{equation}
Obviously the representation matrix is diagonal, and therefore we have
\begin{equation}
\widetilde{\mathcal{N}}_3:~  V_{r}=\left(
\begin{array}{ccc}
 1 & 0 & 0 \\
 0 & 1 & 0 \\
 0 & 0 & 1
\end{array}
\right)
\end{equation}
for $ l\neq{n}\neq{m}$. The eigenvalues shown in Eq.~\eqref{eq:eigenvalues_Sigma3N3} could be partially degenerate such that only a mixing vector can be fixed,
\begin{equation}
\label{eq:degeneracy_Sigma3N3}\begin{split}
v_{r}=(1,0,0)^{T} & ~~~\text{for}~~ m=n, m\neq l,\\
v_{r}=(0,1,0)^{T} & ~~~\text{for}~~ m=l, m\neq n,\\
v_{r}=(0,0,1)^{T} & ~~~\text{for}~~ l=n, m\neq l\,.
\end{split}
\end{equation}
\end{itemize}
If this is the case, the unitary matrix $V_{r}$ can be fully determined through an extension of the residual symmetry $\mathcal{G}_r=\langle a^ma'^na''^l\rangle$ to $\mathcal{G}_r=\langle a^ma'^na''^l\rangle\times\mathcal{G}_2$, where two of three parameters $m$, $n$, and $l$ are equal, and the generator of $\mathcal{G}_2$ should be commutable with $a^ma'^na''^l$. As a consequence, $\mathcal{G}_2$ can only be a subgroup generated by $a^pa'^sa''^t$. If $p\neq{s}\neq{t}$, $V_{r}$ is fixed to be $\widetilde{\mathcal{N}}_3$ by $\mathcal{G}_2$, If any two of $p$, $s$, and $t$ are equal, another mixing vector can be determined, and $V_{r}$ is a unit matrix as well. Here we assume that mixing vectors associated with $\langle a^ma'^na''^l\rangle$ and $\mathcal{G}_2$ are different; otherwise, the above extension procedure should be continued. However, the conclusion of $V_{r}$ being a unit matrix remains. We summarize that the unitary diagonalization matrix $V_{r}$ can only be of the form of $\widetilde{\mathcal{N}}_{1,2,3}$ for $\Sigma(3N^3)$ flavor symmetry. Assuming neutrinos to be Dirac particles, we shall consider all possible combinations of $V_{l}$ and $V_{\nu}$ and the corresponding predictions for the PMNS matrix in the following.

\begin{itemize}[labelindent=-0.7em, leftmargin=1.6em]
\item{$(V_{l}, V_{\nu})=(\widetilde{\mathcal{N}}_3, \widetilde{\mathcal{N}}_3)$}

Obviously the PMNS matrix is a trivial identity matrix in this case,
\begin{equation}
\label{eq:PMNS1_Sigma}U_{\text{PMNS}}=\left(
\begin{array}{ccc}
 1 & 0 & 0 \\
 0 & 1 & 0 \\
 0 & 0 & 1
\end{array}
\right)\,,
\end{equation}
which is not viable.

\item{$(V_{l}, V_{\nu})=(\widetilde{\mathcal{N}}_1, \widetilde{\mathcal{N}}_3)$, $(\widetilde{\mathcal{N}}_3, \widetilde{\mathcal{N}}_1)$, $(\widetilde{\mathcal{N}}_2, \widetilde{\mathcal{N}}_3)$, $(\widetilde{\mathcal{N}}_3, \widetilde{\mathcal{N}}_2)$}

The lepton flavor mixing is found to be the democratic pattern:
\begin{equation}
\label{eq:PMNS2_Sigma}U_{\text{PMNS}}=\frac{1}{\sqrt{3}}\left(
\begin{array}{ccc}
 1 & 1 & 1 \\
 1 & \omega   & \omega^2 \\
 1 & \omega^2 & \omega
\end{array}
\right)\,.
\end{equation}

\item{$(V_{l}, V_{\nu})=(\widetilde{\mathcal{N}}_1, \widetilde{\mathcal{N}}_1)$, $(\widetilde{\mathcal{N}}_1, \widetilde{\mathcal{N}}_2)$, $(\widetilde{\mathcal{N}}_2, \widetilde{\mathcal{N}}_1)$, $(\widetilde{\mathcal{N}}_2, \widetilde{\mathcal{N}}_2)$}

The PMNS matrix is determined to be of the same form as Eq.~\eqref{eq:couple6_gen}, i.e.,
\begin{equation}
\label{eq:PMNS3_Sigma}U_{\text{PMNS}}=\frac{1}{3}\left(
\begin{array}{ccc}
1+2e^{i\delta}\cos\theta &  1-2e^{i\delta}\cos\left(\theta-\frac{\pi}{3}\right)  & 1-2e^{i\delta}\cos\left(\theta+\frac{\pi}{3}\right) \\
1-2e^{i\delta}\cos\left(\theta+\frac{\pi}{3}\right)  &  1+2e^{i\delta}\cos\theta   & 1-2e^{i\delta}\cos\left(\theta-\frac{\pi}{3}\right) \\
1-2e^{i\delta}\cos\left(\theta-\frac{\pi}{3}\right)   &  1-2e^{i\delta}\cos\left(\theta+\frac{\pi}{3}\right)  &  1+2e^{i\delta}\cos\theta
\end{array}
\right)\,,
\end{equation}
where
\begin{eqnarray*}
\nonumber&&\hskip-0.15in\theta=\frac{(l+m-2n-p+2s-t)\pi}{3N}, ~\delta=\frac{(l-m+p-t)\pi}{N},~~\mathrm{for}~~(V_{l}, V_{\nu})=(\widetilde{\mathcal{N}}_1, \widetilde{\mathcal{N}}_1),\\
\nonumber&&\hskip-0.15in\theta=\frac{(l+m-2n+p+s-2t)\pi}{3N}, ~\delta=\frac{(l-m+p-s)\pi}{N},~~\mathrm{for}~~(V_{l}, V_{\nu})=(\widetilde{\mathcal{N}}_1, \widetilde{\mathcal{N}}_2),\\
\nonumber&&\hskip-0.15in\theta=\frac{(-2l+m+n+p-2s+t)\pi}{3N}, ~\delta=-\frac{(m-n-p+t)\pi}{N},~~\mathrm{for}~~(V_{l}, V_{\nu})=(\widetilde{\mathcal{N}}_2, \widetilde{\mathcal{N}}_1),\\
&&\hskip-0.15in\theta=\frac{(-2l+m+n-p-s+2t)\pi}{3N}, ~\delta=-\frac{(m-n-p+s)\pi}{N},~~\mathrm{for}~~(V_{l}, V_{\nu})=(\widetilde{\mathcal{N}}_2, \widetilde{\mathcal{N}}_2)\,.
\end{eqnarray*}
Here we have assumed that the residual symmetries $\mathcal{G}_{l}$ and $\mathcal{G}_{\nu}$ are generated by the elements $b^{k}a^ma'^na''^l$ and $b^{k^{\prime}}a^pa'^sa''^t$, respectively, with $k, k^{\prime}=1, 2$. As shown in Fig.~\ref{fig:theta13_23_N3N3}, the observed values of three leptonic mixing angles cannot be obtained from this mixing pattern. Therefore, the $\Sigma(3N^3)$ group cannot give rise to a viable lepton flavor mixing unless the residual flavor symmetries are partially or completely accidental. If we apply the $\Sigma(3N^3)$ flavor symmetry to the quark sector, the CKM matrix would be of the form of Eqs.~(\ref{eq:PMNS1_Sigma}, \ref{eq:PMNS2_Sigma}, \ref{eq:PMNS3_Sigma}) as well, and the Cabibbo angle cannot be generated.

\end{itemize}

\section{\label{sec:scan_groups}Mixing patterns from a scan of finite discrete groups}

In previous sections, we have performed a comprehensive study of the mixing patterns that can be derived from all the $SU(3)$ subgroups and the known $U(3)$ subgroup series $\Sigma(3N^3)$. Although the $SU(3)$ finite subgroups are known very well, the classification of the finite subgroups of $U(3)$ is far from complete. Only a few series such as $\Sigma(3N^3)$ are known so far. There are still lots of finite groups that are subgroups of $U(3)$ but not subgroups of $SU(3)$. The structures of these groups are not known at present, and therefore the possible mixing patterns derived from these group cannot be studied analytically. The task of classifying all finite subgroups of $U(3)$ is a daunting challenge. Instead of doing that, we shall perform an exhaustive scan over the discrete groups with order less than 2000 with the help of the computer algebra system \texttt{GAP}~\cite{GAP}, and the associated flavor mixings for all possible residual subgroups of each group are discussed. In the following, we shall focus on the groups with faithful three-dimensional irreducible representations to which the three left-handed leptons are assigned.

It is worth mentioning that the groups of orders of 768 and 1536 account for a large part of our scan program. There are 1090235 groups of order 768 and 408641062 groups of order 1536. The groups of order 768 can be straightforwardly investigated one by one. However, it would be rather time consuming to scrutinize through all the groups of order 1536, and the required computational resources are tremendous and beyond our capability. We notice that all of the groups are regularly arranged in the \texttt{GAP} SmallGroup Library~\cite{SmallGroups}. The command \texttt{SmallGroupsInformation(.)} can be utilized to get the information about the arrangement of the groups of a given order. For example, by using command \texttt{SmallGroupsInformation(1536)}, we can obtain a list of the group information about the groups of order 1536.
\begin{verbatim}
 gap>SmallGroupsInformation(1536);

   There are 408641062 groups of order 1536.
      1 - 10494213 are the nilpotent groups.
      10494214 - 408526597 have a normal Sylow 3-subgroup.
      408526598 - 408544625 have a normal Sylow 2-subgroup.
      408544626 - 408641062 have no normal Sylow subgroup.
   This size belongs to layer 8 of the SmallGroups library.
   IdSmallGroup is not available for this size.
\end{verbatim}
From the above output, we know that the groups of order 1536 are arranged according to the properties of their subgroups. The groups of order 768 have the same arrangement pattern, which is given by
\begin{verbatim}
 gap>SmallGroupsInformation(768);

   There are 1090235 groups of order 768.
   They are sorted by normal Sylow subgroups.
      1 - 56092 are the nilpotent groups.
      56093 - 1083472 have a normal Sylow 3-subgroup
                      with centralizer of index 2.
      1083473 - 1085323 have a normal Sylow 2-subgroup.
      1085324 - 1090235 have no normal Sylow subgroup.
   This size belongs to layer 3 of the SmallGroups library.
   IdSmallGroup is available for this size.
\end{verbatim}
When we search for the groups of order 768, we find that the groups belonging to the first two categories do not admit three-dimensional faithful irreducible representations. We speculate that finite nilpotent groups and the groups with normal Sylow 3-subgroup have no three-dimensional faithful irreducible representations. This conjecture is checked to hold true for many other groups although we are not able to prove it in a strict mathematical way now. Under this conjecture, the first 408526597 groups of order 1536 can be skipped, and we only need to handle the remaining 114465 groups. We find that only 11 groups of order 1536 have faithful three-dimensional irreducible representations. $\Delta(1536)$ in the series of $\Delta(6n^2)$ with $n=16$ is known to be a group with faithful three-dimensional representations, it is really found in our results, and the group identification number is $[1536,408544632]$. Our conjecture is verified again.

We have studied all the groups with order less than 2000 as flavor symmetry, and the groups of order 1536 are treated under our conjecture. Only those groups with faithful three-dimensional irreducible representations are considered. Note that the irreducible representations given in \texttt{GAP} are not necessarily unitary, we must first convert the nonunitary \texttt{GAP} representations into unitary representations. The method of how to construct the corresponding unitary representations from a given nonunitary representation is presented in the Appendix~\ref{sec:Appd_unitarize}. Our method is more efficient than the Gram-Schmidt orthogonalization approach in Ref.~\cite{Ludl:2010bj}. Neutrinos are assumed to be Dirac particles, and then the residual subgroups $\mathcal{G}_{\nu}$ in the neutrino sector and $\mathcal{G}_{l, U, D}$ in the charged fermion sectors can be any Abelian subgroups of the postulated flavor symmetry group. We wish to completely fix the mixing matrix through breaking of the flavor symmetry to residual symmetries in this work. As a consequence, for each group, we find all its Abelian subgroups with order greater than two, and the corresponding group structures and generators are extracted. The next step is to determine the unitary diagonalization matrices of these subgroups. If a subgroup has two or more generators, all the representation matrices of the generators should be diagonalized simultaneously. If the eigenvalues of the generators of a subgroup are partially or completely degenerate such that the corresponding diagonalization matrix cannot be determined uniquely (in other words, the residual subgroup cannot distinguish three generations of leptons or quarks), such a subgroup would be discarded. Notice that the diagonalization matrix is determined up to permutations and phases of their column vectors, since no predictions can be made for the fermion masses in the present framework. The first row of the diagonalization matrix can always be set to be real and non-negative by redefining the overall phases of each column, and the three columns can be arranged such that the elements of the first row decrease the column index. A large number of duplicates in the unitary diagonalization matrix  are removed via this rearrangement procedure.

Furthermore, we consider all possible residual subgroups for $\mathcal{G}_{l}$ and $\mathcal{G}_{\nu}$ ($\mathcal{G}_{U}$ and $\mathcal{G}_{D}$ in case of quark mixing) for each finite group with three-dimensional irreducible representations, and then the predictions for the PMNS (or CKM) matrix can easily be obtained from the unitary diagonalization matrices of the chosen remnant subgroups. Using the freedom to exchange the rows and columns for the PMNS matrix, we can order the PMNS matrix in the following ways: the (13) element is chosen to be the smallest one, and the (12) element is smaller than or equal to the (11) element such that the solar mixing angle is smaller than or equal to maximal mixing. Subsequently the
flavor mixing angles $\theta_{12}$, $\theta_{13}$, $\theta_{23}$ and Dirac $CP$ violating phase $\delta_{CP}$ can be straightforwardly extracted.  The exchange of the second and the third rows of the PMNS matrix leaves both $\theta_{12}$ and $\theta_{13}$ invariant, while it transforms $\theta_{23}$ into $\pi/2-\theta_{23}$ and $\delta_{CP}$ to $\pi+\delta_{CP}$. Notice that all the three mixing angles would keep intact and the Dirac CP phase would change from $\delta_{CP}$ to $-\delta_{CP}$ if we embed the three lepton doublet fields into the complex conjugate of the three-dimensional representation. To quantitatively measure whether and how well the predicted mixing pattern can accommodate the current experimental data, we perform a simple $\chi^2$ analysis and the $\chi^2$ function is constructed in the conventional way,
\begin{equation}
\label{Eq:chi2}\chi^2=\sum_{ij=12,13,13}\left(\frac{\left(\sin^2\theta_{ij}\right)^{th}-\left(\sin^2\theta_{ij}\right)^{bf}}{\sigma_{ij}}\right)^2\,,
\end{equation}
where $\left(\sin^2\theta_{ij}\right)^{th}$ denote the mixing angles predicted for different residual subgroups, $\left(\sin^2\theta_{ij}\right)^{bf}$ and $\sigma_{ij}$ refer to the best fit values and $1\sigma$ errors, respectively taken from Ref.~\cite{Gonzalez-Garcia:2014bfa}. Note that $\chi^2$ is very sensitive to the prediction of $\theta_{13}$, because the reactor angle $\theta_{13}$ is quite precisely measured. The Dirac phase $\delta_{CP}$ is weakly constrained by the present data; therefore, we do not include the information of $\delta_{CP}$ in the $\chi^2$ function in Eq.~\eqref{Eq:chi2}.

The explicit forms of the lepton flavor mixing matrices that can be derived from finite groups of order smaller than 2000 have been worked out. Nevertheless, they are too long to be included in the present paper. We have made them available at our Website~\cite{webdata}, where the predictions for mixing parameters, $\chi^2$ values, and the associated remnant subgroups are presented as well.

\subsection{\label{subsec:lepton_scan}Results for lepton mixing}

We find that 90 groups with order less than 2000 can generate the mixing angles compatible with experimental data up to the $3\sigma$ level, if neutrinos are Dirac particles. On the other hand, only 10 groups can give rise to viable lepton mixing for Majorana neutrinos. To our surprise, the PMNS matrix turns out to be of the trimaximal form in all cases, i.e.,
\begin{equation}
\label{eq:PMNS_scan}U_{\text{PMNS}}=\frac{1}{\sqrt{3}} \left(
\begin{array}{ccc}
\sqrt{2}\cos \theta _{\nu } & ~1~ & -\sqrt{2}\sin \theta _{\nu } \\
 -\sqrt{2}\cos \left(\theta _{\nu }-\frac{\pi }{3}\right) & ~1~ & \sqrt{2}\sin \left(\theta _{\nu }-\frac{\pi }{3}\right) \\
 -\sqrt{2}\cos \left(\theta _{\nu }+\frac{\pi }{3}\right) & ~1~ & \sqrt{2}\sin \left(\theta _{\nu }+\frac{\pi }{3}\right)
\end{array}
\right)\,,
\end{equation}
with
\begin{equation}
\label{eq:theta_values}\theta_{\nu}=\pm\frac{\pi}{17},~ \pm\frac{\pi}{18},~ \pm\frac{5\pi}{84},~ \pm\frac{2\pi}{33}\,.
\end{equation}
The general expressions of the mixing angles in terms of $\theta_{\nu}$ are given by Eq.~\eqref{eq:mixing_angles_gen} with the correlations of Eq.~\eqref{eq:correlation_trimaximal}. If we replace $\theta_{\nu}$ with $-\theta_{\nu}$, the solar as well as the reactor mixing angles remain invariant, while the atmospheric mixing angle becomes its complement angle and the Dirac phase $\delta_{CP}$ is shifted by $\pi$. The same results are obtained if we exchange the second and the third rows of the mixing matrix in Eq.~\eqref{eq:PMNS_scan}. For the viable values of $\theta_{\nu}$ in Eq.~\eqref{eq:theta_values}, the predictions for the mixing angles and the $\chi^2$ value can easily be calculated. The numerical results are listed in Table~\ref{tab:theta_nu}. We see that the atmospheric mixing angle $\theta_{23}$ is in the first octant and the second octant for positive and negative $\theta_{\nu}$, respectively. It is worth mentioning that the Dirac $CP$ phase $\delta_{CP}$ is equal to $0$ for $\theta_{\nu}>0$ and $\pi$ for $\theta_{\nu}<0$. Most importantly, all mixing patterns with $\chi^2\leq100$ lead to a trivial Dirac phase, and nontrivial $\delta_{CP}$ can only arise from patterns that do not accommodate the data well; i.e., the $\chi^2$ value is larger than 100.

\begin{table}[t!]
\centering
\begin{tabular}{|c|c|c|c|c|c|c|c|c|}
\hline\hline
 \multirow{2}{*}{$\theta_{\nu}$} & \multirow{2}{*}{$\sin^2\theta_{13}$} & \multirow{2}{*}{$\sin^2\theta_{12}$} &\multicolumn{2}{c|}{ $\sin^2\theta_{23}$} & \multicolumn{2}{c|}{$\chi^2$(NO)}& \multicolumn{2}{c|}{$\chi^2$(IO)} \\ \cline{4-9}
  &   &    &  $\theta_{\nu}>0$  &    $\theta_{\nu}<0$  &  $\theta_{\nu}>0$  &   $\theta_{\nu}<0$  &   $\theta_{\nu}>0$  &   $\theta_{\nu}<0$  \\\hline
 $\pm\frac{\pi}{17}$ & 0.0225 & 0.341 & 0.393 & 0.607 & 13.000 & 17.456 & 33.576 & 9.638 \\\hline
$\pm\frac{\pi}{18}$ & 0.0201 & 0.340 & 0.399 & 0.601 & 14.174 & 18.807 & 34.576 & 11.731 \\\hline
$\pm\frac{5\pi}{84}$ & 0.0230 & 0.341 & 0.392 & 0.608 & 14.314 & 18.724 & 34.795 & 10.605 \\\hline
$\pm\frac{2\pi}{33}$ & 0.0239  & 0.341 & 0.390 & 0.610 & 17.520 & 21.853 & 37.616 & 13.076 \\\hline
\hline\hline
$\pm\frac{\pi}{16}$ & 0.0254  & 0.342 & 0.387 & 0.613 & 26.766 & 30.947 & 45.546 & 20.408 \\\hline
$\pm\frac{5\pi}{96}$ & 0.0177  & 0.339 & 0.406 & 0.594 & 27.034 & 31.787 & 47.092 & 25.495 \\\hline
$\pm\frac{2\pi}{39}$ &  0.0172 & 0.339 & 0.407 & 0.593 & 31.476 & 36.245 & 51.443 & 30.145 \\\hline
$\pm\frac{5\pi}{78}$ & 0.0267 & 0.342 & 0.384 & 0.616 & 38.449 & 42.485 & 55.436 & 29.808 \\\hline
$\pm\frac{\pi}{20}$ & 0.0163  & 0.339 & 0.409 & 0.591 & 39.606 & 44.395 & 59.421 & 38.608 \\\hline
\hline
\end{tabular}
\caption{\label{tab:theta_nu}
Mixing angles predicted by finite groups up to order 2000, where we require $\chi^2$ is less than 50 in the case of NO neutrino mass spectrum. For $\theta_{\nu}=\pm\frac{\pi}{17}$, $\pm\frac{\pi}{18}$,  $\pm\frac{5\pi}{84}$, $\pm\frac{2\pi}{33}$, all  three mixing angles agree within $3\sigma$ with the experimental data~\cite{Gonzalez-Garcia:2014bfa}. For the remaining values $\theta_{\nu}=\pm\frac{\pi}{16}$, $\pm\frac{5\pi}{96}$, $\pm\frac{2\pi}{39}$, $\pm\frac{5\pi}{78}$, $\pm\frac{\pi}{20}$, both solar and atmospheric mixing angles are within the $3\sigma$ intervals while the reactor mixing angle is slightly beyond the allowed range.}
\end{table}

The finite discrete groups that can give rise to the viable mixing pattern in Eq.~\eqref{eq:PMNS_scan} and the corresponding residual subgroups are listed in Table~\ref{tab:lepton_mixing1}, where the notation $[g,j]$ denotes the $j\text{th}$ finite group of order $g$ as classified in \texttt{GAP}'s SmallGroups library~\cite{SmallGroups}. The superscript  *  indicates that the corresponding group comprises Klein subgroups such that neutrinos could be Majorana particles. The subscripts $\Delta$ and $\Delta^{\prime}$ imply that the corresponding groups belong to the series $D_{n,n}^{(0)}\cong\Delta(6n^2)$ and $D_{9n^{\prime},3n^{\prime}}^{(1)}\cong(Z_{9n^{\prime}}\times Z_{3n^{\prime}})\rtimes S_3$, respectively. We see that most of these groups have the structures $(\cdots)\rtimes Z_3\rtimes Z_2$ or $(\cdots)\rtimes Z_3\rtimes Z_4$, and the orders of the groups are divisible by 162 except for the $\Delta(6n^2)$ groups and the groups that are isomorphic to the direct product of $\Delta(6n^2)$ with cyclic groups. Moreover, we notice that different groups may lead to the same PMNS matrix, particularly in the case of $\theta_{\nu}=\pm\pi/18$. For a given flavor symmetry group, there are generally many possible combinations of the residual symmetries $\mathcal{G}_l$ and $\mathcal{G}_{\nu}$ corresponding to the same $U_{\text{PMNS}}$. Regarding the type D group series $D_{n,n}^{(0)}$ and $D_{9n^{\prime},3n^{\prime}}^{(1)}$ with order less than 2000, we find the numerical results obtained here with \texttt{GAP} are exactly the same as those obtained from analytical study in Sec.~\ref{sec:mixing_from_SU(3)}. This is a very convincing check to our results.

It is remarkable that the smallest groups that can generate the leptonic mixing angles within $3\sigma$ region of global fits are
$[162,10]$, $[162,12]$, and $[162,14]$, and neutrinos are required to be Dirac particles. Both $[162,10]$ and $[162,12]$ are $U(3)$ but not $SU(3)$ subgroups, and $[162,14]$ is exactly the type D group $D_{9,3}^{(1)}\cong (Z_9\times Z_3)\rtimes S_3$. Although the structure of both  $[162,12]$ and $[162,14]$ is apparently $(Z_9\times Z_3)\rtimes S_3$, the definitions of the semidirect in the two groups are distinct; therefore, they are not isomorphic. These three groups are also found in Ref.~\cite{Holthausen:2013vba}, where a scan of finite groups up to order 200 has been performed under the assumption of Dirac neutrinos. These three groups $[162,10]$, $[162,12]$, and $[162,14]$ provide a new possibility for flavor model building. It is interesting to construct a model of lepton mixing based on any of these three groups. Such a model is still absent as far as we know. It may be better to construct such models in the framework of extra dimension theory rather than supersymmetry as usual, since the tiny Dirac neutrino masses can be naturally produced in extra dimension through fermion wave-function localization~\cite{Grossman:1999ra,Chen:2009gy}.

If neutrinos are Majorana particles and the residual symmetry $\mathcal{G}_{\nu}$ is constrained to be a Klein subgroup, then ten groups $[648, 259]$, $[648, 260]$, $[648, 266]$, $[1296, 1820]$, $[1296, 1821]$, $[1296, 1827]$, $[1944, 832]$, $[1944, 833]$, $[1944, 849]$, and $[1944, 2363]$ can predict acceptable lepton mixing angles, and they lead to the same PMNS matrix that is of the trimaximal form with $\theta=\pm\pi/18$. The groups of order 648 and 1296 are related as follows:
{\small
\begin{equation}
[1296, 1820]\cong{Z}_2\times[648, 259],~ [1296, 1821]\cong{Z}_2\times[648, 260],~[1296, 1827]\cong{Z}_2\times[648, 266]\,.
\end{equation}}
This is exactly the reason why they can lead to the same mixing pattern. Notice that only two of  the above ten groups are $SU(3)$ subgroups, $[648, 259]$ is a group of type $D_{9n^{\prime},3n^{\prime}}^{(1)}$ with $n^{\prime}=2$, and $[1944, 849]$ is a $\Delta(6n^2)$ group with $n=18$. We see that the results of Sec.~\ref{subsec:numerical_results} for $[648, 259]$ and $[1944, 849]$ are reproduced by the present numerical analysis. The finite discrete groups with order less than 1536 has been considered in Ref.~\cite{Holthausen:2012wt} under the assumption of Majorana neutrinos, only three groups $\Delta(6\times10^2)$, $\left(Z_{18}\times Z_6\right)\rtimes S_3$, and $\Delta(6\times16^2)$ with the identification numbers $[600, 179]$, $[648, 259]$, and $[1536,408544632]$ are found viable, and the reactor mixing angle is predicted to be $\sin^2\theta_{13}=0.0288$, 0.0201, and 0.0254, respectively. All the mixing patterns claimed in~\cite{Holthausen:2012wt} are confirmed by our  comprehensive analysis~\cite{webdata}. However, the values of $\sin^2\theta_{13}=0.0288$ and 0.0254 are outside the present $3\sigma$ interval~\cite{Gonzalez-Garcia:2014bfa}. As a result, the groups $[600, 179]$ as well as $[1536,408544632]$ are omitted in Table~\ref{tab:lepton_mixing1}, and only $[648, 259]$ is left. Comparing our results with those of Ref.~\cite{Holthausen:2012wt}, we find five additional groups $[648, 260]$, $[648, 266]$, $[1296, 1820]$, $[1296, 1821]$, $[1296, 1827]$ that can also give the experimentally favored mixing angles, if we concentrate on groups with order up to 1536. The reason may be that Ref.~\cite{Holthausen:2012wt} only considers the finite subgroups of $SU(3)$ that cannot be written as direct products with cyclic groups.

We have also studied all the possible leptonic mixing patterns that can be obtained from the ``exceptional'' finite subgroups $\Sigma(60)\cong{A}_5$, $\Sigma(60)\times{Z}_3$, $\Sigma(168)\cong{PSL}(2, 7)$, $\Sigma(168)\times{Z}_3$, $\Sigma(36\times3)$, $\Sigma(72\times3)$, $\Sigma(216\times3)$, and $\Sigma(360\times3)$ in our systematic scan. Please see our Web page~\cite{webdata} for details. All the patterns presented in Ref.~\cite{Hagedorn:2013nra} have been validated, and we also obtain other mixing patterns. This is because the requirement that the residual flavor symmetries $\mathcal{G}_{l}$ and $\mathcal{G}_{\nu}$ should generate the whole flavor symmetry group is relaxed in our work. We find that only the group $\Sigma(216\times 3)$ with identification number $[648, 532]$ can generate the experimentally acceptable lepton mixing angles. If we take $\mathcal{G}_{\nu}=Z_{18}$, $\mathcal{G}_{l}=Z_3$, or $\mathcal{G}_{l}=Z_3\times{Z}_3$, the PMNS matrix would be of the trimaximal form shown in Eq.~\eqref{eq:couple7_gen} with $\theta=\pm\pi/18$. However, $\mathcal{G}_{l}$ and $\mathcal{G}_{\nu}$ do not generate the original group
$\Sigma(216\times 3)$, but only a subgroup $D_{9,3}^{(1)}\cong (Z_9\times Z_3)\rtimes S_3$ that is denoted by $[162,14]$ in the SmallGroups library. The same results are found in Ref.~\cite{Hagedorn:2013nra}.

\begin{figure}[hptb!]
\centering
\vskip1.2cm
\includegraphics[width=0.95\textwidth]{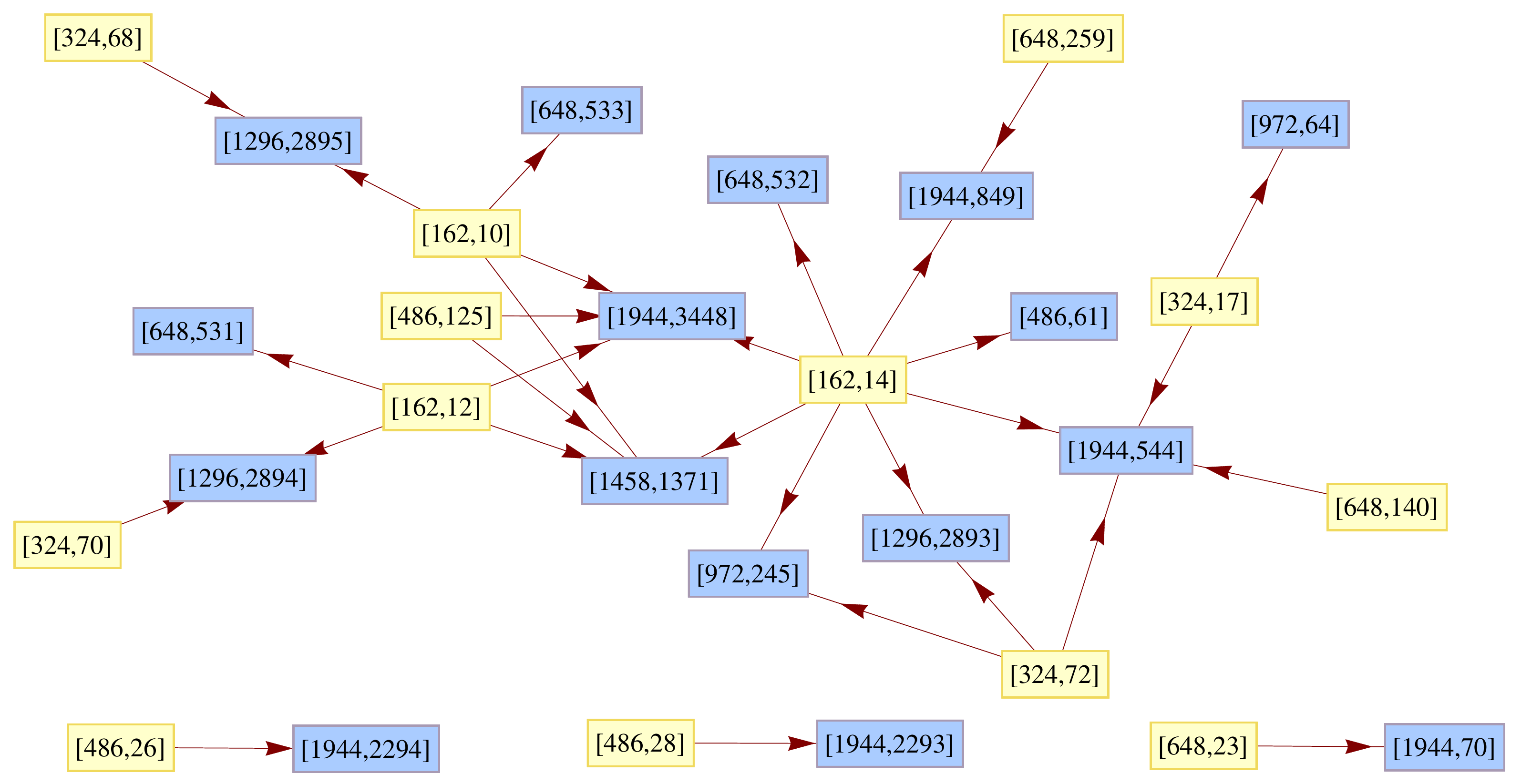}
\caption{\label{fig:group_relations}The groups generated by the residual symmetries $\mathcal{G}_{\nu}$ and $\mathcal{G}_{l}$ in the lepton sector. If a subgroup is generated, it is linked to the parent group by an arrowed line and the arrow points to the parent group. The blue boxes denote that all the possible residual subgroups $\left\{\mathcal{G}_{\nu}, \mathcal{G}_{l}\right\}$ do not generate the entire group. The yellow boxes imply that  the group can be generated by at least one set of the remnant symmetries. The similar figure including all the 90 groups in Table~\ref{tab:lepton_mixing1} looks quite complex, and it is available at our Web site~\cite{webdata}.
}
\end{figure}

If a certain flavor mixing pattern can be generated from a flavor symmetry group $\mathcal{G_F}$, generally the groups containing $\mathcal{G_F}$ as a subgroup can also give rise to the same mixing pattern. It is interesting to investigate whether the viable mixing pattern predicted by a group can also been generated from its subgroups. In other words, we want to discuss whether a mixing pattern is intrinsically associated with a group or only one of its subgroups. For every group listed in Table~\ref{tab:lepton_mixing1}, we see that there are a large number of possible remnant symmetry groups $\mathcal{G}_{\nu}$ and $\mathcal{G}_{l}$ that lead to the same prediction for $\theta_{\nu}$. Each set of $\left\{\mathcal{G}_{\nu}, \mathcal{G}_{l}\right\}$ generates either the parent group or one of its subgroups. In particular, we find 16 groups whose residual subgroups cannot give rise to the entire group. These 16 groups and the groups generated by remnant symmetries are plotted in Fig.~\ref{fig:group_relations}. A group and its subgroups are linked by arrowed lines, and the arrow points to the parent group. For example, the residual subgroups $\mathcal{G}_{\nu}=Z_{18}$, $\mathcal{G}_{l}=Z_3$, or $\mathcal{G}_{l}=Z_3\times Z_3$ of $[486, 61]$, which predict the trimaximal lepton mixing with $\theta_{\nu}=\pm\pi/18$, only generate the subgroup $[162, 14]$ rather than $[486, 61]$ itself. This implies that the group $[486, 61]$ does not give us anything more than $[162, 14]$ as far as lepton mixing is concerned. As a consequence, $[162, 14]$ is preferred over $[486, 61]$ as a flavor symmetry group, since they lead to the same lepton mixing pattern, and yet the former is less cumbersome and more economical from the view of model building. For the remaining 74 groups, the residual symmetries can generate the whole groups as well as the subgroups. The correlations between different groups can be shown in a figure similar to Fig.~\ref{fig:group_relations}. The resulting diagram is too complex to be included in the present paper, and it is available at our Web site~\cite{webdata}.

Note that the above predictions for lepton mixing would generally be modified by the inclusion of higher dimensional operators. Therefore these mixing patterns that slightly deviate from the experimental $3\sigma$ ranges may still be interesting, since accordance with the data could easily be achieved in an explicit model once small subleading corrections are taken into account. First, we find that the mixing patterns disfavored at the $3\sigma$ level with $\chi^2<50$ for NO spectrum are still of the trimaximal form, and the corresponding values of $\theta_{\nu}$ are given by
\begin{equation}
\theta_{\nu} =\pm\frac{\pi}{16}, \pm\frac{5\pi}{96}, \pm\frac{2\pi}{39}, \pm\frac{5\pi}{78}, \pm\frac{\pi}{20}\,.
\end{equation}
The predictions for the three lepton mixing angles are displayed in Table~\ref{tab:theta_nu}. The associated flavor symmetry groups and the residual symmetries are listed in Table~\ref{tab:lepton_mixing2}. One sees that both solar and atmospheric mixing angles are compatible with experimental results while the measured value of the reactor mixing angle cannot be accommodated without small corrections.

The second phenomenologically interesting mixing pattern is the transpose of the trimaximal mixing matrix. The explicit form of the PMNS matrix is given in Eq.~\eqref{eq:couple5_gen}, where the second or the third row of the PMNS matrix is $\left(1, 1, 1\right)/\sqrt{3}$. As we have shown in Sec.~\ref{subsec:PMNS_analytical_Dirac}, in general both $\theta_{12}$ and $\theta_{23}$ have to acquire high order corrections in order to achieve agreement with the experimental data in this case. For instance, these three groups $[162, 10]$, $[162, 12]$, and $[162, 14]$ can predict the PMNS matrix in Eq.~\eqref{eq:couple5_gen} with $\theta=\pi/9$, which gives rise to $\sin^2\theta_{13}\simeq0.0201$, $\sin^2\theta_{12}\simeq0.399$, $\sin^2\theta_{23}\simeq0.340$, or $\sin^2\theta_{23}\simeq0.660$. Obviously, $\theta_{13}$ agrees with the experimental result, and small corrections could bring $\theta_{12}$ and $\theta_{23}$ into the experimentally favored regions. The observed values of $\theta_{13}$ can also be produced by $\Delta(6\times11^2)$, $\Delta(6\times14^2)$, and $\Delta(6\times17^2)$ groups, which give $\theta=7\pi/66$, $3\pi/28$, and $11\pi/102$, respectively.

The third and last interesting mixing pattern takes the following form:
\begin{equation}
\label{eq:PMNS_third}U_{\text{PMNS}}=\frac{1}{2}\left(
\begin{array}{ccc}
\sqrt{\frac{1}{6} \left(9+5 \sqrt{3}\right)} & ~1~ & \sqrt{\frac{1}{6} \left(9-5 \sqrt{3}\right)} \\
-\sqrt{\frac{1}{3} \left(3-\sqrt{3}\right)} & ~\sqrt{2}~ & -\sqrt{\frac{1}{3} \left(3+\sqrt{3}\right)} \\
-\sqrt{\frac{1}{2} \left(3-\sqrt{3}\right)} & ~1~ & \sqrt{\frac{1}{2} \left(3+\sqrt{3}\right)}
\end{array}
\right)\,.
\end{equation}
This mixing pattern can be generated from the ``exception'' finite group $\Sigma(216\times3)$ or the group with identification number $[648, 531]$, if the residual symmetries $\mathcal{G}_{\nu}=Z_{4}$ and $\mathcal{G}_l=Z_{18}$ are preserved. The mixing angles read
\begin{equation}
\sin^2\theta_{13}\simeq0.0142,\quad  \sin^2\theta_{12}\simeq0.254,\quad \sin^2\theta_{23}=0.4\,.
\end{equation}
If the second and the third rows of the PMNS matrix in Eq.~\eqref{eq:PMNS_third} are exchanged, the atmospheric mixing angle becomes $\sin^2\theta_{23}=0.6$. In both cases the atmospheric mixing angle is compatible with the experimental data at the $3\sigma$ level. However, the predicted values of $\theta_{12}$ and $\theta_{13}$ are a bit too small, and  this could be reconciled with the experimental results in a concrete model after small corrections are included.

\LTcapwidth=\textwidth
\begin{center}
\begin{longtable}{|m{0.04\columnwidth}<{\centering}|m{0.13\columnwidth}<{\centering}|m{0.34\columnwidth}<{\centering}|m{0.46\columnwidth}|}
\hline\hline
$\theta_{\nu}$ & GAP-id & Group Structure &~~~~~ Residual Symmetries $~(\mathcal{G}_l~|~\mathcal{G}_\nu)$ \\
\hline
\endfirsthead

\hline
$\theta_{\nu}$ & GAP-id & Group Structure &~~~~~ Residual Symmetries $~(\mathcal{G}_l~|~\mathcal{G}_\nu)$ \\
\hline
\endhead

\hline
\caption{(continued)}\\
\endfoot

\hline\hline

\caption{\label{tab:lepton_mixing1}
Mixing patterns predicted by finite discrete groups up to order 2000, where the mixing angles are required to be in the $3\sigma$ ranges of the global fit~\cite{Gonzalez-Garcia:2014bfa}. The superscript  *  indicates that the corresponding group has Klein subgroups and it allows  neutrinos to be Majorana particles. The subscripts $\vartriangle$ and $\vartriangle^{\prime}$ denote the groups that belong to the $D_{n,n}^{(0)}\cong\Delta(6n^2)$ and $D_{9n^{\prime},3n^{\prime}}^{(1)}\cong(Z_{9n^{\prime}}\times Z_{3n^{\prime}})\rtimes S_3$ group series, respectively. The notation $(\mathcal{G}_l~|~\mathcal{G}_\nu)$ denotes all possible combinations between the subgroups in the $\mathcal{G}_l$ list and the subgroups in the $\mathcal{G}_{\nu}$ list. The explicit forms of the residual subgroups can be found at our Web site~\cite{webdata}.
}
\endlastfoot
$\pm\frac{\pi }{17}$ & $~~[1734, 5]_{\vartriangle}$ &  $ (Z_{17} \times Z_{17}) \rtimes Z_{3} \rtimes Z_{2} $  & $($$Z_3$ $|$ $Z_{34}$$)$ \\ \hline
\multirow{45}{*}{$\pm\frac{\pi}{18}$} & $[162, 10]$ &  $ (Z_{3} \times Z_{3} \times Z_{3}) \rtimes Z_{3} \rtimes Z_{2} $  & $($$Z_9$ $|$ $Z_3\times Z_6$, $Z_6$$)$ \\ \cline{2-4}
 & $[162, 12]$ &  $ (Z_{9} \times Z_{3}) \rtimes Z_{3} \rtimes Z_{2} $  & $($$Z_9$ $|$ $Z_{18}$$)$ \\ \cline{2-4}& $~~[162, 14]_{\vartriangle'}$ &  $ (Z_{9} \times Z_{3}) \rtimes Z_{3} \rtimes Z_{2} $  & $($$Z_3$, $Z_3^2$ $|$ $Z_{18}$$)$ \\ \cline{2-4}& $[324, 13]$ &  $ (Z_{3} \times Z_{3}) \rtimes Z_{3} \rtimes Z_{4} \rtimes Z_{3} $  & $($$Z_9$, $Z_{18}$ $|$ $Z_3\times Z_{12}$, $Z_{12}$$)$ \\ \cline{2-4}& $[324, 15]$ &  $ (Z_{9} \times Z_{3}) \rtimes Z_{3} \rtimes Z_{4} $  & $($$Z_9$, $Z_{18}$ $|$ $Z_{36}$$)$ \\ \cline{2-4}& $[324, 17]$ &  $ (Z_{3} \times Z_{3}) \rtimes Z_{3} \rtimes Z_{4} \rtimes Z_{3} $  & $($$Z_3\times Z_6$, $Z_3$, $Z_3^2$, $Z_6$ $|$ $Z_{36}$$)$ \\ \cline{2-4}& $[324, 68]$ &  $ Z_{2} \times ((Z_{3} \times Z_{3} \times Z_{3}) \rtimes Z_{3} \rtimes Z_{2}) $  & $($$Z_9$, $Z_{18}$ $|$ $Z_2\times Z_6$, $Z_3\times Z_6$, $Z_6$, $Z_6^2$$)$ \\ \cline{2-4}& $[324, 70]$ &  $ Z_{2} \times ((Z_{9} \times Z_{3}) \rtimes Z_{3} \rtimes Z_{2}) $  & $($$Z_9$, $Z_{18}$ $|$ $Z_2\times Z_{18}$, $Z_{18}$$)$ \\ \cline{2-4}& $[324, 72]$ &  $ Z_{2} \times ((Z_{9} \times Z_{3}) \rtimes Z_{3} \rtimes Z_{2}) $  & $($$Z_3\times Z_6$, $Z_3$, $Z_3^2$, $Z_6$ $|$ $Z_2\times Z_{18}$, $Z_{18}$$)$ \\ \cline{2-4}& $[486, 26]$ &  $ (Z_{27} \times Z_{3}) \rtimes Z_{3} \rtimes Z_{2} $  & $($$Z_{27}$ $|$ $Z_{54}$$)$ \\ \cline{2-4}& $[486, 28]$ &  $ (Z_{27} \times Z_{3}) \rtimes Z_{3} \rtimes Z_{2} $  & $($$Z_{27}$ $|$ $Z_{54}$$)$ \\ \cline{2-4}& $~~[486, 61]_{\vartriangle}$ &  $ (Z_{9} \times Z_{9}) \rtimes Z_{3} \rtimes Z_{2} $  & $($$Z_3$, $Z_3^2$ $|$ $Z_{18}$$)$ \\ \cline{2-4}& $[486, 125]$ &  $ (Z_{9} \times Z_{3} \times Z_{3}) \rtimes Z_{3} \rtimes Z_{2} $  & $($$Z_3\times Z_9$, $Z_3$, $Z_3^2$, $Z_9$ $|$ $Z_3\times Z_6$, $Z_3\times Z_{18}$, $Z_6$, $Z_{18}$$)$ \\ \cline{2-4}& $[648, 19]$ &  $ (Z_{3} \times Z_{3}) \rtimes Z_{3} \rtimes Z_{8} \rtimes Z_{3} $  & $($$Z_9$, $Z_{18}$, $Z_{36}$ $|$ $Z_3\times Z_{24}$, $Z_{24}$$)$ \\ \cline{2-4}& $[648, 21]$ &  $ (Z_{9} \times Z_{3}) \rtimes Z_{3} \rtimes Z_{8} $  & $($$Z_9$, $Z_{18}$, $Z_{36}$ $|$ $Z_{72}$$)$ \\ \cline{2-4}& $[648, 23]$ &  $ (Z_{3} \times Z_{3}) \rtimes Z_{3} \rtimes Z_{8} \rtimes Z_{3} $  & $($$Z_3\times Z_6$, $Z_3\times Z_{12}$, $Z_3$, $Z_3^2$, $Z_6$, $Z_{12}$ $|$ $Z_{72}$$)$ \\ \cline{2-4}& $[648, 128]$ &  $ Z_{4} \times ((Z_{3} \times Z_{3} \times Z_{3}) \rtimes Z_{3} \rtimes Z_{2}) $  & $($$Z_9$, $Z_{18}$, $Z_{36}$ $|$ $Z_2\times Z_6$, $Z_2\times Z_{12}$, $Z_3\times Z_6$, $Z_3\times Z_{12}$, $Z_6\times Z_{12}$, $Z_6$, $Z_6^2$, $Z_{12}$$)$ \\ \cline{2-4}& $[648, 134]$ &  $ Z_{4} \times ((Z_{9} \times Z_{3}) \rtimes Z_{3} \rtimes Z_{2}) $  & $($$Z_9$, $Z_{18}$, $Z_{36}$ $|$ $Z_2\times Z_{18}$, $Z_2\times Z_{36}$, $Z_{18}$, $Z_{36}$$)$ \\ \cline{2-4}& $[648, 140]$ &  $ Z_{4} \times ((Z_{9} \times Z_{3}) \rtimes Z_{3} \rtimes Z_{2}) $  & $($$Z_3\times Z_6$, $Z_3\times Z_{12}$, $Z_3$, $Z_3^2$, $Z_6$, $Z_{12}$ $|$ $Z_2\times Z_{18}$, $Z_2\times Z_{36}$, $Z_{18}$, $Z_{36}$$)$ \\ \cline{2-4}& $~~[648, 259]^*_{\vartriangle'}$ &  $ (Z_{18} \times Z_{6}) \rtimes Z_{3} \rtimes Z_{2} $  & $($$Z_3$, $Z_3^2$ $|$ $Z_2\times Z_6$, $Z_2\times Z_{18}$, $Z_2^2$, $Z_{18}$$)$ \\ \cline{2-4}& $~[648, 260]^*$ &  $ (Z_{18} \times Z_{6}) \rtimes Z_{3} \rtimes Z_{2} $  & $($$Z_9$ $|$ $Z_2\times Z_6$, $Z_2\times Z_{18}$, $Z_2^2$, $Z_{18}$$)$ \\ \cline{2-4}& $~[648, 266]^*$ &  $ (Z_{6} \times Z_{6} \times Z_{3}) \rtimes Z_{3} \rtimes Z_{2} $  & $($$Z_9$ $|$ $Z_2\times Z_6$, $Z_3\times Z_6$, $Z_2^2$, $Z_6$, $Z_6^2$$)$ \\ \cline{2-4}& $[648, 531]$ &  $ Z_{3} \cdot ((Z_{3} \times Z_{3}) \rtimes Q_{8} \rtimes Z_{3})=((Z_{3} \times Z_{3}) \rtimes Z_{3} \rtimes Q_{8}) \cdot Z_{3} $  & $($$Z_9$ $|$ $Z_{18}$$)$ \\ \cline{2-4}& $[648, 532]$ &  $ (Z_{3} \times Z_{3}) \rtimes Z_{3} \rtimes Q_{8} \rtimes Z_{3} $  & $($$Z_3$, $Z_3^2$ $|$ $Z_{18}$$)$ \\ \cline{2-4}& $[648, 533]$ &  $ (Z_{3} \times Z_{3}) \rtimes Z_{3} \rtimes Q_{8} \rtimes Z_{3} $  & $($$Z_9$ $|$ $Z_3\times Z_6$, $Z_6$$)$ \\ \cline{2-4}& $[810, 21]$ &  $ Z_{5} \times ((Z_{3} \times Z_{3} \times Z_{3}) \rtimes Z_{3} \rtimes Z_{2}) $  & $($$Z_9$, $Z_{45}$ $|$ $Z_3\times Z_6$, $Z_3\times Z_{30}$, $Z_6$, $Z_{30}$$)$ \\ \cline{2-4}& $[810, 23]$ &  $ Z_{5} \times ((Z_{9} \times Z_{3}) \rtimes Z_{3} \rtimes Z_{2}) $  & $($$Z_9$, $Z_{45}$ $|$ $Z_{18}$, $Z_{90}$$)$ \\ \cline{2-4}& $[810, 25]$ &  $ Z_{5} \times ((Z_{9} \times Z_{3}) \rtimes Z_{3} \rtimes Z_{2}) $  & $($$Z_3\times Z_{15}$, $Z_3$, $Z_3^2$, $Z_{15}$ $|$ $Z_{18}$, $Z_{90}$$)$ \\ \cline{2-4}& $[972, 29]$ &  $ (Z_{27} \times Z_{3}) \rtimes Z_{3} \rtimes Z_{4} $  & $($$Z_{27}$, $Z_{54}$ $|$ $Z_{108}$$)$ \\ \cline{2-4}& $[972, 31]$ &  $ (Z_{27} \times Z_{3}) \rtimes Z_{3} \rtimes Z_{4} $  & $($$Z_{27}$, $Z_{54}$ $|$ $Z_{108}$$)$ \\ \cline{2-4}& $[972, 64]$ &  $ (Z_{9} \times Z_{9}) \rtimes Z_{3} \rtimes Z_{4} $  & $($$Z_3\times Z_6$, $Z_3$, $Z_3^2$, $Z_6$ $|$ $Z_{36}$$)$ \\ \cline{2-4}& $[972, 210]$ &  $ Z_{2} \times ((Z_{27} \times Z_{3}) \rtimes Z_{3} \rtimes Z_{2}) $  & $($$Z_{27}$, $Z_{54}$ $|$ $Z_2\times Z_{54}$, $Z_{54}$$)$ \\ \cline{2-4}& $[972, 212]$ &  $ Z_{2} \times ((Z_{27} \times Z_{3}) \rtimes Z_{3} \rtimes Z_{2}) $  & $($$Z_{27}$, $Z_{54}$ $|$ $Z_2\times Z_{54}$, $Z_{54}$$)$ \\ \cline{2-4}& $[972, 245]$ &  $ Z_{2} \times ((Z_{9} \times Z_{9}) \rtimes Z_{3} \rtimes Z_{2}) $  & $($$Z_3\times Z_6$, $Z_3$, $Z_3^2$, $Z_6$ $|$ $Z_2\times Z_{18}$, $Z_{18}$$)$ \\ \cline{2-4}& $[972, 309]$ &  $ (Z_{9} \times Z_{3}) \rtimes Z_{3} \rtimes Z_{4} \rtimes Z_{3} $  & $($$Z_3\times Z_6$, $Z_3\times Z_9$, $Z_3\times Z_{18}$, $Z_3$, $Z_3^2$, $Z_6$, $Z_9$, $Z_{18}$ $|$ $Z_3\times Z_{12}$, $Z_3\times Z_{36}$, $Z_{12}$, $Z_{36}$$)$ \\ \cline{2-4}& $[972, 626]$ &  $ Z_{2} \times ((Z_{9} \times Z_{3} \times Z_{3}) \rtimes Z_{3} \rtimes Z_{2}) $  & $($$Z_3\times Z_6$, $Z_3\times Z_9$, $Z_3\times Z_{18}$, $Z_3$, $Z_3^2$, $Z_6$, $Z_9$, $Z_{18}$ $|$ $Z_2\times Z_6$, $Z_2\times Z_{18}$, $Z_3\times Z_6$, $Z_3\times Z_{18}$, $Z_6\times Z_{18}$, $Z_6$, $Z_6^2$, $Z_{18}$$)$ \\ \cline{2-4}& $[1134, 95]$ &  $ Z_{7} \times ((Z_{3} \times Z_{3} \times Z_{3}) \rtimes Z_{3} \rtimes Z_{2}) $  & $($$Z_9$, $Z_{63}$ $|$ $Z_3\times Z_6$, $Z_3\times Z_{42}$, $Z_6$, $Z_{42}$$)$ \\ \cline{2-4}& $[1134, 97]$ &  $ Z_{7} \times ((Z_{9} \times Z_{3}) \rtimes Z_{3} \rtimes Z_{2}) $  & $($$Z_9$, $Z_{63}$ $|$ $Z_{18}$, $Z_{126}$$)$ \\ \cline{2-4}& $[1134, 99]$ &  $ Z_{7} \times ((Z_{9} \times Z_{3}) \rtimes Z_{3} \rtimes Z_{2}) $  & $($$Z_3\times Z_{21}$, $Z_3$, $Z_3^2$, $Z_{21}$ $|$ $Z_{18}$, $Z_{126}$$)$ \\ \cline{2-4}
 \multirow{47}{*}{$\pm\frac{\pi}{18}$}  & $[1296, 35]$ &  $ (Z_{3} \times Z_{3}) \rtimes Z_{3} \rtimes Z_{16} \rtimes Z_{3} $  & $($$Z_9$, $Z_{18}$, $Z_{36}$, $Z_{72}$ $|$ $Z_3\times Z_{48}$, $Z_{48}$$)$ \\ \cline{2-4}& $[1296, 37]$ &  $ (Z_{9} \times Z_{3}) \rtimes Z_{3} \rtimes Z_{16} $  & $($$Z_9$, $Z_{18}$, $Z_{36}$, $Z_{72}$ $|$ $Z_{144}$$)$ \\ \cline{2-4}& $[1296, 39]$ &  $ (Z_{3} \times Z_{3}) \rtimes Z_{3} \rtimes Z_{16} \rtimes Z_{3} $  & $($$Z_3\times Z_6$, $Z_3\times Z_{12}$, $Z_3\times Z_{24}$, $Z_3$, $Z_3^2$, $Z_6$, $Z_{12}$, $Z_{24}$ $|$ $Z_{144}$$)$ \\ \cline{2-4}& $[1296, 274]$ &  $ Z_{8} \times ((Z_{3} \times Z_{3} \times Z_{3}) \rtimes Z_{3} \rtimes Z_{2}) $  & $($$Z_9$, $Z_{18}$, $Z_{36}$, $Z_{72}$ $|$ $Z_2\times Z_6$, $Z_2\times Z_{12}$, $Z_2\times Z_{24}$, $Z_3\times Z_6$, $Z_3\times Z_{12}$, $Z_3\times Z_{24}$, $Z_6\times Z_{12}$, $Z_6\times Z_{24}$, $Z_6$, $Z_6^2$, $Z_{12}$, $Z_{24}$$)$ \\ \cline{2-4}& $[1296, 284]$ &  $ Z_{8} \times ((Z_{9} \times Z_{3}) \rtimes Z_{3} \rtimes Z_{2}) $  & $($$Z_9$, $Z_{18}$, $Z_{36}$, $Z_{72}$ $|$ $Z_2\times Z_{18}$, $Z_2\times Z_{36}$, $Z_2\times Z_{72}$, $Z_{18}$, $Z_{36}$, $Z_{72}$$)$ \\ \cline{2-4}& $[1296, 294]$ &  $ Z_{8} \times ((Z_{9} \times Z_{3}) \rtimes Z_{3} \rtimes Z_{2}) $  & $($$Z_3\times Z_6$, $Z_3\times Z_{12}$, $Z_3\times Z_{24}$, $Z_3$, $Z_3^2$, $Z_6$, $Z_{12}$, $Z_{24}$ $|$ $Z_2\times Z_{18}$, $Z_2\times Z_{36}$, $Z_2\times Z_{72}$, $Z_{18}$, $Z_{36}$, $Z_{72}$$)$ \\ \cline{2-4}& $[1296, 688]$ &  $ (Z_{6} \times Z_{6}) \rtimes Z_{3} \rtimes Z_{4} \rtimes Z_{3} $  & $($$Z_3\times Z_6$, $Z_3$, $Z_3^2$, $Z_6$ $|$ $Z_2\times Z_4$, $Z_2\times Z_{12}$, $Z_2\times Z_{36}$, $Z_{36}$$)$ \\ \cline{2-4}& $[1296, 689]$ &  $ (Z_{18} \times Z_{6}) \rtimes Z_{3} \rtimes Z_{4} $  & $($$Z_9$, $Z_{18}$ $|$ $Z_2\times Z_4$, $Z_2\times Z_{12}$, $Z_2\times Z_{36}$, $Z_{36}$$)$ \\ \cline{2-4}& $[1296, 699]$ &  $ (Z_{6} \times Z_{6}) \rtimes Z_{3} \rtimes Z_{4} \rtimes Z_{3} $  & $($$Z_9$, $Z_{18}$ $|$ $Z_2\times Z_4$, $Z_2\times Z_{12}$, $Z_3\times Z_{12}$, $Z_6\times Z_{12}$, $Z_{12}$$)$ \\ \cline{2-4}& $~[1296, 1820]^*$ &  $ Z_{2} \times ((Z_{18} \times Z_{6}) \rtimes Z_{3} \rtimes Z_{2}) $  & $($$Z_3\times Z_6$, $Z_3$, $Z_3^2$, $Z_6$ $|$ $Z_2\times Z_6$, $Z_2\times Z_{18}$, $Z_2^2\times Z_6$, $Z_2^2\times Z_{18}$, $Z_2^2$, $Z_2^3$, $Z_{18}$$)$ \\ \cline{2-4}& $~[1296, 1821]^*$ &  $ Z_{2} \times ((Z_{18} \times Z_{6}) \rtimes Z_{3} \rtimes Z_{2}) $  & $($$Z_9$, $Z_{18}$ $|$ $Z_2\times Z_6$, $Z_2\times Z_{18}$, $Z_2^2\times Z_6$, $Z_2^2\times Z_{18}$, $Z_2^2$, $Z_2^3$, $Z_{18}$$)$ \\ \cline{2-4}& $~[1296, 1827]^*$ &  $ Z_{2} \times ((Z_{6} \times Z_{6} \times Z_{3}) \rtimes Z_{3} \rtimes Z_{2}) $  & $($$Z_9$, $Z_{18}$ $|$ $Z_2\times Z_6$, $Z_2\times Z_6^2$, $Z_2^2\times Z_6$, $Z_3\times Z_6$, $Z_2^2$, $Z_2^3$, $Z_6$, $Z_6^2$$)$ \\ \cline{2-4}& $[1296, 2893]$ &  $ Z_{2} \times ((Z_{3} \times Z_{3}) \rtimes Z_{3} \rtimes Q_{8} \rtimes Z_{3}) $  & $($$Z_3\times Z_6$, $Z_3$, $Z_3^2$, $Z_6$ $|$ $Z_2\times Z_{18}$, $Z_{18}$$)$ \\ \cline{2-4}& $[1296, 2894]$ &  $ Z_{2} \times (Z_{3} \cdot ((Z_{3} \times Z_{3}) \rtimes Q_{8} \rtimes Z_{3})=((Z_{3} \times Z_{3}) \rtimes Z_{3} \rtimes Q_{8}) \cdot Z_{3}) $  & $($$Z_9$, $Z_{18}$ $|$ $Z_2\times Z_{18}$, $Z_{18}$$)$ \\ \cline{2-4}& $[1296, 2895]$ &  $ Z_{2} \times ((Z_{3} \times Z_{3}) \rtimes Z_{3} \rtimes Q_{8} \rtimes Z_{3}) $  & $($$Z_9$, $Z_{18}$ $|$ $Z_2\times Z_6$, $Z_3\times Z_6$, $Z_6$, $Z_6^2$$)$ \\ \cline{2-4}& $[1458, 615]$ &  $ (Z_{81} \times Z_{3}) \rtimes Z_{3} \rtimes Z_{2} $  & $($$Z_{81}$ $|$ $Z_{162}$$)$ \\ \cline{2-4}& $[1458, 618]$ &  $ (Z_{81} \times Z_{3}) \rtimes Z_{3} \rtimes Z_{2} $  & $($$Z_{81}$ $|$ $Z_{162}$$)$ \\ \cline{2-4}& $~~[1458, 659]_{\vartriangle'}$ &  $ (Z_{27} \times Z_{9}) \rtimes Z_{3} \rtimes Z_{2} $  & $($$Z_3$, $Z_3^2$ $|$ $Z_{18}$, $Z_{54}$$)$ \\ \cline{2-4}& $[1458, 663]$ &  $ (Z_{27} \times Z_{9}) \rtimes Z_{3} \rtimes Z_{2} $  & $($$Z_3$, $Z_3^2$ $|$ $Z_{18}$, $Z_{54}$$)$ \\ \cline{2-4}& $[1458, 666]$ &  $ (Z_{27} \times Z_{9}) \rtimes Z_{3} \rtimes Z_{2} $  & $($$Z_3$, $Z_3^2$ $|$ $Z_{18}$, $Z_{54}$$)$ \\ \cline{2-4}& $[1458, 1095]$ &  $ (Z_{27} \times Z_{3} \times Z_{3}) \rtimes Z_{3} \rtimes Z_{2} $  & $($$Z_3\times Z_9$, $Z_3\times Z_{27}$, $Z_3$, $Z_3^2$, $Z_9$, $Z_{27}$ $|$ $Z_3\times Z_6$, $Z_3\times Z_{18}$, $Z_3\times Z_{54}$, $Z_6$, $Z_{18}$, $Z_{54}$$)$ \\ \cline{2-4}& $[1458, 1371]$ &  $ (Z_{9} \times Z_{9} \times Z_{3}) \rtimes Z_{3} \rtimes Z_{2} $  & $($$Z_3\times Z_9$, $Z_3$, $Z_3^2$, $Z_9$ $|$ $Z_3\times Z_6$, $Z_3\times Z_{18}$, $Z_6$, $Z_{18}$$)$ \\ \cline{2-4}& $[1620, 29]$ &  $ Z_{5} \times ((Z_{3} \times Z_{3}) \rtimes Z_{3} \rtimes Z_{4} \rtimes Z_{3}) $  & $($$Z_9$, $Z_{18}$, $Z_{45}$, $Z_{90}$ $|$ $Z_3\times Z_{12}$, $Z_3\times Z_{60}$, $Z_{12}$, $Z_{60}$$)$ \\ \cline{2-4}& $[1620, 31]$ &  $ Z_{5} \times ((Z_{9} \times Z_{3}) \rtimes Z_{3} \rtimes Z_{4}) $  & $($$Z_9$, $Z_{18}$, $Z_{45}$, $Z_{90}$ $|$ $Z_{36}$, $Z_{180}$$)$ \\ \cline{2-4}& $[1620, 33]$ &  $ Z_{5} \times ((Z_{3} \times Z_{3}) \rtimes Z_{3} \rtimes Z_{4} \rtimes Z_{3}) $  & $($$Z_3\times Z_6$, $Z_3\times Z_{15}$, $Z_3\times Z_{30}$, $Z_3$, $Z_3^2$, $Z_6$, $Z_{15}$, $Z_{30}$ $|$ $Z_{36}$, $Z_{180}$$)$ \\ \cline{2-4}& $[1620, 182]$ &  $ Z_{10} \times ((Z_{3} \times Z_{3} \times Z_{3}) \rtimes Z_{3} \rtimes Z_{2}) $  & $($$Z_9$, $Z_{18}$, $Z_{45}$, $Z_{90}$ $|$ $Z_2\times Z_6$, $Z_2\times Z_{30}$, $Z_3\times Z_6$, $Z_3\times Z_{30}$, $Z_6\times Z_{30}$, $Z_6$, $Z_6^2$, $Z_{30}$$)$ \\ \cline{2-4}& $[1620, 184]$ &  $ Z_{10} \times ((Z_{9} \times Z_{3}) \rtimes Z_{3} \rtimes Z_{2}) $  & $($$Z_9$, $Z_{18}$, $Z_{45}$, $Z_{90}$ $|$ $Z_2\times Z_{18}$, $Z_2\times Z_{90}$, $Z_{18}$, $Z_{90}$$)$ \\ \cline{2-4}& $[1620, 186]$ &  $ Z_{10} \times ((Z_{9} \times Z_{3}) \rtimes Z_{3} \rtimes Z_{2}) $  & $($$Z_3\times Z_6$, $Z_3\times Z_{15}$, $Z_3\times Z_{30}$, $Z_3$, $Z_3^2$, $Z_6$, $Z_{15}$, $Z_{30}$ $|$ $Z_2\times Z_{18}$, $Z_2\times Z_{90}$, $Z_{18}$, $Z_{90}$$)$ \\ \cline{2-4} & $[1782, 21]$ &  $ Z_{11} \times ((Z_{3} \times Z_{3} \times Z_{3}) \rtimes Z_{3} \rtimes Z_{2}) $  & $($$Z_9$, $Z_{99}$ $|$ $Z_3\times Z_6$, $Z_3\times Z_{66}$, $Z_6$, $Z_{66}$$)$ \\ \cline{2-4}
 \multirow{33}{*}{$\pm\frac{\pi}{18}$} & $[1782, 23]$ &  $ Z_{11} \times ((Z_{9} \times Z_{3}) \rtimes Z_{3} \rtimes Z_{2}) $  & $($$Z_9$, $Z_{99}$ $|$ $Z_{18}$, $Z_{198}$$)$ \\ \cline{2-4}& $[1782, 25]$ &  $ Z_{11} \times ((Z_{9} \times Z_{3}) \rtimes Z_{3} \rtimes Z_{2}) $  & $($$Z_3\times Z_{33}$, $Z_3$, $Z_3^2$, $Z_{33}$ $|$ $Z_{18}$, $Z_{198}$$)$ \\ \cline{2-4}& $[1944, 35]$ &  $ (Z_{27} \times Z_{3}) \rtimes Z_{3} \rtimes Z_{8} $  & $($$Z_{27}$, $Z_{54}$, $Z_{108}$ $|$ $Z_{216}$$)$ \\ \cline{2-4}& $[1944, 37]$ &  $ (Z_{27} \times Z_{3}) \rtimes Z_{3} \rtimes Z_{8} $  & $($$Z_{27}$, $Z_{54}$, $Z_{108}$ $|$ $Z_{216}$$)$ \\ \cline{2-4}& $[1944, 70]$ &  $ (Z_{9} \times Z_{9}) \rtimes Z_{3} \rtimes Z_{8} $  & $($$Z_3\times Z_6$, $Z_3\times Z_{12}$, $Z_3$, $Z_3^2$, $Z_6$, $Z_{12}$ $|$ $Z_{72}$$)$ \\ \cline{2-4}& $[1944, 362]$ &  $ Z_{4} \times ((Z_{27} \times Z_{3}) \rtimes Z_{3} \rtimes Z_{2}) $  & $($$Z_{27}$, $Z_{54}$, $Z_{108}$ $|$ $Z_2\times Z_{54}$, $Z_2\times Z_{108}$, $Z_{54}$, $Z_{108}$$)$ \\ \cline{2-4}& $[1944, 368]$ &  $ Z_{4} \times ((Z_{27} \times Z_{3}) \rtimes Z_{3} \rtimes Z_{2}) $  & $($$Z_{27}$, $Z_{54}$, $Z_{108}$ $|$ $Z_2\times Z_{54}$, $Z_2\times Z_{108}$, $Z_{54}$, $Z_{108}$$)$ \\ \cline{2-4}& $[1944, 544]$ &  $ Z_{4} \times ((Z_{9} \times Z_{9}) \rtimes Z_{3} \rtimes Z_{2}) $  & $($$Z_3\times Z_6$, $Z_3\times Z_{12}$, $Z_3$, $Z_3^2$, $Z_6$, $Z_{12}$ $|$ $Z_2\times Z_{18}$, $Z_2\times Z_{36}$, $Z_{18}$, $Z_{36}$$)$ \\ \cline{2-4}& $[1944, 707]$ &  $ (Z_{9} \times Z_{3}) \rtimes Z_{3} \rtimes Z_{8} \rtimes Z_{3} $  & $($$Z_3\times Z_6$, $Z_3\times Z_9$, $Z_3\times Z_{12}$, $Z_3\times Z_{18}$, $Z_3\times Z_{36}$, $Z_3$, $Z_3^2$, $Z_6$, $Z_9$, $Z_{12}$, $Z_{18}$, $Z_{36}$ $|$ $Z_3\times Z_{24}$, $Z_3\times Z_{72}$, $Z_{24}$, $Z_{72}$$)$ \\ \cline{2-4}& $~[1944, 832]^*$ &  $ (Z_{54} \times Z_{6}) \rtimes Z_{3} \rtimes Z_{2} $  & $($$Z_{27}$ $|$ $Z_2\times Z_6$, $Z_2\times Z_{18}$, $Z_2\times Z_{54}$, $Z_2^2$, $Z_{54}$$)$ \\ \cline{2-4}& $~[1944, 833]^*$ &  $ (Z_{54} \times Z_{6}) \rtimes Z_{3} \rtimes Z_{2} $  & $($$Z_{27}$ $|$ $Z_2\times Z_6$, $Z_2\times Z_{18}$, $Z_2\times Z_{54}$, $Z_2^2$, $Z_{54}$$)$ \\ \cline{2-4}& $~~[1944, 849]^*_{\vartriangle}$ &  $ (Z_{18} \times Z_{18}) \rtimes Z_{3} \rtimes Z_{2} $  & $($$Z_3$, $Z_3^2$ $|$ $Z_2\times Z_6$, $Z_2\times Z_{18}$, $Z_2^2$, $Z_{18}$$)$ \\ \cline{2-4}& $[1944, 1771]$ &  $ Z_{4} \times ((Z_{9} \times Z_{3} \times Z_{3}) \rtimes Z_{3} \rtimes Z_{2}) $  & $($$Z_3\times Z_6$, $Z_3\times Z_9$, $Z_3\times Z_{12}$, $Z_3\times Z_{18}$, $Z_3\times Z_{36}$, $Z_3$, $Z_3^2$, $Z_6$, $Z_9$, $Z_{12}$, $Z_{18}$, $Z_{36}$ $|$ $Z_2\times Z_6$, $Z_2\times Z_{12}$, $Z_2\times Z_{18}$, $Z_2\times Z_{36}$, $Z_3\times Z_6$, $Z_3\times Z_{12}$, $Z_3\times Z_{18}$, $Z_3\times Z_{36}$, $Z_6\times Z_{12}$, $Z_6\times Z_{18}$, $Z_6\times Z_{36}$, $Z_6$, $Z_6^2$, $Z_{12}$, $Z_{18}$, $Z_{36}$$)$ \\ \cline{2-4}& $[1944, 2293]$ &  $ Z_{9} \cdot ((Z_{3} \times Z_{3}) \rtimes Q_{8} \rtimes Z_{3})=((Z_{3} \times Z_{3}) \rtimes Z_{3} \rtimes Q_{8}) \cdot Z_{9} $  & $($$Z_{27}$ $|$ $Z_{54}$$)$ \\ \cline{2-4}& $[1944, 2294]$ &  $ Z_{9} \cdot ((Z_{3} \times Z_{3}) \rtimes Q_{8} \rtimes Z_{3})=((Z_{3} \times Z_{3}) \rtimes Z_{3} \rtimes Q_{8}) \cdot Z_{9} $  & $($$Z_{27}$ $|$ $Z_{54}$$)$ \\ \cline{2-4}& $~[1944, 2363]^*$ &  $ (Z_{18} \times Z_{6} \times Z_{3}) \rtimes Z_{3} \rtimes Z_{2} $  & $($$Z_3\times Z_9$, $Z_3$, $Z_3^2$, $Z_9$ $|$ $Z_2\times Z_6$, $Z_2\times Z_{18}$, $Z_3\times Z_6$, $Z_3\times Z_{18}$, $Z_6\times Z_{18}$, $Z_2^2$, $Z_6$, $Z_6^2$, $Z_{18}$$)$ \\ \cline{2-4}& $[1944, 3448]$ &  $ (Z_{9} \times Z_{3}) \rtimes Z_{3} \rtimes Q_{8} \rtimes Z_{3} $  & $($$Z_3\times Z_9$, $Z_3$, $Z_3^2$, $Z_9$ $|$ $Z_3\times Z_6$, $Z_3\times Z_{18}$, $Z_6$, $Z_{18}$$)$ \\ \hline$\pm\frac{5 \pi }{84}$ & $~~[1176, 243]_{\vartriangle}$ &  $ (Z_{14} \times Z_{14}) \rtimes Z_{3} \rtimes Z_{2} $  & $($$Z_3$ $|$ $Z_4$, $Z_{28}$$)$ \\ \hline\multirow{3}{*}{$\pm\frac{2 \pi }{33}$} & $~~[726, 5]_{\vartriangle}$ &  $ (Z_{11} \times Z_{11}) \rtimes Z_{3} \rtimes Z_{2} $  & $($$Z_3$ $|$ $Z_{22}$$)$ \\ \cline{2-4}& $[1452, 11]$ &  $ (Z_{11} \times Z_{11}) \rtimes Z_{3} \rtimes Z_{4} $  & $($$Z_3$, $Z_6$ $|$ $Z_{44}$$)$ \\ \cline{2-4}& $[1452, 23]$ &  $ Z_{2} \times ((Z_{11} \times Z_{11}) \rtimes Z_{3} \rtimes Z_{2}) $  & $($$Z_3$, $Z_6$ $|$ $Z_2\times Z_{22}$, $Z_{22}$$)$ \\ \hline
\end{longtable}
\end{center}

\begin{table}[hptb]
\centering
\begin{tabular}{|m{0.04\columnwidth}<{\centering}|m{0.20\columnwidth}<{\centering}|m{0.28\columnwidth}<{\centering}|m{0.4\columnwidth}|}

\hline
\hline
$\theta_{\nu}$ & GAP-id  & Group Structure &~~~~~ Residual Symmetries $~(\mathcal{G}_l~|~\mathcal{G}_\nu)$ \\
\hline
\multirow{8}{*}{$\pm\frac{\pi }{16}$} & $~~[384, 568]_{\vartriangle}$ &  $ (Z_{8} \times Z_{8}) \rtimes Z_{3} \rtimes Z_{2} $  & $($$Z_3$ $|$ $Z_{16}$$)$ \\ \cline{2-4}& $[768, 1085335]$ &  $ (Z_{8} \times Z_{8}) \rtimes Z_{3} \rtimes Z_{4} $  & $($$Z_3$, $Z_6$ $|$ $Z_2\times Z_{16}$, $Z_{16}$$)$ \\ \cline{2-4}& $[768, 1085727]$ &  $ Z_{2} \times ((Z_{8} \times Z_{8}) \rtimes Z_{3} \rtimes Z_{2}) $  & $($$Z_3$, $Z_6$ $|$ $Z_2\times Z_{16}$, $Z_{16}$$)$ \\ \cline{2-4}& $[1152, 154124]$ &  $ Z_{3} \times ((Z_{8} \times Z_{8}) \rtimes Z_{3} \rtimes Z_{2}) $  & $($$Z_3$, $Z_3^2$ $|$ $Z_{16}$, $Z_{48}$$)$ \\ \cline{2-4}& $~~[1536, 408544632]^*_{\vartriangle}$ &  $ (Z_{16} \times Z_{16}) \rtimes Z_{3} \rtimes Z_{2} $  & $($$Z_3$ $|$ $Z_2\times Z_4$, $Z_2\times Z_8$, $Z_2\times Z_{16}$, $Z_2^2$, $Z_4$, $Z_8$, $Z_{16}$$)$ \\ \cline{2-4}& $[1536, 408544641]$ &  $ (Z_{8} \times Z_{8}) \rtimes Z_{3} \rtimes Z_{8} $  & $($$Z_3$, $Z_6$, $Z_{12}$ $|$ $Z_2\times Z_{16}$, $Z_4\times Z_{16}$, $Z_{16}$$)$ \\ \cline{2-4}& $[1536, 408544715]$ &  $ Z_{4} \times ((Z_{8} \times Z_{8}) \rtimes Z_{3} \rtimes Z_{2}) $  & $($$Z_3$, $Z_6$, $Z_{12}$ $|$ $Z_2\times Z_{16}$, $Z_4\times Z_{16}$, $Z_{16}$$)$ \\ \cline{2-4}& $[1920, 236349]$ &  $ Z_{5} \times ((Z_{8} \times Z_{8}) \rtimes Z_{3} \rtimes Z_{2}) $  & $($$Z_3$, $Z_{15}$ $|$ $Z_{16}$, $Z_{80}$$)$ \\ \hline$\pm\frac{5 \pi }{96}$ & $~~[1536, 408544632]_{\vartriangle}$ &  $ (Z_{16} \times Z_{16}) \rtimes Z_{3} \rtimes Z_{2} $  & $($$Z_3$ $|$ $Z_{32}$$)$ \\ \hline$\pm\frac{2 \pi }{39}$ & $~~[1014, 7]_{\vartriangle}$ &  $ (Z_{13} \times Z_{13}) \rtimes Z_{3} \rtimes Z_{2} $  & $($$Z_3$ $|$ $Z_{26}$$)$ \\ \hline$\pm\frac{5 \pi }{78}$ & $~~[1014, 7]_{\vartriangle}$ &  $ (Z_{13} \times Z_{13}) \rtimes Z_{3} \rtimes Z_{2} $  & $($$Z_3$ $|$ $Z_{26}$$)$ \\ \hline\multirow{4}{*}{$\pm\frac{\pi }{20}$} & $~~[600, 179]_{\vartriangle}$ &  $ (Z_{10} \times Z_{10}) \rtimes Z_{3} \rtimes Z_{2} $  & $($$Z_3$ $|$ $Z_4$, $Z_{20}$$)$ \\ \cline{2-4}& $[1200, 682]$ &  $ (Z_{10} \times Z_{10}) \rtimes Z_{3} \rtimes Z_{4} $  & $($$Z_3$, $Z_6$ $|$ $Z_2\times Z_4$, $Z_2\times Z_{20}$, $Z_4$, $Z_{20}$$)$ \\ \cline{2-4}& $[1200, 1011]$ &  $ Z_{2} \times ((Z_{10} \times Z_{10}) \rtimes Z_{3} \rtimes Z_{2}) $  & $($$Z_3$, $Z_6$ $|$ $Z_2\times Z_4$, $Z_2\times Z_{20}$, $Z_4$, $Z_{20}$$)$ \\ \cline{2-4}& $[1800, 687]$ &  $ Z_{3} \times ((Z_{10} \times Z_{10}) \rtimes Z_{3} \rtimes Z_{2}) $  & $($$Z_3$, $Z_3^2$ $|$ $Z_4$, $Z_{12}$, $Z_{20}$, $Z_{60}$$)$ \\ \hline\hline
\end{tabular}
\caption{\label{tab:lepton_mixing2}The mixing patterns with $\chi^2$ less than 50 for NO spectrum generated by finite discrete groups up to order 2000. The patterns within the $3\sigma$ ranges of data~\cite{Gonzalez-Garcia:2014bfa} are not included since they have been collected in Table~\ref{tab:lepton_mixing1}. The superscript * indicates that the corresponding group has Klein subgroups and it allows  neutrinos to be Majorana particles. The subscripts $\vartriangle$ and $\vartriangle^{\prime}$ denote the groups that belong to the $D_{n,n}^{(0)}\cong\Delta(6n^2)$ and $D_{9n^{\prime},3n^{\prime}}^{(1)}\cong(Z_{9n^{\prime}}\times Z_{3n^{\prime}})\rtimes S_3$ group series, respectively. The notation $(\mathcal{G}_l~|~\mathcal{G}_\nu)$ denotes all possible combinations between the subgroups in the $\mathcal{G}_l$ list and the subgroups in the $\mathcal{G}_{\nu}$ list. The explicit forms of the residual subgroups can be found on our Web site~\cite{webdata}.}
\end{table}

\subsection{\label{subsec:quark_scan}Results for quark mixing}

The quark mixing matrix has been rather accurately measured. The fit results for the magnitudes of all nine CKM matrix elements are~\cite{pdg}
\begin{equation}
\label{eq:CKM_expe}|V_{\text{CKM}}|=\left(\begin{array}{ccc}
0.97427\pm0.00014 ~ &~   0.22536\pm0.00061  ~&~  0.00355\pm0.00015  \\
0.22522\pm0.00061  ~& ~  0.97343\pm0.00015  ~&~  0.0414\pm0.0012  \\
0.00886^{+0.00033}_{-0.00032}  ~ & ~ 0.0405^{+0.0011}_{-0.0012}   ~& ~ 0.99914\pm0.00005
\end{array}
\right)\,.
\end{equation}
In this subsection, we shall investigate whether the observed quark mixing in Eq.~\eqref{eq:CKM_expe} can be derived from a flavor symmetry group. Similar to the lepton sector, the left-handed quark doublets are assigned to an irreducible triplet representation of the group. We have scanned all the Abelian subgroups of each finite group with order less than 2000. However, we do not find any group that can give rise to the correct order of magnitude of the CKM matrix shown in Eq.~\eqref{eq:CKM_expe}. Furthermore, we relax the constraints on each CKM matrix element, and we only require that $0.1\leq |V_{us}|\leq0.3$ and $|V_{ub}|\leq |V_{cb}|<|V_{us}|$ should be fulfilled. Surprisingly, the admissible CKM matrices turn out to uniquely take the following form:
\begin{equation}
V_{\text{CKM}}=\left(\begin{array}{ccc}
\cos\theta_q  &~ -\sin\theta_q &~ 0 \\
\sin\theta_q  &~ \cos\theta_q  &~ 0 \\
 0 &~ 0 &~ 1
\end{array}
\right)\,,
\end{equation}
where the values of the rotation angle $\theta_{q}$ are
\begin{equation}
\label{eq:theta_q}
\theta_q\in\left\{\frac{\pi}{30}, \frac{\pi}{28}, \frac{\pi}{26}, \frac{\pi}{24}, \frac{\pi }{22},\frac{\pi}{20}, \frac{\pi}{18}, \frac{\pi}{17}, \frac{\pi}{16}, \frac{\pi}{15}, \frac{\pi}{14}, \frac{\pi}{13}, \frac{\pi}{12}, \frac{3\pi}{34}, \frac{\pi}{11}, \frac{3\pi}{32}\right\}\,.
\end{equation}
\begin{figure}[t!]
\centering
\includegraphics[width=0.70\textwidth]{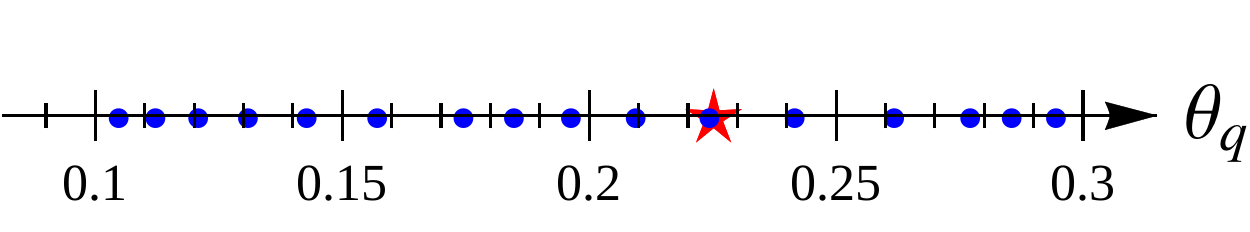}
\caption{\label{fig:quark_mixing}
The values of the Cabibbo mixing angle predicted by discrete flavor symmetry groups with order less than 2000. The blue dots represent the results shown in Eq.~\eqref{eq:theta_q}, and the red star refers to the experimental best fit value.}
\end{figure}
That is to say, the Cabibbo mixing angle can be naturally produced from a finite group. This point has already been seen from the analysis of the $SU(3)$ subgroups in Sec.~\ref{subsec:quark_mixing_analytical}. The Cabibbo mixing is a good leading order approximation to the measured value of the CKM matrix, as the other two smaller quark mixing angles could arise from higher order corrections. For comparison with experimental data, the predicted values of the Cabibbo mixing angle are plotted in Fig.~\ref{fig:quark_mixing}. It is remarkable that the value $\theta_{q}=\pi/14$ is very close to the experimental result. The groups that can generate the discrete set of Cabibbo angles in Eq.~\eqref{eq:theta_q} and the associated residual subgroups $\mathcal{G}_{U}$ and $\mathcal{G}_D$ are summarized in Table~\ref{tab:quark_mixing}. Notice that interchanging the roles of the subgroups $\mathcal{G}_U$ and $\mathcal{G}_D$ would transform the CKM matrix into its Hermitian conjugate, and therefore the same Cabibbo angle would be obtained. Generally the residual symmetries $\mathcal{G}_U$ and $\mathcal{G}_D$ do not give rise to the entire flavor symmetry group.  It is interesting that some groups can lead to two values of $\theta_{q}$ for distinct residual symmetries. For example, the group $\Delta(6\times11^2)$ with identification number $[726, 5]$ can predict $\theta_q=\pi/22$ or $\pi/11$. We see that the group of the smallest order is $[216,95]$, it leads to $\theta_q=\pi/12$, which deviates a bit from the experimental value, and this insignificant discrepancy is expected to be resolved by subleading corrections. What is more, the most favored value $\theta_q=\pi/14$ can be generated from the group $\Delta(6\times7^2)$ with the group identification $[294,7]$. This result is consistent with our analytical study performed in Sec.~\ref{subsec:quark_mixing_analytical}.
$\Delta(6\times7^2)$ provides a new flavor symmetry for constructing models of quark flavor mixing.

Since the third generation quark is much heavier than the first two generations, it is assumed that the first and the second generations of the left-handed quark fields transform as a doublet while the third generation is a singlet in the well-known $U(2)$ flavor symmetry theory~\cite{Barbieri:1995uv}. The doublet plus singlet assignment for the quark sector is also exploited by scanning the finite groups with faithful two-dimensional irreducible representation up to order 1000. All the possible remnant subgroups $\mathcal{G}_{U}$ and $\mathcal{G}_{D}$ are considered for each permissible flavor symmetry group. We find that the observed Cabibbo mixing angle can easily be  accommodated. The phenomenologically viable value of $\theta_{q}$, which can be achieved in this way, are presented in Table~\ref{tab:quark_mixing_2plus1} where the constraint $0.223\leq|V_{us}|\leq0.227$ is imposed. The complete list of the admissible values of $\theta_{q}$ and the flavor symmetry groups as well as the residual symmetries are provided at the Web site~\cite{webdata}. It is remarkable that the dihedral group $D_{14}$ with \texttt{GAP} id $[14, 1]$ can give rise to the interesting Cabibbo angle $\theta_{q}=\pi/14$.

\begin{longtable}{|m{0.02\columnwidth}<{\centering}|m{0.20\columnwidth}<{\centering}|m{0.7\columnwidth}|}

\hline
\hline
$\theta_q$ & GAP-id & ~~~~~~~~~~~Residual Symmetries$~(\mathcal{G}_U~|~\mathcal{G}_D)$ or $~(\mathcal{G}_D~|~\mathcal{G}_U)$\\
\hline
\endfirsthead

\hline
$\theta_q$ & GAP-id & ~~~~~~~~~~~Residual Symmetries $~(\mathcal{G}_U~|~\mathcal{G}_D)$ or $~(\mathcal{G}_D~|~\mathcal{G}_U)$\\
\hline
\endhead
\hline
\caption[]{(continued)}\\
\endfoot

\hline
\hline
\caption{\label{tab:quark_mixing}The Cabibbo mixing angle predicted by finite discrete groups up to order 2000, where we require $0.1\leq |V_{us}|\leq0.3$ and $|V_{ub}|\leq |V_{cb}|<|V_{us}|$. The three left-handed quark doublets are assumed to transform as a faithful irreducible triplet of the flavor symmetry group. The superscript  *  indicates that the corresponding group has Klein subgroups and it allows  neutrinos to be Majorana particles. The subscripts $\vartriangle$ and $\vartriangle^{\prime}$ denote the groups that belong to the $D_{n,n}^{(0)}\cong\Delta(6n^2)$ and $D_{9n^{\prime},3n^{\prime}}^{(1)}\cong(Z_{9n^{\prime}}\times Z_{3n^{\prime}})\rtimes S_3$ group series, respectively. The notation $(\mathcal{G}_U~|~\mathcal{G}_D)$ denotes all possible combinations between the subgroups in the $\mathcal{G}_U$ list and the subgroups in the $\mathcal{G}_{D}$ list. The same prediction for the Cabibbo angle would be obtained if the roles of $\mathcal{G}_{U}$ and $\mathcal{G}_{D}$ are interchanged. The explicit forms of the residual subgroups can be found on our Web site~\cite{webdata}.}\\
\endlastfoot
$\frac{\pi }{30}$ & $~~[1350, 46]_{\vartriangle}$ & $($$Z_{10}$, $Z_{30}$ $|$ $Z_{10}$, $Z_{30}$$)$ \\ \hline$\frac{\pi }{28}$ & $~~[1176, 243]_{\vartriangle}$ & $($$Z_2\times Z_{14}$, $Z_2^2$, $Z_{14}$ $|$ $Z_4$, $Z_{28}$$)$ \\ \hline$\frac{\pi }{26}$ & $~~[1014, 7]_{\vartriangle}$ & $($$Z_{26}$ $|$ $Z_{26}$$)$ \\ \hline\multirow{5}{*}{$\frac{\pi }{24}$} & $~~[864, 701]_{\vartriangle}$ & $($$Z_2\times Z_4$, $Z_2\times Z_6$, $Z_2\times Z_{12}$, $Z_2^2$, $Z_4$, $Z_{12}$ $|$ $Z_8$, $Z_{24}$$)$ \\ \cline{2-3} & $[1728, 2847]$ & $($$Z_2\times Z_4$, $Z_2\times Z_{12}$, $Z_4\times Z_{12}$, $Z_4$, $Z_4^2$, $Z_{12}$ $|$ $Z_2\times Z_8$, $Z_2\times Z_{24}$, $Z_8$, $Z_{24}$$)$ \\ \cline{2-3} & $[1728, 13052]$ & $($$Z_2\times Z_4$, $Z_2\times Z_6$, $Z_2\times Z_{12}$, $Z_2^2\times Z_4$, $Z_2^2\times Z_6$, $Z_2^2\times Z_{12}$, $Z_2^2$, $Z_2^3$, $Z_4$, $Z_{12}$ $|$ $Z_2\times Z_8$, $Z_2\times Z_{24}$, $Z_8$, $Z_{24}$$)$ \\ \hline\multirow{3}{*}{$\frac{\pi }{22}$} & $~~[726, 5]_{\vartriangle}$ & $($$Z_{22}$ $|$ $Z_{22}$$)$ \\ \cline{2-3} & $[1452, 11]$ & $($$Z_{44}$ $|$ $Z_{44}$$)$ \\ \cline{2-3} & $[1452, 23]$ & $($$Z_2\times Z_{22}$, $Z_{22}$ $|$ $Z_2\times Z_{22}$, $Z_{22}$$)$ \\ \hline\multirow{4}{*}{$\frac{\pi }{20}$} & $~~[600, 179]_{\vartriangle}$ & $($$Z_2\times Z_{10}$, $Z_2^2$, $Z_{10}$ $|$ $Z_4$, $Z_{20}$$)$ \\ \cline{2-3} & $[1200, 682]$ & $($$Z_2\times Z_4$, $Z_2\times Z_{20}$, $Z_{20}$ $|$ $Z_2\times Z_4$, $Z_2\times Z_{20}$, $Z_4$, $Z_{20}$$)$ \\ \cline{2-3} & $[1200, 1011]$ & $($$Z_2\times Z_4$, $Z_2\times Z_{20}$, $Z_4$, $Z_{20}$ $|$ $Z_2\times Z_{10}$, $Z_2^2\times Z_{10}$, $Z_2^2$, $Z_2^3$, $Z_{10}$$)$ \\ \cline{2-3} & $[1800, 687]$ & $($$Z_2\times Z_6$, $Z_2\times Z_{10}$, $Z_2\times Z_{30}$, $Z_2^2$, $Z_{10}$, $Z_{30}$ $|$ $Z_4$, $Z_{12}$, $Z_{20}$, $Z_{60}$$)$ \\ \hline\multirow{11}{*}{$\frac{\pi }{18}$} & $~~[486, 61]_{\vartriangle}$ & $($$Z_{18}$ $|$ $Z_{18}$$)$ \\ \cline{2-3} & $[972, 64]$ & $($$Z_{36}$ $|$ $Z_{36}$$)$ \\ \cline{2-3} & $[972, 245]$ & $($$Z_2\times Z_{18}$, $Z_{18}$ $|$ $Z_2\times Z_{18}$, $Z_{18}$$)$ \\ \cline{2-3} & $~~[1458, 659]_{\vartriangle'}$ & $($$Z_{18}$, $Z_{54}$ $|$ $Z_{18}$, $Z_{54}$$)$ \\ \cline{2-3} & $[1458, 663]$ & $($$Z_{18}$, $Z_{54}$ $|$ $Z_{18}$, $Z_{54}$$)$ \\ \cline{2-3} & $[1458, 666]$ & $($$Z_{18}$, $Z_{54}$ $|$ $Z_{18}$, $Z_{54}$$)$ \\ \cline{2-3} & $[1458, 1371]$ & $($$Z_3\times Z_6$, $Z_3\times Z_{18}$, $Z_6$, $Z_{18}$ $|$ $Z_3\times Z_6$, $Z_3\times Z_{18}$, $Z_6$, $Z_{18}$$)$ \\ \cline{2-3} & $[1944, 70]$ & $($$Z_{72}$ $|$ $Z_{72}$$)$ \\ \cline{2-3} & $[1944, 544]$ & $($$Z_2\times Z_{18}$, $Z_2\times Z_{36}$, $Z_{18}$, $Z_{36}$ $|$ $Z_2\times Z_{18}$, $Z_2\times Z_{36}$, $Z_{18}$, $Z_{36}$$)$ \\ \cline{2-3} & $~~[1944, 849]^*_{\vartriangle}$ & $($$Z_2\times Z_6$, $Z_2\times Z_{18}$, $Z_2^2$, $Z_{18}$ $|$ $Z_2\times Z_6$, $Z_2\times Z_{18}$, $Z_2^2$, $Z_{18}$$)$, $($$Z_4$, $Z_{12}$, $Z_{36}$ $|$ $Z_4$, $Z_{12}$, $Z_{36}$$)$ \\ \hline$\frac{\pi }{17}$ & $~~[1734, 5]_{\vartriangle}$ & $($$Z_{34}$ $|$ $Z_{34}$$)$ \\ \hline\multirow{4}{*}{$\frac{\pi }{16}$} & $~~[384, 568]_{\vartriangle}$ & $($$Z_2\times Z_4$, $Z_2\times Z_8$, $Z_2^2$, $Z_4$, $Z_8$ $|$ $Z_{16}$$)$ \\ \cline{2-3} & $[768, 1085335]$ & $($$Z_2\times Z_4$, $Z_2\times Z_8$, $Z_4\times Z_8$, $Z_4$, $Z_4^2$, $Z_8$ $|$ $Z_2\times Z_{16}$, $Z_{16}$$)$ \\ \cline{2-3} & $[768, 1085727]$ & $($$Z_2\times Z_4$, $Z_2\times Z_8$, $Z_2^2\times Z_4$, $Z_2^2\times Z_8$, $Z_2^2$, $Z_2^3$, $Z_4$, $Z_8$ $|$ $Z_2\times Z_{16}$, $Z_{16}$$)$ \\ \cline{2-3}
\multirow{8}{*}{$\frac{\pi }{16}$} & $[1152, 154124]$ & $($$Z_2\times Z_4$, $Z_2\times Z_6$, $Z_2\times Z_8$, $Z_2\times Z_{12}$, $Z_2\times Z_{24}$, $Z_2^2$, $Z_4$, $Z_8$, $Z_{12}$, $Z_{24}$ $|$ $Z_{16}$, $Z_{48}$$)$ \\ \cline{2-3} & $~~[1536, 408544632]_{\vartriangle}$ & $($$Z_2\times Z_4$, $Z_2\times Z_8$, $Z_2\times Z_{16}$, $Z_2^2$, $Z_4$, $Z_8$, $Z_{16}$ $|$ $Z_2\times Z_4$, $Z_2\times Z_8$, $Z_2\times Z_{16}$, $Z_2^2$, $Z_4$, $Z_8$, $Z_{16}$$)$, $($$Z_{32}$ $|$ $Z_{32}$$)$ \\ \cline{2-3} & $[1536, 408544641]$ & $($$Z_2\times Z_8$, $Z_4\times Z_8$, $Z_8$, $Z_8^2$ $|$ $Z_2\times Z_{16}$, $Z_4\times Z_{16}$, $Z_{16}$$)$ \\ \cline{2-3} & $[1536, 408544715]$ & $($$Z_2\times Z_4$, $Z_2\times Z_4^2$, $Z_2\times Z_8$, $Z_2^2\times Z_4$, $Z_2^2\times Z_8$, $Z_4\times Z_8$, $Z_2\times Z_4\times Z_8$, $Z_2^2$, $Z_2^3$, $Z_4$, $Z_4^2$, $Z_8$ $|$ $Z_2\times Z_{16}$, $Z_4\times Z_{16}$, $Z_{16}$$)$ \\ \cline{2-3} & $[1920, 236349]$ & $($$Z_2\times Z_4$, $Z_2\times Z_8$, $Z_2\times Z_{10}$, $Z_2\times Z_{20}$, $Z_2\times Z_{40}$, $Z_2^2$, $Z_4$, $Z_8$, $Z_{20}$, $Z_{40}$ $|$ $Z_{16}$, $Z_{80}$$)$ \\ \hline$\frac{\pi }{15}$ & $~~[1350, 46]_{\vartriangle}$ & $($$Z_{10}$, $Z_{30}$ $|$ $Z_{10}$, $Z_{30}$$)$\\ \hline
\multirow{3}{*}{} & $~~[294, 7]_{\vartriangle}$ & $($$Z_{14}$ $|$ $Z_{14}$$)$ \\ \cline{2-3} & $[588, 16]$ & $($$Z_{28}$ $|$ $Z_{28}$$)$ \\ \cline{2-3} & $[588, 39]$ & $($$Z_2\times Z_{14}$, $Z_{14}$ $|$ $Z_2\times Z_{14}$, $Z_{14}$$)$ \\ \cline{2-3}
\multirow{7}{*}[14pt]{$\frac{\pi }{14}$}& $[882, 38]$ & $($$Z_{14}$, $Z_{42}$ $|$ $Z_{14}$, $Z_{42}$$)$ \\ \cline{2-3} & $[1176, 57]$ & $($$Z_{56}$ $|$ $Z_{56}$$)$ \\ \cline{2-3} & $[1176, 99]$ & $($$Z_2\times Z_{14}$, $Z_2\times Z_{28}$, $Z_{14}$, $Z_{28}$ $|$ $Z_2\times Z_{14}$, $Z_2\times Z_{28}$, $Z_{14}$, $Z_{28}$$)$ \\ \cline{2-3} & $~~[1176, 243]_{\vartriangle}$ & $($$Z_2\times Z_{14}$, $Z_2^2$, $Z_{14}$ $|$ $Z_2\times Z_{14}$, $Z_2^2$, $Z_{14}$$)$, $($$Z_4$, $Z_{28}$ $|$ $Z_4$, $Z_{28}$$)$ \\ \cline{2-3} & $[1470, 13]$ & $($$Z_{14}$, $Z_{70}$ $|$ $Z_{14}$, $Z_{70}$$)$ \\ \cline{2-3} & $[1764, 102]$ & $($$Z_{28}$, $Z_{84}$ $|$ $Z_{28}$, $Z_{84}$$)$ \\ \cline{2-3} & $[1764, 171]$ & $($$Z_2\times Z_{14}$, $Z_2\times Z_{42}$, $Z_{14}$, $Z_{42}$ $|$ $Z_2\times Z_{14}$, $Z_2\times Z_{42}$, $Z_{14}$, $Z_{42}$$)$ \\ \hline$\frac{\pi }{13}$ & $~~[1014, 7]_{\vartriangle}$ & $($$Z_{26}$ $|$ $Z_{26}$$)$ \\ \hline\multirow{26}{*}{$\frac{\pi }{12}$} & $~~[216, 95]_{\vartriangle}$ & $($$Z_2\times Z_6$, $Z_2^2$ $|$ $Z_4$, $Z_{12}$$)$ \\ \cline{2-3} & $[432, 260]$ & $($$Z_2\times Z_4$, $Z_2\times Z_{12}$ $|$ $Z_2\times Z_4$, $Z_2\times Z_{12}$, $Z_4$, $Z_{12}$$)$ \\ \cline{2-3} & $[432, 538]$ & $($$Z_2\times Z_4$, $Z_2\times Z_{12}$, $Z_4$, $Z_{12}$ $|$ $Z_2\times Z_6$, $Z_2^2\times Z_6$, $Z_2^2$, $Z_2^3$$)$ \\ \cline{2-3} & $~~[648, 259]^*_{\vartriangle'}$ & $($$Z_2\times Z_6$, $Z_2\times Z_{18}$, $Z_2^2$, $Z_{18}$ $|$ $Z_4$, $Z_{12}$, $Z_{36}$$)$ \\ \cline{2-3} & $~[648, 260]^*$ & $($$Z_2\times Z_6$, $Z_2\times Z_{18}$, $Z_2^2$, $Z_{18}$ $|$ $Z_4$, $Z_{12}$, $Z_{36}$$)$ \\ \cline{2-3} & $~[648, 266]^*$ & $($$Z_2\times Z_6$, $Z_3\times Z_6$, $Z_2^2$, $Z_6$, $Z_6^2$ $|$ $Z_3\times Z_{12}$, $Z_4$, $Z_{12}$$)$ \\ \cline{2-3} & $[648, 563]$ & $($$Z_2\times Z_6$, $Z_2\times Z_{18}$, $Z_2^2$ $|$ $Z_4$, $Z_{12}$, $Z_{36}$$)$ \\ \cline{2-3} & $~~[864, 701]_{\vartriangle}$ & $($$Z_2\times Z_4$, $Z_2\times Z_6$, $Z_2\times Z_{12}$, $Z_2^2$, $Z_4$, $Z_{12}$ $|$ $Z_2\times Z_4$, $Z_2\times Z_6$, $Z_2\times Z_{12}$, $Z_2^2$, $Z_4$, $Z_{12}$$)$, $($$Z_8$, $Z_{24}$ $|$ $Z_8$, $Z_{24}$$)$ \\ \cline{2-3} & $[864, 703]$ & $($$Z_2\times Z_8$, $Z_2\times Z_{24}$ $|$ $Z_2\times Z_8$, $Z_2\times Z_{24}$, $Z_8$, $Z_{24}$$)$ \\ \cline{2-3} & $[864, 2348]$ & $($$Z_2\times Z_4$, $Z_2\times Z_{12}$ $|$ $Z_2\times Z_4$, $Z_2\times Z_6$, $Z_2\times Z_{12}$, $Z_2^2\times Z_4$, $Z_2^2\times Z_6$, $Z_2^2\times Z_{12}$, $Z_4\times Z_{12}$, $Z_2^2$, $Z_2^3$, $Z_4$, $Z_4^2$, $Z_{12}$$)$, $($$Z_2\times Z_6$, $Z_2^2\times Z_4$, $Z_2^2\times Z_6$, $Z_2^2\times Z_{12}$, $Z_2^2$, $Z_2^3$ $|$ $Z_4\times Z_{12}$, $Z_4$, $Z_4^2$, $Z_{12}$$)$ \\ \cline{2-3} & $[1080, 298]$ & $($$Z_2\times Z_6$, $Z_2\times Z_{10}$, $Z_2\times Z_{30}$, $Z_2^2$ $|$ $Z_4$, $Z_{12}$, $Z_{20}$, $Z_{60}$$)$ \\ \cline{2-3} & $[1296, 688]$ & $($$Z_2\times Z_4$, $Z_2\times Z_{12}$, $Z_2\times Z_{36}$, $Z_{36}$ $|$ $Z_2\times Z_4$, $Z_2\times Z_{12}$, $Z_2\times Z_{36}$, $Z_4$, $Z_{12}$, $Z_{36}$$)$ \\ \cline{2-3} & $[1296, 689]$ & $($$Z_2\times Z_4$, $Z_2\times Z_{12}$, $Z_2\times Z_{36}$, $Z_{36}$ $|$ $Z_2\times Z_4$, $Z_2\times Z_{12}$, $Z_2\times Z_{36}$, $Z_4$, $Z_{12}$, $Z_{36}$$)$ \\ \cline{2-3} & $[1296, 699]$ & $($$Z_2\times Z_4$, $Z_2\times Z_{12}$, $Z_3\times Z_{12}$, $Z_6\times Z_{12}$, $Z_{12}$ $|$ $Z_2\times Z_4$, $Z_2\times Z_{12}$, $Z_3\times Z_{12}$, $Z_6\times Z_{12}$, $Z_4$, $Z_{12}$$)$ \\ \cline{2-3} & $~[1296, 1820]^*$ & $($$Z_2\times Z_4$, $Z_2\times Z_{12}$, $Z_2\times Z_{36}$, $Z_4$, $Z_{12}$, $Z_{36}$ $|$ $Z_2\times Z_6$, $Z_2\times Z_{18}$, $Z_2^2\times Z_6$, $Z_2^2\times Z_{18}$, $Z_2^2$, $Z_2^3$, $Z_{18}$$)$ \\ \cline{2-3} & $~[1296, 1821]^*$ & $($$Z_2\times Z_4$, $Z_2\times Z_{12}$, $Z_2\times Z_{36}$, $Z_4$, $Z_{12}$, $Z_{36}$ $|$ $Z_2\times Z_6$, $Z_2\times Z_{18}$, $Z_2^2\times Z_6$, $Z_2^2\times Z_{18}$, $Z_2^2$, $Z_2^3$, $Z_{18}$$)$ \\ \cline{2-3} & $~[1296, 1827]^*$ & $($$Z_2\times Z_4$, $Z_2\times Z_{12}$, $Z_3\times Z_{12}$, $Z_6\times Z_{12}$, $Z_4$, $Z_{12}$ $|$ $Z_2\times Z_6$, $Z_2\times Z_6^2$, $Z_2^2\times Z_6$, $Z_3\times Z_6$, $Z_2^2$, $Z_2^3$, $Z_6$, $Z_6^2$$)$ \\ \cline{2-3}
\multirow{22}{*}{$\frac{\pi }{12}$}& $[1296, 2113]$ & $($$Z_2\times Z_4$, $Z_2\times Z_{12}$, $Z_2\times Z_{36}$ $|$ $Z_2\times Z_4$, $Z_2\times Z_{12}$, $Z_2\times Z_{36}$, $Z_4$, $Z_{12}$, $Z_{36}$$)$ \\ \cline{2-3} & $[1296, 2993]$ & $($$Z_2\times Z_4$, $Z_2\times Z_{12}$, $Z_2\times Z_{36}$, $Z_4$, $Z_{12}$, $Z_{36}$ $|$ $Z_2\times Z_6$, $Z_2\times Z_{18}$, $Z_2^2\times Z_6$, $Z_2^2\times Z_{18}$, $Z_2^2$, $Z_2^3$$)$ \\ \cline{2-3} & $[1512, 500]$ & $($$Z_2\times Z_6$, $Z_2\times Z_{14}$, $Z_2\times Z_{42}$, $Z_2^2$ $|$ $Z_4$, $Z_{12}$, $Z_{28}$, $Z_{84}$$)$ \\ \cline{2-3} & $[1728, 2847]$ & $($$Z_2\times Z_4$, $Z_2\times Z_{12}$, $Z_4\times Z_{12}$, $Z_4$, $Z_4^2$, $Z_{12}$ $|$ $Z_2\times Z_4$, $Z_2\times Z_{12}$, $Z_4\times Z_{12}$, $Z_4$, $Z_4^2$, $Z_{12}$$)$, $($$Z_2\times Z_8$, $Z_2\times Z_{24}$, $Z_8$, $Z_{24}$ $|$ $Z_2\times Z_8$, $Z_2\times Z_{24}$, $Z_8$, $Z_{24}$$)$ \\ \cline{2-3} & $[1728, 2855]$ & $($$Z_2\times Z_{16}$, $Z_2\times Z_{48}$ $|$ $Z_2\times Z_{16}$, $Z_2\times Z_{48}$, $Z_{16}$, $Z_{48}$$)$ \\ \cline{2-3} & $[1728, 13052]$ & $($$Z_2\times Z_4$, $Z_2\times Z_6$, $Z_2\times Z_{12}$, $Z_2^2\times Z_4$, $Z_2^2\times Z_6$, $Z_2^2\times Z_{12}$, $Z_2^2$, $Z_2^3$, $Z_4$, $Z_{12}$ $|$ $Z_2\times Z_4$, $Z_2\times Z_6$, $Z_2\times Z_{12}$, $Z_2^2\times Z_4$, $Z_2^2\times Z_6$, $Z_2^2\times Z_{12}$, $Z_2^2$, $Z_2^3$, $Z_4$, $Z_{12}$$)$, $($$Z_2\times Z_8$, $Z_2\times Z_{24}$, $Z_8$, $Z_{24}$ $|$ $Z_2\times Z_8$, $Z_2\times Z_{24}$, $Z_8$, $Z_{24}$$)$ \\ \cline{2-3}& $[1728, 13083]$ & $($$Z_2\times Z_4$, $Z_2\times Z_8$, $Z_2\times Z_{12}$, $Z_2\times Z_{24}$ $|$ $Z_2\times Z_4$, $Z_2\times Z_6$, $Z_2\times Z_8$, $Z_2\times Z_{12}$, $Z_2\times Z_{24}$, $Z_2^2\times Z_4$, $Z_2^2\times Z_6$, $Z_2^2\times Z_8$, $Z_2^2\times Z_{12}$, $Z_2^2\times Z_{24}$, $Z_4\times Z_8$, $Z_4\times Z_{12}$, $Z_4\times Z_{24}$, $Z_2^2$, $Z_2^3$, $Z_4$, $Z_4^2$, $Z_8$, $Z_{12}$, $Z_{24}$$)$, $($$Z_2\times Z_6$, $Z_2^2\times Z_4$, $Z_2^2\times Z_6$, $Z_2^2\times Z_8$, $Z_2^2\times Z_{12}$, $Z_2^2\times Z_{24}$, $Z_2^2$, $Z_2^3$ $|$ $Z_4\times Z_8$, $Z_4\times Z_{12}$, $Z_4\times Z_{24}$, $Z_4$, $Z_4^2$, $Z_8$, $Z_{12}$, $Z_{24}$$)$ \\ \cline{2-3} & $~[1944, 832]^*$ & $($$Z_2\times Z_6$, $Z_2\times Z_{18}$, $Z_2\times Z_{54}$, $Z_2^2$, $Z_{54}$ $|$ $Z_4$, $Z_{12}$, $Z_{36}$, $Z_{108}$$)$ \\ \cline{2-3} & $~[1944, 833]^*$ & $($$Z_2\times Z_6$, $Z_2\times Z_{18}$, $Z_2\times Z_{54}$, $Z_2^2$, $Z_{54}$ $|$ $Z_4$, $Z_{12}$, $Z_{36}$, $Z_{108}$$)$ \\ \cline{2-3} & $~~[1944, 849]^*_{\vartriangle}$ & $($$Z_2\times Z_6$, $Z_2\times Z_{18}$, $Z_2^2$, $Z_{18}$ $|$ $Z_4$, $Z_{12}$, $Z_{36}$$)$ \\ \cline{2-3} & $~[1944, 2363]^*$ & $($$Z_2\times Z_6$, $Z_2\times Z_{18}$, $Z_3\times Z_6$, $Z_3\times Z_{18}$, $Z_6\times Z_{18}$, $Z_2^2$, $Z_6$, $Z_6^2$, $Z_{18}$ $|$ $Z_3\times Z_{12}$, $Z_3\times Z_{36}$, $Z_4$, $Z_{12}$, $Z_{36}$$)$ \\ \cline{2-3} & $[1944, 2415]$ & $($$Z_2\times Z_6$, $Z_2\times Z_{18}$, $Z_2\times Z_{54}$, $Z_2^2$ $|$ $Z_4$, $Z_{12}$, $Z_{36}$, $Z_{108}$$)$ \\ \hline$\frac{3 \pi }{34}$ & $~~[1734, 5]_{\vartriangle}$ & $($$Z_{34}$ $|$ $Z_{34}$$)$ \\ \hline\multirow{3}{*}{$\frac{\pi }{11}$} & $~~[726, 5]_{\vartriangle}$ & $($$Z_{22}$ $|$ $Z_{22}$$)$ \\ \cline{2-3} & $[1452, 11]$ & $($$Z_{44}$ $|$ $Z_{44}$$)$ \\ \cline{2-3} & $[1452, 23]$ & $($$Z_2\times Z_{22}$, $Z_{22}$ $|$ $Z_2\times Z_{22}$, $Z_{22}$$)$ \\ \hline$\frac{3 \pi }{32}$ & $~~[1536, 408544632]_{\vartriangle}$ & $($$Z_2\times Z_4$, $Z_2\times Z_8$, $Z_2\times Z_{16}$, $Z_2^2$, $Z_4$, $Z_8$, $Z_{16}$ $|$ $Z_{32}$$)$ \\ \hline
\end{longtable}

\begin{longtable}[c]{|m{0.04\columnwidth}<{\centering}|m{0.4\columnwidth}<{\centering}||m{0.04\columnwidth}<{\centering}|m{0.4\columnwidth}<{\centering}|}

\hline
\hline
$\theta_q$ & GAP-id & $\theta_q$ & GAP-id \\
\hline
\endfirsthead

\hline
$\theta_q$ & GAP-id & $\theta_q$ & GAP-id  \\
\hline
\endhead
\hline
\caption[]{(continued)}\\
\endfoot

\hline
\hline
\caption{\label{tab:quark_mixing_2plus1}The Cabibbo mixing angle predicted by finite discrete groups up to order 1000, where we require $0.223\leq |V_{us}|\leq0.227$. The explicit form of the residual subgroups $\mathcal{G}_{U,D}$ can be found at our Web site~\cite{webdata}. Here the first and the second generations of the quark doublet fields are assigned to a faithful doublet of the flavor symmetry group, and the third generation transforms as a singlet. }\\
\endlastfoot

$\frac{32 \pi }{447}$ & $[894, 3]$ & $\frac{31 \pi }{433}$ & $[866, 1]$  \\ \hline $\frac{30 \pi }{419}$ & $[838, 1]$ & $\frac{29 \pi }{405}$ & $[810, 3]$  \\ \hline $\frac{28 \pi }{391}$ & $[782, 3]$ & $\frac{27 \pi }{377}$ & $[754, 3]$  \\ \hline $\frac{26 \pi }{363}$ & $[726, 3]$ & $\frac{25 \pi }{349}$ & $[698, 1]$  \\ \hline $\frac{24 \pi }{335}$ & $[670, 3]$ & $\frac{23 \pi }{321}$ & $[642, 3]$  \\ \hline $\frac{22 \pi }{307}$ & $[614, 1]$ & $\frac{21 \pi }{293}$ & $[586, 1]$  \\ \hline $\frac{20 \pi }{279}$ & $[558, 5]$ & $\frac{19 \pi }{265}$ & $[530, 3]$  \\ \hline $\frac{18 \pi }{251}$ & $[502, 1]$ & $\frac{71 \pi }{990}$ & $[990, 11]$  \\ \hline $\frac{35 \pi }{488}$ & $[976, 6]$, $[976, 7]$, $[976, 8]$ & $\frac{69 \pi }{962}$ & $[962, 3]$  \\ \hline $\frac{17 \pi }{237}$ & $[474, 5]$, $[948, 5]$, $[948, 14]$ & $\frac{67 \pi }{934}$ & $[934, 1]$  \\ \hline $\frac{33 \pi }{460}$ & $[920, 24]$, $[920, 26]$, $[920, 28]$ & $\frac{65 \pi }{906}$ & $[906, 5]$  \\ \hline $\frac{16 \pi }{223}$ & $[446, 1]$, $[892, 1]$, $[892, 3]$ & $\frac{63 \pi }{878}$ & $[878, 1]$  \\ \hline $\frac{31 \pi }{432}$ & $[864, 4]$, $[864, 7]$, $[864, 8]$ & $\frac{61 \pi }{850}$ & $[850, 3]$  \\ \hline $\frac{15 \pi }{209}$ & $[418, 3]$, $[836, 3]$, $[836, 8]$ & $\frac{59 \pi }{822}$ & $[822, 3]$  \\ \hline $\frac{29 \pi }{404}$ & $[808, 4]$, $[808, 6]$, $[808, 8]$ & $\frac{57 \pi }{794}$ & $[794, 1]$  \\ \hline $\frac{14 \pi }{195}$ & $[390, 11]$, $[780, 11]$, $[780, 52]$ & $\frac{55 \pi }{766}$ & $[766, 1]$  \\ \hline $\frac{27 \pi }{376}$ & $[752, 5]$, $[752, 6]$, $[752, 7]$ & $\frac{53 \pi }{738}$ & $[738, 3]$  \\ \hline $\frac{13 \pi }{181}$ & $[362, 1]$, $[724, 1]$, $[724, 4]$ & $\frac{51 \pi }{710}$ & $[710, 5]$  \\ \hline $\frac{25 \pi }{348}$ & $[696, 25]$, $[696, 27]$, $[696, 29]$ & $\frac{49 \pi }{682}$ & $[682, 3]$  \\ \hline $\frac{12 \pi }{167}$ & $[334, 1]$, $[668, 1]$, $[668, 3]$ & $\frac{47 \pi }{654}$ & $[654, 5]$  \\ \hline $\frac{35 \pi }{487}$ & $[974, 1]$ & $\frac{23 \pi }{320}$ & $[640, 6]$, $[640, 7]$, $[640, 8]$  \\ \hline $\frac{34 \pi }{473}$ & $[946, 3]$ & $\frac{45 \pi }{626}$ & $[626, 1]$  \\ \hline $\frac{11 \pi }{153}$ & $[306, 3]$, $[612, 3]$, $[612, 11]$, $[918, 3]$, $[918, 12]$ & $\frac{43 \pi }{598}$ & $[598, 3]$  \\ \hline $\frac{32 \pi }{445}$ & $[890, 3]$ & $\frac{21 \pi }{292}$ & $[584, 4]$, $[584, 6]$, $[584, 8]$  \\ \hline $\frac{31 \pi }{431}$ & $[862, 1]$ & $\frac{41 \pi }{570}$ & $[570, 11]$  \\ \hline $\frac{10 \pi }{139}$ & $[278, 1]$, $[556, 1]$, $[556, 3]$, $[834, 4]$, $[834, 5]$ & $\frac{39 \pi }{542}$ & $[542, 1]$  \\ \hline $\frac{29 \pi }{403}$ & $[806, 3]$ & $\frac{19 \pi }{264}$ & $[528, 65]$, $[528, 66]$, $[528, 67]$  \\ \hline $\frac{28 \pi }{389}$ & $[778, 1]$ & $\frac{37 \pi }{514}$ & $[514, 1]$  \\ \hline $\frac{9 \pi }{125}$ & $[250, 1]$, $[500, 1]$, $[500, 4]$, $[750, 2]$, $[750, 3]$, $[1000, 1]$, $[1000, 4]$, $[1000, 5]$, $[1000, 6]$, $[1000, 8]$ & $\frac{71 \pi }{986}$ & $[986, 3]$  \\ \hline $\frac{35 \pi }{486}$ & $[486, 1]$, $[972, 1]$, $[972, 4]$ & $\frac{26 \pi }{361}$ & $[722, 1]$  \\ \hline $\frac{69 \pi }{958}$ & $[958, 1]$ & $\frac{17 \pi }{236}$ & $[472, 3]$, $[472, 5]$, $[472, 7]$, $[944, 4]$, $[944, 5]$, $[944, 6]$, $[944, 7]$, $[944, 9]$, $[944, 30]$  \\ \hline $\frac{67 \pi }{930}$ & $[930, 17]$ & $\frac{25 \pi }{347}$ & $[694, 1]$  \\ \hline $\frac{33 \pi }{458}$ & $[458, 1]$, $[916, 1]$, $[916, 4]$ & $\frac{65 \pi }{902}$ & $[902, 3]$  \\ \hline $\frac{8 \pi }{111}$ & $[222, 5]$, $[444, 5]$, $[444, 17]$, $[666, 5]$, $[666, 15]$, $[888, 5]$, $[888, 37]$, $[888, 38]$, $[888, 39]$, $[888, 41]$ & $\frac{63 \pi }{874}$ & $[874, 3]$  \\ \hline $\frac{31 \pi }{430}$ & $[430, 3]$, $[860, 3]$, $[860, 10]$ & $\frac{23 \pi }{319}$ & $[638, 3]$  \\ \hline $\frac{61 \pi }{846}$ & $[846, 3]$ & $\frac{15 \pi }{208}$ & $[416, 6]$, $[416, 7]$, $[416, 8]$, $[832, 6]$, $[832, 7]$, $[832, 8]$, $[832, 64]$, $[832, 68]$, $[832, 531]$  \\ \hline $\frac{59 \pi }{818}$ & $[818, 1]$ & $\frac{22 \pi }{305}$ & $[610, 5]$  \\ \hline $\frac{29 \pi }{402}$ & $[402, 5]$, $[804, 5]$, $[804, 14]$ & $\frac{36 \pi }{499}$ & $[998, 1]$  \\ \hline $\frac{57 \pi }{790}$ & $[790, 3]$ & $\frac{7 \pi }{97}$ & $[194, 1]$, $[388, 1]$, $[388, 4]$, $[582, 4]$, $[582, 5]$, $[776, 1]$, $[776, 4]$, $[776, 5]$, $[776, 6]$, $[776, 8]$, $[970, 2]$, $[970, 3]$  \\ \hline $\frac{55 \pi }{762}$ & $[762, 5]$ & $\frac{34 \pi }{471}$ & $[942, 5]$  \\ \hline $\frac{27 \pi }{374}$ & $[374, 3]$, $[748, 3]$, $[748, 10]$ & $\frac{20 \pi }{277}$ & $[554, 1]$  \\ \hline $\frac{53 \pi }{734}$ & $[734, 1]$ & $\frac{33 \pi }{457}$ & $[914, 1]$  \\ \hline $\frac{13 \pi }{180}$ & $[360, 25]$, $[360, 27]$, $[360, 29]$, $[720, 65]$, $[720, 67]$, $[720, 68]$, $[720, 69]$, $[720, 71]$, $[720, 178]$ & $\frac{32 \pi }{443}$ & $[886, 1]$  \\ \hline $\frac{51 \pi }{706}$ & $[706, 1]$ & $\frac{19 \pi }{263}$ & $[526, 1]$  \\ \hline $\frac{25 \pi }{346}$ & $[346, 1]$, $[692, 1]$, $[692, 4]$ & $\frac{31 \pi }{429}$ & $[858, 11]$  \\ \hline $\frac{49 \pi }{678}$ & $[678, 3]$ & $\frac{6 \pi }{83}$ & $[166, 1]$, $[332, 1]$, $[332, 3]$, $[498, 2]$, $[498, 3]$, $[664, 1]$, $[664, 3]$, $[664, 4]$, $[664, 5]$, $[664, 7]$, $[830, 2]$, $[830, 3]$, $[996, 2]$, $[996, 3]$, $[996, 7]$, $[996, 9]$  \\ \hline $\frac{71 \pi }{982}$ & $[982, 1]$ & $\frac{47 \pi }{650}$ & $[650, 3]$  \\ \hline $\frac{35 \pi }{484}$ & $[968, 3]$, $[968, 5]$, $[968, 7]$ & $\frac{29 \pi }{401}$ & $[802, 1]$  \\ \hline $\frac{23 \pi }{318}$ & $[318, 3]$, $[636, 3]$, $[636, 11]$, $[954, 3]$, $[954, 7]$ & $\frac{17 \pi }{235}$ & $[470, 3]$, $[940, 3]$, $[940, 10]$  \\ \hline $\frac{45 \pi }{622}$ & $[622, 1]$ & $\frac{28 \pi }{387}$ & $[774, 5]$  \\ \hline $\frac{67 \pi }{926}$ & $[926, 1]$ & $\frac{11 \pi }{152}$ & $[304, 5]$, $[304, 6]$, $[304, 7]$, $[608, 5]$, $[608, 6]$, $[608, 7]$, $[608, 11]$, $[608, 25]$, $[608, 99]$, $[912, 60]$, $[912, 61]$, $[912, 62]$, $[912, 92]$, $[912, 93]$, $[912, 94]$  \\ \hline $\frac{65 \pi }{898}$ & $[898, 1]$ & $\frac{27 \pi }{373}$ & $[746, 1]$  \\ \hline $\frac{43 \pi }{594}$ & $[594, 3]$ & $\frac{16 \pi }{221}$ & $[442, 3]$, $[884, 3]$, $[884, 14]$  \\ \hline $\frac{21 \pi }{290}$ & $[290, 3]$, $[580, 3]$, $[580, 14]$, $[870, 3]$, $[870, 7]$ & $\frac{26 \pi }{359}$ & $[718, 1]$  \\ \hline $\frac{31 \pi }{428}$ & $[856, 3]$, $[856, 5]$, $[856, 7]$ & $\frac{36 \pi }{497}$ & $[994, 5]$  \\ \hline $\frac{41 \pi }{566}$ & $[566, 1]$ & $\frac{61 \pi }{842}$ & $[842, 1]$  \\ \hline $\frac{5 \pi }{69}$ & $[138, 3]$, $[276, 3]$, $[276, 9]$, $[414, 3]$, $[414, 7]$, $[552, 3]$, $[552, 23]$, $[552, 24]$, $[552, 25]$, $[552, 27]$, $[690, 5]$, $[690, 7]$, $[828, 3]$, $[828, 9]$, $[828, 13]$, $[828, 27]$, $[966, 9]$, $[966, 11]$ & $\frac{59 \pi }{814}$ & $[814, 3]$  \\ \hline $\frac{39 \pi }{538}$ & $[538, 1]$ & $\frac{34 \pi }{469}$ & $[938, 3]$  \\ \hline $\frac{29 \pi }{400}$ & $[800, 6]$, $[800, 7]$, $[800, 8]$ & $\frac{24 \pi }{331}$ & $[662, 1]$  \\ \hline $\frac{19 \pi }{262}$ & $[262, 1]$, $[524, 1]$, $[524, 3]$, $[786, 2]$, $[786, 3]$ & $\frac{33 \pi }{455}$ & $[910, 7]$  \\ \hline $\frac{14 \pi }{193}$ & $[386, 1]$, $[772, 1]$, $[772, 4]$ & $\frac{37 \pi }{510}$ & $[510, 7]$  \\ \hline $\frac{23 \pi }{317}$ & $[634, 1]$ & $\frac{55 \pi }{758}$ & $[758, 1]$  \\ \hline $\frac{32 \pi }{441}$ & $[882, 5]$ & $\frac{9 \pi }{124}$ & $[248, 3]$, $[248, 5]$, $[248, 7]$, $[496, 4]$, $[496, 5]$, $[496, 6]$, $[496, 7]$, $[496, 9]$, $[496, 30]$, $[744, 24]$, $[744, 26]$, $[744, 28]$, $[744, 34]$, $[744, 36]$, $[744, 38]$, $[992, 4]$, $[992, 5]$, $[992, 6]$, $[992, 7]$, $[992, 11]$, $[992, 18]$, $[992, 25]$, $[992, 96]$, $[992, 99]$  \\ \hline $\frac{71 \pi }{978}$ & $[978, 5]$ & $\frac{31 \pi }{427}$ & $[854, 3]$  \\ \hline $\frac{53 \pi }{730}$ & $[730, 3]$ & $\frac{22 \pi }{303}$ & $[606, 3]$  \\ \hline $\frac{35 \pi }{482}$ & $[482, 1]$, $[964, 1]$, $[964, 4]$ & $\frac{13 \pi }{179}$ & $[358, 1]$, $[716, 1]$, $[716, 3]$  \\ \hline $\frac{69 \pi }{950}$ & $[950, 3]$ & $\frac{30 \pi }{413}$ & $[826, 3]$  \\ \hline $\frac{17 \pi }{234}$ & $[234, 5]$, $[468, 5]$, $[468, 17]$, $[702, 5]$, $[702, 36]$, $[936, 5]$, $[936, 37]$, $[936, 38]$, $[936, 39]$, $[936, 41]$ & $\frac{21 \pi }{289}$ & $[578, 1]$  \\ \hline $\frac{67 \pi }{922}$ & $[922, 1]$ & $\frac{25 \pi }{344}$ & $[688, 5]$, $[688, 6]$, $[688, 7]$  \\ \hline $\frac{29 \pi }{399}$ & $[798, 23]$ & $\frac{33 \pi }{454}$ & $[454, 1]$, $[908, 1]$, $[908, 3]$  \\ \hline $\frac{49 \pi }{674}$ & $[674, 1]$ & $\frac{65 \pi }{894}$ & $[894, 3]$  \\ \hline $\frac{4 \pi }{55}$ & $[110, 5]$, $[220, 5]$, $[220, 14]$, $[330, 7]$, $[330, 11]$, $[440, 5]$, $[440, 34]$, $[440, 35]$, $[440, 36]$, $[440, 38]$, $[550, 5]$, $[550, 13]$, $[660, 7]$, $[660, 11]$, $[660, 35]$, $[660, 39]$, $[770, 9]$, $[770, 11]$, $[880, 5]$, $[880, 90]$, $[880, 91]$, $[880, 92]$, $[880, 93]$, $[880, 94]$, $[880, 96]$, $[880, 199]$, $[990, 7]$, $[990, 11]$, $[990, 25]$ & $\frac{63 \pi }{866}$ & $[866, 1]$  \\ \hline $\frac{47 \pi }{646}$ & $[646, 3]$ & $\frac{35 \pi }{481}$ & $[962, 3]$  \\ \hline $\frac{31 \pi }{426}$ & $[426, 3]$, $[852, 3]$, $[852, 9]$ & $\frac{27 \pi }{371}$ & $[742, 3]$  \\ \hline $\frac{23 \pi }{316}$ & $[632, 3]$, $[632, 5]$, $[632, 7]$ & $\frac{61 \pi }{838}$ & $[838, 1]$  \\ \hline $\frac{19 \pi }{261}$ & $[522, 3]$ & $\frac{34 \pi }{467}$ & $[934, 1]$  \\ \hline $\frac{15 \pi }{206}$ & $[206, 1]$, $[412, 1]$, $[412, 3]$, $[618, 4]$, $[618, 5]$, $[824, 1]$, $[824, 3]$, $[824, 4]$, $[824, 5]$, $[824, 7]$ & $\frac{26 \pi }{357}$ & $[714, 11]$  \\ \hline $\frac{59 \pi }{810}$ & $[810, 3]$ & $\frac{11 \pi }{151}$ & $[302, 1]$, $[604, 1]$, $[604, 3]$, $[906, 4]$, $[906, 5]$  \\ \hline $\frac{29 \pi }{398}$ & $[398, 1]$, $[796, 1]$, $[796, 3]$ & $\frac{18 \pi }{247}$ & $[494, 3]$, $[988, 3]$, $[988, 10]$  \\ \hline $\frac{43 \pi }{590}$ & $[590, 3]$ & $\frac{25 \pi }{343}$ & $[686, 1]$  \\ \hline $\frac{57 \pi }{782}$ & $[782, 3]$ & &  \\

\end{longtable}

\subsection{\label{subsec:lepton_quark_scan}Combined results for lepton and quark mixings}

\begin{table}[t!]
\centering
\begin{tabular}{|m{0.17\columnwidth}<{\centering}|m{0.06\columnwidth}<{\centering}|m{0.06\columnwidth}<{\centering}||m{0.17\columnwidth}<{\centering}|m{0.06\columnwidth}<{\centering}|m{0.06\columnwidth}<{\centering}|}

\hline
\hline
GAP-id & $\theta_{\nu}$ & $\theta_q$ & GAP-id & $\theta_{\nu}$ & $\theta_q$\\
\cline{1-3}\cline{4-6}
$~~[486, 61]_{\vartriangle}$ & $\pm\frac{\pi }{18}$ & $\frac{\pi }{18}$ & $[1452, 11]$ & $\pm\frac{2 \pi }{33}$ & $\frac{\pi }{22},\frac{\pi }{11}$\\ \cline{1-3}\cline{4-6}$~~[648, 259]^*_{\vartriangle'}$ & $\pm\frac{\pi }{18}$ & $\frac{\pi }{12}$ & $[1452, 23]$ & $\pm\frac{2 \pi }{33}$ & $\frac{\pi }{22},\frac{\pi }{11}$\\ \cline{1-3}\cline{4-6}$~[648, 260]^*$ & $\pm\frac{\pi }{18}$ & $\frac{\pi }{12}$ & $~~[1458, 659]_{\vartriangle'}$ & $\pm\frac{\pi }{18}$ & $\frac{\pi }{18}$\\ \cline{1-3}\cline{4-6}$~[648, 266]^*$ & $\pm\frac{\pi }{18}$ & $\frac{\pi }{12}$ & $[1458, 663]$ & $\pm\frac{\pi }{18}$ & $\frac{\pi }{18}$\\ \cline{1-3}\cline{4-6}$~~[726, 5]_{\vartriangle}$ & $\pm\frac{2 \pi }{33}$ & $\frac{\pi }{22},\frac{\pi }{11}$ & $[1458, 666]$ & $\pm\frac{\pi }{18}$ & $\frac{\pi }{18}$\\ \cline{1-3}\cline{4-6}$[972, 64]$ & $\pm\frac{\pi }{18}$ & $\frac{\pi }{18}$ & $[1458, 1371]$ & $\pm\frac{\pi }{18}$ & $\frac{\pi }{18}$\\ \cline{1-3}\cline{4-6}$[972, 245]$ & $\pm\frac{\pi }{18}$ & $\frac{\pi }{18}$ & $~~[1734, 5]_{\vartriangle}$ & $\pm\frac{\pi }{17}$ & $\frac{\pi }{17},\frac{3 \pi }{34}$\\ \cline{1-3}\cline{4-6}$~~[1176, 243]_{\vartriangle}$ & $\pm\frac{5 \pi }{84}$ & $\frac{\pi }{14},\frac{\pi }{28}$ & $[1944, 70]$ & $\pm\frac{\pi }{18}$ & $\frac{\pi }{18}$\\ \cline{1-3}\cline{4-6}$[1296, 688]$ & $\pm\frac{\pi }{18}$ & $\frac{\pi }{12}$ & $[1944, 544]$ & $\pm\frac{\pi }{18}$ & $\frac{\pi }{18}$\\ \cline{1-3}\cline{4-6}$[1296, 689]$ & $\pm\frac{\pi }{18}$ & $\frac{\pi }{12}$ & $~[1944, 832]^*$ & $\pm\frac{\pi }{18}$ & $\frac{\pi }{12}$\\ \cline{1-3}\cline{4-6}$[1296, 699]$ & $\pm\frac{\pi }{18}$ & $\frac{\pi }{12}$ & $~[1944, 833]^*$ & $\pm\frac{\pi }{18}$ & $\frac{\pi }{12}$\\ \cline{1-3}\cline{4-6}$~[1296, 1820]^*$ & $\pm\frac{\pi }{18}$ & $\frac{\pi }{12}$ & $~~[1944, 849]^*_{\vartriangle}$ & $\pm\frac{\pi }{18}$ & $\frac{\pi }{12},\frac{\pi }{18}$\\ \cline{1-3}\cline{4-6}$~[1296, 1821]^*$ & $\pm\frac{\pi }{18}$ & $\frac{\pi }{12}$ & $~[1944, 2363]^*$ & $\pm\frac{\pi }{18}$ & $\frac{\pi }{12}$\\ \cline{1-3}\cline{4-6}$~[1296, 1827]^*$ & $\pm\frac{\pi }{18}$ & $\frac{\pi }{12}$ &  &  &  \\ \hline \hline
\end{tabular}
\caption{\label{tab:lepton_quark_combined}Lepton flavor mixing and Cabibbo mixing angle predicted by finite discrete groups with order less than 2000, where both left-handed lepton and quark doublets are assumed to transform as irreducible three-dimensional representations of the flavor symmetry group. We require the lepton mixing angles to be compatible with experimental data at the $3\sigma$ level and the CKM entries should fulfill $0.1\leq |V_{us}|\leq0.3$ and $|V_{ub}|\leq |V_{cb}|<|V_{us}|$.  The superscript * indicates that the corresponding group has Klein subgroups and it allows  neutrinos to be Majorana particles. The subscripts $\vartriangle$ and $\vartriangle^{\prime}$ denote the groups that belong to the $D_{n,n}^{(0)}\cong\Delta(6n^2)$ and $D_{9n^{\prime},3n^{\prime}}^{(1)}\cong(Z_{9n^{\prime}}\times Z_{3n^{\prime}})\rtimes S_3$ group series, respectively.}
\end{table}

\begin{table}[t!]
\centering
\begin{tabular}{|m{0.15\columnwidth}<{\centering}|m{0.06\columnwidth}<{\centering}|m{0.24\columnwidth}<{\centering}||m{0.06\columnwidth}<{\centering}|m{0.35\columnwidth}<{\centering}|}

\hline
\hline
GAP-id & $\theta_{\nu}$  & $\mathcal{G}'_{3}$ & $\theta_q$ & $\mathcal{G}'_{2}$ \\
\hline
$[486, 19]$ & $\pm \frac{\pi }{18}$ & $[162, 10]$ & $\frac{\pi }{18}$ & $[18, 1]$,~$[54, 3]$  \\ \hline$[486, 21]$ & $\pm \frac{\pi }{18}$ & $[162, 12]$ & $\frac{\pi }{18}$ & $[18, 1]$,~$[54, 3]$  \\ \hline$[486, 23]$ & $\pm \frac{\pi }{18}$ & $~~[162, 14]_{\vartriangle'}$ & $\frac{\pi }{18}$ & $[18, 1]$,~$[54, 3]$  \\ \hline$[648, 126]$ & $\pm \frac{\pi }{18}$ & $[162, 10]$,~$[324, 68]$ & $\frac{\pi }{12}$ & $[24, 4]$,~$[72, 26]$  \\ \hline$[648, 129]$ & $\pm \frac{\pi }{18}$ & $[162, 10]$,~$[324, 68]$ & $\frac{\pi }{12}$ & $[24, 6]$,~$[72, 28]$  \\ \hline$[648, 132]$ & $\pm \frac{\pi }{18}$ & $[162, 12]$,~$[324, 70]$ & $\frac{\pi }{12}$ & $[24, 4]$,~$[72, 26]$  \\ \hline$[648, 135]$ & $\pm \frac{\pi }{18}$ & $[162, 12]$,~$[324, 70]$ & $\frac{\pi }{12}$ & $[24, 6]$,~$[72, 28]$  \\ \hline$[648, 138]$ & $\pm \frac{\pi }{18}$ & $~~[162, 14]_{\vartriangle'}$,~$[324, 72]$ & $\frac{\pi }{12}$ & $[24, 4]$,~$[72, 26]$  \\ \hline$[648, 141]$ & $\pm \frac{\pi }{18}$ & $~~[162, 14]_{\vartriangle'}$,~$[324, 72]$ & $\frac{\pi }{12}$ & $[24, 6]$,~$[72, 28]$  \\ \hline$[648, 160]$ & $\pm \frac{\pi }{18}$ & $[162, 10]$,~$[324, 68]$ & $\frac{\pi }{12}$ & $[24, 8]$,~$[72, 30]$  \\ \hline$[648, 164]$ & $\pm \frac{\pi }{18}$ & $[162, 12]$,~$[324, 70]$ & $\frac{\pi }{12}$ & $[24, 8]$,~$[72, 30]$  \\ \hline$[648, 168]$ & $\pm \frac{\pi }{18}$ & $~~[162, 14]_{\vartriangle'}$,~$[324, 72]$ & $\frac{\pi }{12}$ & $[24, 8]$,~$[72, 30]$  \\ \hline$[810, 34]$ & $\pm \frac{\pi }{18}$ & $[162, 10]$ & $\frac{\pi }{15},\frac{\pi }{30}$ & $[30, 3]$,~$[90, 7]$  \\ \hline$[810, 36]$ & $\pm \frac{\pi }{18}$ & $[162, 12]$ & $\frac{\pi }{15},\frac{\pi }{30}$ & $[30, 3]$,~$[90, 7]$  \\ \hline$[810, 38]$ & $\pm \frac{\pi }{18}$ & $~~[162, 14]_{\vartriangle'}$ & $\frac{\pi }{15},\frac{\pi }{30}$ & $[30, 3]$,~$[90, 7]$  \\ \hline$[972, 22]$ & $\pm \frac{\pi }{18}$ & $[162, 10]$,~$[324, 13]$ & $\frac{\pi }{18}$ & $[18, 1]$,~$[36, 1]$,~$[54, 3]$,~$[108, 6]$  \\ \hline$[972, 24]$ & $\pm \frac{\pi }{18}$ & $[162, 12]$,~$[324, 15]$ & $\frac{\pi }{18}$ & $[18, 1]$,~$[36, 1]$,~$[54, 3]$,~$[108, 6]$  \\ \hline$[972, 26]$ & $\pm \frac{\pi }{18}$ & $~~[162, 14]_{\vartriangle'}$,~$[324, 17]$ & $\frac{\pi }{18}$ & $[18, 1]$,~$[36, 1]$,~$[54, 3]$,~$[108, 6]$  \\ \hline$[972, 203]$ & $\pm \frac{\pi }{18}$ & $[162, 10]$,~$[324, 68]$ & $\frac{\pi }{18}$ & $[18, 1]$,~$[36, 4]$,~$[54, 3]$,~$[108, 23]$  \\ \hline$[972, 205]$ & $\pm \frac{\pi }{18}$ & $[162, 12]$,~$[324, 70]$ & $\frac{\pi }{18}$ & $[18, 1]$,~$[36, 4]$,~$[54, 3]$,~$[108, 23]$  \\ \hline$[972, 207]$ & $\pm \frac{\pi }{18}$ & $~~[162, 14]_{\vartriangle'}$,~$[324, 72]$ & $\frac{\pi }{18}$ & $[18, 1]$,~$[36, 4]$,~$[54, 3]$,~$[108, 23]$  \\ \hline\hline
\end{tabular}
\caption{\label{tab:lepton_quark_combined_v2}Lepton flavor mixing and Cabibbo mixing angle predicted by finite discrete groups with order less than 1000, where the left-handed lepton and quark fields are assigned to triplet and doublet plus singlet of the flavor symmetry group, respectively. $\mathcal{G}^{\prime}_3$ and $\mathcal{G}^{\prime}_2$ refer to the subgroups generated by the three-dimensional and two-dimensional representations, respectively, to which lepton and quark fields are assigned. We require the lepton mixing angles should be compatible with experimental data at the $3\sigma$ level and the CKM matrix elements should fulfill $0.1\leq |V_{us}|\leq0.3$. The subscripts $\vartriangle^{\prime}$ denote the groups that belong to $D_{9n^{\prime},3n^{\prime}}^{(1)}\cong(Z_{9n^{\prime}}\times Z_{3n^{\prime}})\rtimes S_3$ group series.}
\end{table}

If flavor symmetry acts on both lepton and quark sectors, from Table~\ref{tab:lepton_mixing1} and \ref{tab:quark_mixing}, we know that some finite groups can generate a lepton mixing pattern in the $3\sigma$ region together with an acceptable Cabibbo angle. These interesting groups and the corresponding predictions for $\theta_{\nu}$ and $\theta_{q}$ are presented in Table~\ref{tab:lepton_quark_combined}. The associated residual symmetries $\mathcal{G}_{\nu, l, U, D}$ can be read from Tables~\ref{tab:lepton_mixing1} and \ref{tab:quark_mixing} and our Web page~\cite{webdata}. We see that the smallest group is $\Delta(6\times9^2)$, which gives rise to the viable trimaximal lepton mixing with $\theta_{\nu}=\pm\pi/18$ and the Cabibbo mixing angle $\theta_{q}=\pi/18$. The neutrino should be Dirac particles in this case, since $\Delta(6\times9^2)$ does not possess Klein subgroups. In addition, the predicted Cabibbo angle is a little smaller than the experimental best fit value. However, higher order corrections are typically expected to be of order $\theta_q^2\sim0.05$ in concrete models~\cite{Altarelli:2010gt,King:2013eh,King:2014nza}. Such small corrections could pull the Cabibbo angle into the experimentally preferred ranges, and the correct order of magnitude of the CKM matrix elements $V_{cb}$ and $V_{ts}$ can be achieved.

The second-smallest groups are $[648, 259]$, $[648, 260]$, and $[648, 266]$, where the first one is exactly the type D group $D_{9n^{\prime},3n^{\prime}}^{(1)}\cong(Z_{9n^{\prime}}\times Z_{3n^{\prime}})\rtimes S_3$ with $n^{\prime}=2$. All these three groups lead to $\theta_{\nu}=\pm\pi/18$ and $\theta_q=\pi/12$. In contrast with the smallest group $\Delta(6\times9^2)$, neutrinos can be either Dirac or Majorana particles in this case, and the Cabibbo angle is predicted to be slightly larger than the measured value. This could easily be reconciled with the experimental data in models with small corrections. On the other hand, if one would like to obtain a Cabibbo angle $\theta_q=\pi/14$, which is already consistent with experimental results at leading order, then the group $\Delta(6\times14^2)$ with identification number $[1176, 243]$ is a promising flavor symmetry. The PMNS matrix would be of trimaximal form with $\theta_{\nu}=\pm5\pi/84$, and all three lepton mixing angles are in the experimentally preferred range for Dirac neutrinos. However, subleading corrections to $\theta_{13}$ are necessary in order to achieve agreement with the data if neutrinos are Majorana particles, as shown in Sec.~\ref{subsec:numerical_results}.

\begin{figure}[t!]
\centering
\includegraphics[width=0.99\textwidth]{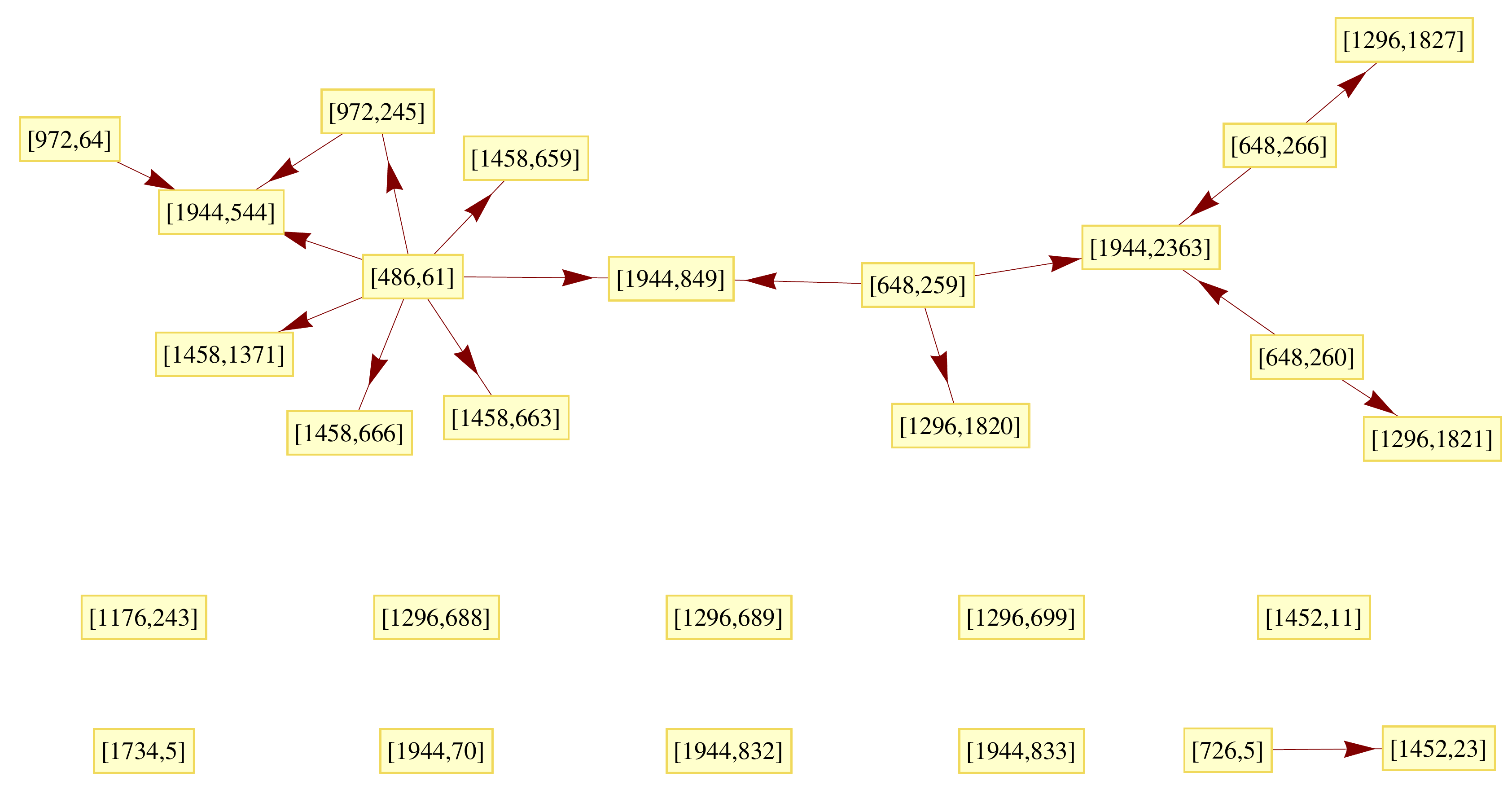}
\caption{\label{fig:group_chains_quark_lepton}
The groups generated by the residual symmetries $\left\{\mathcal{G}_{\nu}, \mathcal{G}_{l}, \mathcal{G}_{U}, \mathcal{G}_{D}\right\}$ for the discrete flavor symmetry groups listed in table~\ref{tab:lepton_quark_combined}. If a subgroup is generated, it is linked to the parent group by an arrowed line and the arrow points to the parent group.}
\end{figure}

From Table~\ref{tab:lepton_quark_combined}, we see that the values of the parameters $\theta_{\nu}$ and $\theta_{q}$ are tightly constrained and limited, and they can only take the eight possible values: $\left(\theta_{\nu}, \theta_{q}\right)=$$\left(\pm\pi/18, \pi/18\right)$, $\left(\pm\pi/18, \pi/12\right)$, $\left(\pm2\pi/33, \pi/22\right)$, $\left(\pm2\pi/33, \pi/11\right)$, $\left(\pm5\pi/84, \pi/14\right)$, $\left(\pm5\pi/84, \pi/28\right)$, $\left(\pm\pi/17, \pi/17\right)$ and $\left(\pm\pi/17, 3\pi/34\right)$. In particular, a multitude of different groups gives rise to the same values $\theta_{\nu}$ and $\theta_{q}$. For instance, all the groups $[486, 61]$, $[972, 64]$, $[972, 245]$, $[1458, 659]$, $[1458, 663]$, $[1458, 666]$, $[1458, 1371]$, $[1944, 70]$, $[1944, 544]$, and $[1944, 849]$ predict $\left(\theta_{\nu}, \theta_{q}\right)=$$\left(\pm\pi/18, \pi/18\right)$.
For each group, generally many distinct residual symmetries $\mathcal{G}_{\nu}$, $\mathcal{G}_{l}$, $\mathcal{G}_{U}$, and $\mathcal{G}_{D}$ can lead to the same prediction for $\theta_{\nu}$ and $\theta_q$. The group structures generated by every set of $\left\{\mathcal{G}_{\nu}, \mathcal{G}_{l}, \mathcal{G}_{U}, \mathcal{G}_{D}\right\}$ are studied, we find the original group can always be generated, and certain subgroups can also be generated in some cases. All the groups generated by the residual subgroups $\left\{\mathcal{G}_{\nu}, \mathcal{G}_{l}, \mathcal{G}_{U}, \mathcal{G}_{D}\right\}$ are plotted in Fig.~\ref{fig:group_chains_quark_lepton}. Notice that no new groups besides those listed in Table~\ref{tab:lepton_quark_combined} are found. A group and its subgroups are linked by arrowed lines and the arrow points to the parent group. As an example, from this figure, we can read that the many residual subgroups of $[972, 245]$ for $\left(\theta_{\nu}, \theta_{q}\right)=$$\left(\pm\pi/18, \pi/18\right)$ give rise to not only the whole group $[972, 245]$ but also its subgroup $[486, 61]$. This is exactly the reason why $[972, 245]$ and $[486, 61]$ lead to the same predictions for $\theta_{\nu}$ and $\theta_{q}$. The same values of $\theta_{\nu}$ and $\theta_{q}$ obtained from other different groups can easily be understood in this way. From model building perspective, $[486, 61]$ is preferred over $[972, 245]$, since $[486, 61]$ is smaller in size than $[972, 245]$ and consequently it is less cumbersome. However, the larger group $[972, 245]$ provides more possible symmetry breaking patterns such that it should be more flexible to construct models.

In view of the above appealing model-independent results, it would be interesting to construct a model of quark and lepton mixing based on the flavor symmetry $\Delta(6\times9^2)$, $[648, 259]$, $[648, 260]$, $[648, 266]$, $\Delta(6\times14^2)$, or other groups listed in Table~\ref{tab:lepton_quark_combined}. For instance, guided by the remnant symmetries in Tables~\ref{tab:lepton_mixing1} and \ref{tab:quark_mixing}, we could introduce auxiliary cyclic symmetries, some flavon fields, and driving fields in an appropriate way within the context of supersymmetry~\cite{Altarelli:2010gt,King:2013eh,King:2014nza}, such that the postulated flavor symmetry is spontaneously broken down to the residual subgroups $\mathcal{G}_{\nu, l}$ in the lepton sector and $\mathcal{G}_{U, D}$ in the quark sector at leading order. Then the desired lepton and quark mixing patterns in Table~\ref{tab:lepton_quark_combined} would be produced at leading order, and small corrections could appear if subleading order contributions are included.

On the other hand, if the left-handed lepton and quark fields are assigned to triplet and doublet plus singlet representations of the flavor symmetry group, respectively, the experimentally acceptable lepton and quark mixing patterns can be produced as well, as shown in Table~\ref{tab:lepton_quark_combined_v2}. Here only finite groups with order less than 1000 have been exploited. We see that both the three-dimensional and the two-dimensional representations to which the lepton and quark fields are assigned only generate some subgroups of the flavor symmetry group $\mathcal{G_F}$ since both of them are not faithful representations of $\mathcal{G_F}$. It is remarkable that the viable trimaximal lepton mixing with $\theta_{\nu}=\pm\pi/18$ and the Cabibbo mixing angle $\theta_{q}=\pi/18$, $\pi/12$, $\pi/15$ can be produced. The smallest groups $\left[496, 19\right]$, $\left[496, 21\right]$, $\left[496, 23\right]$ as well as the second-smallest groups $\left[648, 126\right]$, $\left[648, 129\right]$, $\left[648, 132\right]$, $\left[648, 135\right]$, $\left[648, 138\right]$, $\left[648, 141\right]$, $\left[648, 160\right]$, $\left[648, 164\right]$, and $\left[648, 168\right]$ are interesting flavor symmetry groups that predict acceptable lepton and quark flavor mixing at leading order.

\section{\label{sec:conclusion}Conclusions}

Flavor symmetry and its spontaneous breaking is an interesting approach to explain the observed lepton or even quark flavor mixing. It has been shown that the mixing patterns that can be derived from the discrete flavor symmetry group are rather restricted for neutrino being Majorana particles, only the trimaximal mixing pattern can be in agreement with experimental data, and Dirac $CP$ phase is trivial~\cite{Fonseca:2014koa}. In addition, it is found that the flavor symmetry group should be of order 648 or larger in order to obtain the experimentally preferred mixing angles~\cite{Holthausen:2012wt,Fonseca:2014koa}. Notice that the Majorana neutrino is only \textit{a priori} assumption; so far we have no sufficient evidence that neutrinos are Majorana particles. In this work, we discussed whether new mixing patterns can be obtained and whether the order of the required flavor symmetry groups can be degraded under the assumption that neutrinos are Dirac particles

All the finite subgroups of $SU(3)$ have been identified by three mathematicians: Miller, Blichfeldt and Dickson~\cite{Miller_book}. They can be classified into five categories: type A, type B, type C, type D, and type E. Since both type A and type B groups do not have an irreducible faithful three-dimensional representation, they are not suitable as flavor symmetry groups. We perform a systematical and comprehensive analysis of lepton mixing patterns that can be obtained from type C and type D groups. We find that type D groups can give rise to lepton mixing patterns compatible with the experimental data at the $3\sigma$ level while type C groups cannot. The lepton mixing matrix is still found to be of the trimaximal form, and the Dirac phase remains trivial. As a consequence, the resulting PMNS matrix shown in Eq.~\eqref{eq:couple7_gen} depends on a single real parameter $\theta$, and the solar and atmospheric mixing angles are predicted to be $\sin^2\theta_{12}\simeq0.341$, $\sin^2\theta_{23}\simeq0.395$ or 0.605. These decisive predictions can be tested by precisely measuring $\theta_{12}$ and $\theta_{23}$ at future neutrino oscillation experiments such as JUNO~\cite{JUNO}, LBNE~\cite{Adams:2013qkq}, and LBNO~\cite{::2013kaa}. Moreover, forthcoming long baseline neutrino
experiments LBNE~\cite{Adams:2013qkq}, LBNO~\cite{::2013kaa}, and
Hyper-Kamiokande~\cite{Kearns:2013lea} can measure the Dirac $CP$ phase with a certain precision. If leptonic CP violation is discovered, the present paradigm would be excluded.

The type D subgroup of $SU(3)$ has two independent series, $D_{n,n}^{(0)}\cong \Delta(6n^2)$ and $D_{9n^{\prime}, 3n^{\prime}}^{(1)}\cong(Z_{9n^{\prime}}\times Z_{3n^{\prime}})\rtimes S_3$.
The parameter $\theta$ can only take a series of discrete values depending on the values of $n$ and $n^{\prime}$.
The analytical expressions of $\theta$ are presented for both Dirac and Majorana neutrinos. Note that Majorana neutrinos require the flavor symmetry group contains a Klein group, such that $n$ and $n^{\prime}$ should be even. We find that only half of the $\theta$ values allowed for Dirac neutrinos can be achievable in the case of Majorana neutrinos. Moreover, we present the predictions for the mixing angles as a function of $n$ and $n^{\prime}$. It is amazing that the $D_{9n^{\prime}, 3n^{\prime}}^{(1)}$ group for any $n^{\prime}$ allows all three mixing angles to be within the experimental $3\sigma$ ranges, and the smallest one is $D_{9,3}^{(1)}\cong(Z_{9}\times Z_{3})\rtimes S_3$ with order 162. In contrast, for the $\Delta(6n^2)$ group series, $\Delta(6\times9^2)$ with order 486 is the smallest group that can accommodate the experimental data.

We further extend the flavor symmetry paradigm to the quark sector. Similar to lepton flavor mixing, the CKM matrix arises from the mismatched remnant symmetries $\mathcal{G}_{U}$ and $\mathcal{G}_D$ to which a flavor symmetry group is broken in the up type and down type quark sectors. It is a pity that no $SU(3)$ subgroups can predict the hierarchical mixing pattern among the three quarks although the three generations of the left-handed quark doublets are embedded into an irreducible three-dimensional representation of the flavor symmetry group. However, we can obtain a good leading order mixing pattern where the Cabibbo mixing angle is generated. For instance, both groups $\Delta(6\times7^2)$ and $D_{9n^{\prime},3n^{\prime}}^{(1)}\cong(Z_{9n^{\prime}}\times Z_{3n^{\prime}})\rtimes S_3$  with $n^{\prime}=7$ can predict the Cabibbo angle to fulfill $|V_{us}|=\sin\frac{\pi}{14}\simeq0.2225$, which is quite close to the best fit value. In addition, the group $\Delta(6\times14^2)$ can generate both acceptable Cabibbo angle $|V_{us}|=\sin\frac{\pi}{14}$ and lepton mixing angles compatible with the data.

Usually the flavor symmetry is taken to be a subgroup of $U(3)$ in practice, although generally it is not necessary. As has been shown in Ref.~\cite{Grimus:2011fk}, the flavor symmetry $\mathcal{G_F}$ can possibly be not an $U(3)$ subgroup. However, none of the three-dimensional representations of $\mathcal{G_F}$ would be faithful in this case. The same analysis procedures presented in Secs.~\ref{sec:mixing_from_SU(3)}, \ref{sec:mixing_from_Sigma} and \ref{sec:scan_groups} can be performed, and then essentially the mixing patterns originate from the subgroups of $\mathcal{G_F}$ that are generated through the triplets. Notice that the groups generated by the triplet representation of $\mathcal{G_F}$ are $U(3)$ subgroups. It is sufficient to take the flavor symmetry group to be finite subgroups of $U(3)$, and no additional mixing patterns would be obtained even if we consider the most general flavor symmetry groups of finite order. From the model building perspective, at present we still do not know the advantage of flavor symmetry that is not $U(3)$ subgroups.

Although the structures of the $SU(3)$ subgroups have been clearly solved in mathematics, a complete classification of all $U(3)$ subgroups is not available so far. As a consequence, it seems unfeasible to analytically discuss all the possible flavor mixing patterns arising from the $U(3)$ subgroups, as we have done for the $SU(3)$ subgroups. Given the fact that a $U(3)$ subgroup series $\Sigma(3N^3)$ is known, the mixing patterns that can be derived from $\Sigma(3N^3)$ are studied. However, no mixing pattern in the experimentally preferred $3\sigma$ range is found if lepton mixing is completely determined by residual symmetries that are subgroups of $\Sigma(3N^3)$. Furthermore, we perform a systematical and exhaustive scan over the finite discrete groups up to order 2000 with the help of the computer algebra system \texttt{GAP}. Since there are a huge number of groups with order 1536, the groups of order 1536 are treated under our conjecture that both finite nilpotent groups and the groups with normal Sylow 3-subgroup have no faithful three-dimensional irreducible representations. This conjecture has been checked to hold true for many finite groups, although we cannot prove it now. All possible residual subgroups and the corresponding mixing matrices for each group have been worked out. In particular, the mixing patterns that can be obtained from the $SU(3)$ subgroups of type E are studied. The abundant results of our analysis are available at the Web site~\cite{webdata}. We find that the mixing pattern in accordance with experimental data is still of the trimaximal form even if neutrinos are Dirac particles, and all mixing patterns with $\chi^2\leq100$ predict a trivial Dirac phase. A nontrivial $\delta_{CP}$ can only arise from patterns that do not accommodate the data well.

We find that 90 groups with order less than 2000 can generate the mixing angles compatible with experimental data at the $3\sigma$ level, if neutrinos are Dirac particles. Nevertheless, only 10 groups can give rise to viable lepton mixing for Majorana neutrinos. The smallest groups that can predict the leptonic mixing angles within the $3\sigma$ region are $[162,10]$, $[162,12]$, $[162,14]$ and $[648, 259]$, $[648, 260]$, $[648, 266]$ for Dirac and Majorana neutrinos, respectively, where $[162,14]$ and $[648,259]$ are $SU(3)$ subgroups of type $D_{9n^{\prime},3n^{\prime}}^{(1)}$ with $n^{\prime}=1, 2$, and the others are $U(3)$ but not $SU(3)$ subgroups. Therefore the order of the flavor symmetry group decreases 4 times by relaxing the condition of Majorana neutrinos, although no new mixing pattern distinct from trimaximal mixing is obtained. These three groups of order 162 and the associated symmetry breaking patterns provide new starting points for constructing neutrino mass models.

Then we investigate whether it is also possible to derive quark mixing in the same fashion. The left-handed quark fields are assumed to transform as a triplet of $\mathcal{G_F}$. We impose quite loose constraints on the CKM matrix elements $0.1\leq |V_{us}|\leq0.3$ and $|V_{ub}|\leq |V_{cb}|<|V_{us}|$. To our surprise, only the Cabibbo mixing between the first two generations of quarks is generated and the other two small mixing angles are predicted to be vanishing. We confirm that the group $\Delta(6\times7^2)$ can generate the Cabibbo angle $\theta_q=\pi/14$, which is close to the best fit value, and the smaller group $[216, 95]$ can give $\theta_q=\pi/12$. Moreover, the doublet plus singlet assignment for the left-handed quark fields is investigated. We perform a scan of all finite groups with faithful two-dimensional irreducible representation up to order 1000. The experimentally acceptable Cabibbo mixing angles can easily be generated. For instance, the dihedral group $D_{14}$ can give rise to a Cabibbo angle $\theta_{q}=\pi/14$, which is quite close to the measured value.

Furthermore, we have considered the scenario in which the flavor symmetry acts on both lepton and quark sectors. If both the left-handed lepton and quark fields transform as an irreducible three-dimensional representation of the flavor symmetry group, we see that the smallest group is $\Delta(6\times9^2)$, which gives rise to the phenomenologically viable trimaximal lepton mixing with $\theta_{\nu}=\pm\pi/18$ and the Cabibbo mixing angle $\theta_{q}=\pi/18$. The second-smallest groups are $[648, 259]$, $[648, 260]$, and $[648, 266]$, which predict $\theta_{\nu}=\pm\pi/18$ and $\theta_q=\pi/12$. In both cases, the predicted Cabibbo angles slightly deviate from the experimental value. However, this insignificant discrepancy is expected to be eliminated by higher order corrections in a concrete model. What is more, a larger group $\Delta(6\times14^2)$ can produce both experimentally favored lepton mixing angles and acceptable Cabibbo angle $\theta_{q}=\pi/14$ at leading order. Therefore $\Delta(6\times9^2)$, $[648, 259]$, $[648, 260]$, $[648, 266]$, $\Delta(6\times14^2)$, and other groups in Table~\ref{tab:lepton_quark_combined} provide intriguing opportunity for constructing a complete model of quark and lepton, where the observed quark and lepton mixing angles can be naturally explained. On the other hand, if the left-handed lepton and quark fields are assigned to triplet and doublet plus singlet, respectively, the trimaximal lepton mixing with $\theta_{\nu}=\pm\pi/18$ together with the Cabibbo angle $\theta_{q}=\pi/18$ can also be generated by the flavor symmetry groups $\left[496, 19\right]$, $\left[496, 21\right]$, and $\left[496, 23\right]$.

Scrutinizing all our results~\cite{webdata} carefully, we find that the flavor mixing patterns arising from finite discrete groups seem quite restricted and limited even if neutrinos are Dirac particles. It is interesting to perform a complete classification of all the mixing patterns achievable in this setup, and such an investigation is left for future studies.

\section*{Acknowledgements}

We are grateful to Peng Long for his participation in the early stage of this work. We acknowledge Martin Holthausen for helpful discussions about the computer program \texttt{GAP} during FLASY2013 in Niigata, Japan. This work is supported by the National Natural Science Foundation of China under Grants No. 11275188 and No. 11179007.

\newpage

\appendix

\section{\label{sec:Appd_unitarize}Converting a non-unitary representation into a unitary representation}

Since \texttt{GAP} employs only rational and cyclotomic numbers, the irreducible representations given in \texttt{GAP} may not be unitary. We have to obtain the corresponding unitary representations before we proceed to determine the mixing matrix from the postulated residual flavor symmetries. We can construct the unitary representations from the nonunitary \texttt{GAP} representations in the following way. Let $\rho$ be a linear representation of a finite group $\mathcal{G}$. First, we define
\begin{equation}
P\equiv\sum_{g\in{G}}\rho^{\dagger}(g)\rho(g)\,.
\end{equation}
Obviously $P$ is a Hermitian matrix. According to the rearrangement theorem, we have
\begin{equation}
\rho^{\dagger}(h)P\rho(h)=\sum_{g\in{G}}\rho^{\dagger}(gh)\rho(gh)=P\,.
\end{equation}
Following the standard procedures in linear algebra, the matrix $P$ can be diagonalized as
\begin{equation}
P=UDU^{-1}\,,
\end{equation}
where $U$ is a unitary matrix and it is composed of the normalized eigenvectors of $P$. $D$ is diagonal with entries being the eigenvalues of $P$. Furthermore, we can show that the diagonal elements of $D$ fulfill
\begin{equation}
D_{kk}=\sum_{g\in{G}}\sum_{j}\big(\rho^{\dagger}_{U}(g)\big)_{kj}\big(\rho_{U}(g)\big)_{jk}=\sum_{g\in{G}}\sum_{j}\big|\big(\rho_{U}(g)\big)_{jk}\big|^2\geq \big|\big(\rho_{U}(1)\big)_{kk}\big|^2=1\,,
\end{equation}
where $\rho_{U}(g)=U^{-1}\rho(g)U$, and therefore $\rho_{U}$ is a representation of $\mathcal{G}$ equivalent to $\rho$. The matrix $P$ can be decomposed into the square of a Hermitian matrix $S$, i.e.,
\begin{equation}
 P=S^2,\quad S=UD^{1/2}U^{-1},\quad S=S^{\dagger}\,,
\end{equation}
where $D^{1/2}\equiv\text{diag}\left(\sqrt{D_{11}}\,, \sqrt{D_{22}}\,, \ldots\right)$. Then we can straightforwardly check that
\begin{equation}
\rho^{\prime}(h)=S\rho(h)S^{-1}
\end{equation}
is exactly the desired unitary representation, because it satisfies
\begin{equation}
\rho^{\prime\dagger}(h)\rho^{\prime}(h)=S^{-1}\left(\rho^{\dagger}(h)SS\rho(h)\right)S^{-1}=S^{-1}S^2S^{-1}=1\,.
\end{equation}

\end{document}